\documentclass[fleqn,usenatbib]{mnras}

\usepackage[T1]{fontenc}
\usepackage{ae,aecompl}
\usepackage[flushleft]{threeparttable}

\usepackage{graphicx}
\usepackage{amsmath,mathrsfs}
\usepackage{amssymb}
\usepackage{newtxtext}
\usepackage{mathtools}
\usepackage{pdflscape}	
\usepackage{subcaption}
\usepackage{booktabs}
\usepackage[dvipsnames]{xcolor}

\usepackage{rotating}
\usepackage{tabularx}
\usepackage[labelfont=bf]{caption} 

\usepackage{ragged2e}
\usepackage{lipsum}

\newcommand{\numpsr}{300}
\newcommand{\numcp}{280}
\newcommand{\numred}{110}
\newcommand{\numredcanon}{112}
\newcommand{\numwtn}{168}
\newcommand{\numfddot}{one}
\newcommand{\sigmaRN}{\sigma_{\mathrm{RN}}}
\newcommand{\sigmaRNi}{\sigma_{\mathrm{RN},i}}
\newcommand{\chiRN}{\chi_{\mathrm{RN}}}
\newcommand{\chiRNi}{\chi_{\mathrm{RN},i}}

\usepackage{enumitem}
\title[The UTMOST pulsar timing programme II]{The UTMOST pulsar timing programme II: Timing noise across the pulsar population}

\author[M. E. Lower et al.]{\parbox{\textwidth}{M. E. Lower,$^{1,2}$\thanks{E-mail: mlower@swin.edu.au}
M. Bailes,$^{1,3}$
R. M. Shannon,$^{1,3}$
S. Johnston,$^{2}$
C. Flynn,$^{1,4}$
S. Os{\l}owski,$^{1}$
V. Gupta,$^{1}$
W. Farah,$^{1}$
T. Bateman,$^{5}$
A. J. Green,$^{5}$
R. Hunstead,$^{5}$
A. Jameson,$^{1,4}$\\
F. Jankowski,$^{6}$
A. Parthasarathy,$^{1,2}$
D. C. Price,$^{1, 7}$
A. Sutherland$,^{5}$ 
D. Temby,$^{5}$
and \\
V. Venkatraman Krishnan$^{8,1,3}$
}
\\ \\
$^{1}$Centre for Astrophysics and Supercomputing, Swinburne University of Technology, PO Box 218, Hawthorn, VIC 3122, Australia\\
$^{2}$CSIRO Astronomy and Space Science, Australia Telescope National Facility, Epping, NSW 1710, Australia\\
$^{3}$OzGrav: The ARC Centre of Excellence for Gravitational-wave Discovery, Hawthorn VIC 3122, Australia\\
$^{4}$ARC Centre of Excellence for All-sky Astrophysics (CAASTRO)\\
$^{5}$Sydney Institute for Astronomy (SIfA), School of Physics, The University of Sydney, NSW 2006, Australia\\
$^{6}$Jodrell Bank Centre for Astrophysics, Department of Physics and Astronomy, The University of Manchester, Manchester M13 9PL, UK\\
$^{7}$Department of Astronomy, University of California Berkeley, 501 Campbell Hall, Berkeley CA 94720\\
$^{8}$Max-Planck-Institut f{\"u}r Radioastronomie, Auf dem H{\"u}gel 69, D-53121 Bonn, Germany
}

\date{Accepted XXX. Received YYY; in original form ZZZ}

\begin{document}
\label{firstpage}
\pagerange{\pageref{firstpage}--\pageref{lastpage}}
\maketitle

\begin{abstract}
\begin{raggedright}
While pulsars possess exceptional rotational stability, large scale timing studies have revealed at least two distinct types of irregularities in their rotation: red timing noise and glitches.
Using modern Bayesian techniques, we investigated the timing noise properties of {\numpsr} bright southern-sky radio pulsars that have been observed over 1.0-4.8\,years by the upgraded Molonglo Observatory Synthesis Telescope (MOST).
We reanalysed the spin and spin-down changes associated with nine previously reported pulsar glitches, report the discovery of three new glitches and four unusual glitch-like events in the rotational evolution of PSR J1825$-$0935.
We develop a refined Bayesian framework for determining how red noise strength scales with pulsar spin frequency ($\nu$) and spin-down frequency ($\dot{\nu}$), which we apply to a sample of $\numcp$ non-recycled pulsars. With this new method and a simple power-law scaling relation, we show that red noise strength scales across the non-recycled pulsar population as $\nu^{a} |\dot{\nu}|^{b}$, where $a = -0.84^{+0.47}_{-0.49}$ and $b = 0.97^{+0.16}_{-0.19}$. This method can be easily adapted to utilise more complex, astrophysically motivated red noise models.
Lastly, we highlight our timing of the double neutron star PSR J0737$-$3039, and the rediscovery of a bright radio pulsar originally found during the first Molonglo pulsar surveys with an incorrectly catalogued position.
\end{raggedright}
\end{abstract}



\section{Introduction}

The pulsar timing programme of the UTMOST\footnote{Not an acronym.} project~\citep{Bailes2017} monitors more than 400 pulsars using the upgraded Molonglo Observatory Synthesis telescope. This programme runs in parallel with searches for undiscovered pulsars and single pulses from rotating radio transients (RRATs), and Fast Radio Bursts (FRBs). These searches have already led to the discovery of thirteen FRBs~\citep{Caleb2017, Farah2018a, Farah2019, Gupta2019a, Gupta2019b, Gupta2019c, Gupta2019d} and the intermittent pulsar candidate PSR J1659$-$54~\citep{VVK2020}. As part of the timing programme, updated rotational models for 205 pulsars were published in the first paper of this series~\citep{Jankowski2019} and 9 pulsar glitches have been reported so far~\citep{Jankowski2015aATel, Jankowski2015bATel, Jankowski2016ATel, Jankowski2017ATel, Lower2018, Lower2019}.

While pulsars are renowned for their capacity to be used as astrophysical clocks, many have been observed to exhibit an intrinsic `timing noise' in the measured arrival times of their pulses. 
Timing noise -- stochastic wandering in observed pulse arrival times -- manifests as either a `white' noise processes if the power is distributed normally across all fluctuation frequencies, or `red' noise if the timing residuals are dominated by low-fluctuation frequency structures.
White timing noise can arise from instrumental artefacts, unaccounted radio frequency interference (RFI) and pulse-to-pulse shape variations, often referred to as pulse jitter~\citep{Staelin1968, Jenet1998}.
While fluctuations in the density of the interstellar medium along the line of sight~\citep{Keith2013}, and the passage of nanohertz frequency gravitational waves~\citep{Detweiler1979, Hellings1983} manifest themselves as red noise in pulsar timing residuals, the dominant source of intrinsic red noise over long timescales is believed to arise from irregularities in pulsar rotation. 
One of two primary sources of rotational irregularities are pulsar glitches, discrete spin-up events that often decay exponentially over the following months to years.
Glitches are posited to originate from either the release of built up tension within the neutron star's crust via starquakes~\citep{Ruderman1969, Baym1969}, or the unpinning of superfluid vortices from the crustal lattice~\citep{Anderson1975, Alpar1985, Melatos2008}.
The other main type of rotational irregularity is `spin noise', long-term variations in pulsar spins characterized by a red power spectrum (hereafter referred to as red noise). While red noise is distinct from glitches, it may be possible that glitch recoveries and switching between emission/spin-down states contributes to the overall red noise seen in some pulsars. The nature of the relationship between glitches and red noise is also open for debate~\citep{Hobbs2010, Parthasarathy2019}.

While the precise mechanism behind pulsar red noise is unknown, potential external causes include fluctuations in the spin-down torque~\citep{Cheng1987a}, free-precession~\citep{Stairs2000, Brook2014, Kerr2016}, asteroid belts or debris disks interacting with the pulsar magnetic field~\citep{Cordes2008}, undetected planetary companions~\citep{Kerr2015}, changes in pulse shape~\citep{Brook2016} and discrete quasi-periodic magnetospheric state switching~\citep{Kramer2006a, Lyne2010}. 
Internal mechanisms such as the outward exchange of angular momentum due to coupling between the outer crust and superfluid interior~\citep{Jones1990}, undetected micro-glitches~\citep{Cheng1987b}, vortex re-pinning during glitch recovery~\citep{Melatos2008} and superfluid turbulence~\citep{Greenstein1970, Link2012, Cordes2008} have also been proposed as mechanisms behind red noise.
Long-term studies of large samples of pulsars by~\citet{Hobbs2005} and~\citet{Hobbs2010} found red noise is common across the pulsar population, and claimed pure random walks in pulsar phase, spin frequency or spin-down rate cannot accurately model the observed structures in the timing residuals~\citep{Cordes1985}. However,~\citet{Shannon2010} showed that if the random walk step-sizes are drawn from a power-law function (rather than a Gaussian), then most red noise structures can be replicated.

In this paper we undertake a study of the rotational properties of $\numpsr$ pulsars that have been observed over the past 1.0-4.8 years by UTMOST. This includes a full reanalysis of nine previously reported glitches~\citep{Jankowski2015aATel, Jankowski2015bATel, Jankowski2016ATel, Jankowski2017ATel, Espinoza2011, Sarkissian2017, Palfreyman2018, Lower2018, Liu2018, Sarkissian2019ATel, Kerr2019ATel}, while accounting for the effects of timing noise and the discovery of two new glitches. To parameterize the effects of red noise on the timing residuals, and to avoid biasing our measurements of pulsar spin and spin-down, we employ the Bayesian pulsar timing software {\sc TempoNest}~\citep{Lentati2014}. We search for correlations between pulsar properties and red noise strength, in addition to how it scales as a function of spin and spin-down frequencies across the population.

In Section~\ref{sec:observations} we outline the observing and data processing strategies. In Section~\ref{sec:methods} we describe the phenomenology behind characterising pulsar timing noise and the statistical framework we use to parameterise red timing noise and perform simultaneous measurements of pulsar spin properties. 
We report on our updated timing models and present the results of our red noise search and glitch analysis in Section~\ref{sec:results}.  
In Section~\ref{sec:discussion} we examine potential links between red noise strength and pulsar properties, in addition to outlining and comparing a new, robust method for determining how timing noise scales across the population. 
Lastly, in Section~\ref{sec:conclusion} we draw our conclusions and comment on future applications of our Bayesian framework.

\section{Observations}\label{sec:observations}

\subsection{System overview}

The UTMOST project began with the backend upgrade to the refurbished Molonglo Observatory Synthesis Telescope~\citep{Bailes2017}. MOST is a Mills-Cross design aperture synthesis telescope situated approximately 35\,km South-East of Canberra, Australia. It is comprised of two 778\,m long East-West arms that can be slewed in the north-south direction, and a static North-South arm, that is being re-engineered as part of the UTMOST-2D project (Day et al. in prep.). The telescope operates at a central frequency of 835\,MHz\footnote{The sensitivity of the MOST peaks at $\sim$843\,MHz, as this is where the resonant cavities are tuned to.} covering a bandwidth of 31.25\,MHz. The ring-shaped design of the antenna elements means the instrument is mainly sensitive to right-hand circularly polarized emission.

For the first two years of the timing programme we were capable of mechanically tracking sources in hour angle on the sky. However, maintenance issues and an associated degradation of performance ultimately led us to convert the telescope into a meridian transit instrument in June 2017. While we are no longer able to track sources mechanically, we are able to electronically track up to four pulsars simultaneously as they transit the telescope's $4\degr \times 2\degr$\, primary beam. 
A typical timing observation lasts between 5-20 minutes depending on the brightness of the pulsar and whether it displays interesting behaviours (e.g. emission state-switching).
Observations are usually performed autonomously via the scheduler developed for the UTMOST multi-epoch Survey of Magnetars, Intermittent pulsars, RRATs and FRBs~\citep[SMIRF:][]{VVK2020}, which has improved the efficiency of the timing programme since its June 2017 introduction. Manual observations of targets of interest, phase calibrators and long FRB transit searches are usually performed using the {\sc automatic mode} scheduler detailed in~\citet{Jankowski2019}.

\subsection{Radio frequency interference}
Observations conducted by UTMOST are often contaminated by radio frequency interference (RFI) as its frequency band is shared by radio transmissions from two Australian mobile telecommunications providers. As the telescope is an array, voltage addition in phase only occurs for radio emitting sources that are more than a Fresnel scale away ($\sim 10000$\,km) from the telescope, while anything closer is attenuated. Although this does reduce the overall amount of observed RFI, it is still prevalent in a significant fraction of observations. 
Removal of RFI is performed by passing the data through an excision pipeline that involves spectral kurtosis prior to folding of the raw data with {\sc dspsr}\footnote{\href{http://dspsr.sourceforge.net}{dspsr.sourceforge.net}}~\citep{vanStraten2011}, followed by median filtering of the folded archives via the tools in {\sc psrchive}\footnote{\href{http://psrchive.sourceforge.net}{psrchive.sourceforge.net}}~\citep{Hotan2004, vanStraten2012}. Manual RFI removal with {\sc psrchive}'s interactive {\sc pazi} tool is undertaken when necessary.
More recently, we have modified the RFI cleaning pipeline to use {\sc clfd}\footnote{\href{https://github.com/v-morello/clfd}{github.com/v-morello/clfd}}~\citep{Morello2018}, which uses Tukey's rule to find and zero-weight data corresponding to outliers in the standard deviation, peak-to-peak difference and second bin of the Fourier transform of each sub-integration and channel of a folded observation. This alone has improved the timing accuracy of many slow pulsars we observe by a factor of two. The amount of data lost to RFI excision is typically on the order of 5 percent, but can be as high as 10 to 15 percent during times of high road traffic (and hence an increased number of mobile handsets) near the telescope.

\subsection{Pulsar-timing dataset}

\begin{figure}
    \centering
    \includegraphics[width=\linewidth]{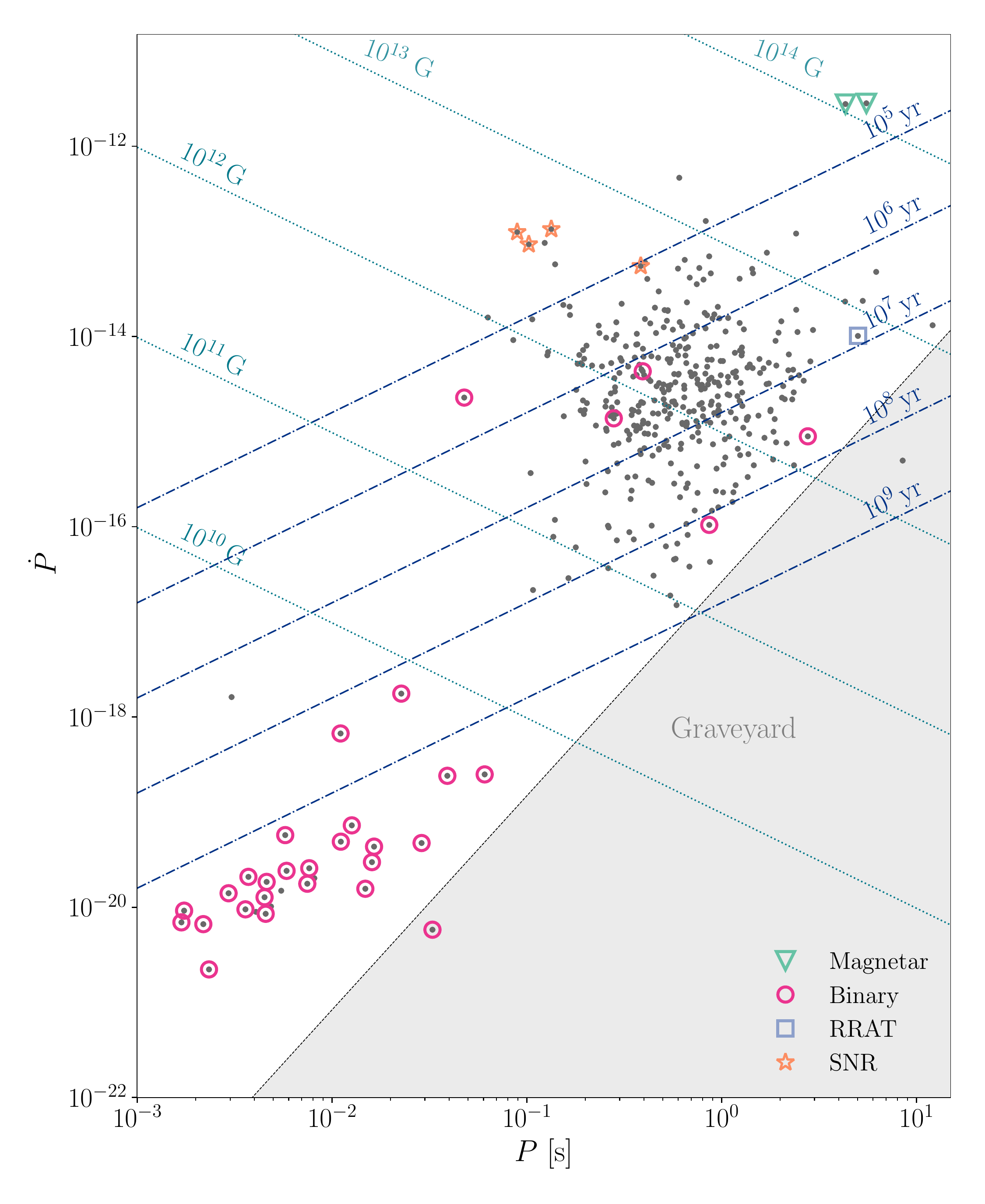}
    \caption{Period, period-derivative diagram for all pulsars regularly monitored by UTMOST. Pulsars residing in binary systems are highlighted by red circles. The RRAT PSR J2033+0042 is indicated by a blue square and the radio-loud magnetars PSR J1622$-$4950 and XTE J1810$-$197 by green triangles. Pulsars associated with supernova remnants are highlighted with stars. Lines of constant characteristic age are defined by the dash-dotted lines and constant surface magnetic field strength by dotted lines.}
    \label{fig:ppdot}
\end{figure}

We began the pulsar-timing programme during October 2015 after phasing of the telescope became routine. Limited pulsar observations prior to this date were undertaken while the telescope was still undergoing upgrades and commissioning, but are largely of lower quality when compared to more recent data.
A general overview of the UTMOST timing programme can be found in \citet{Jankowski2019}, which includes the first scientific results of the timing programme: a study of pulsar proper motions, transverse velocities, pulse duty cycles and flux densities at 843\,MHz, and updated rotational and astrometric parameters for 205 pulsars. 
Currently we perform regular radio monitoring and timing of 412 pulsars, each of which was selected from an initial list of every pulsar for which an observation had been attempted by UTMOST. This includes monitoring the pulsed radio emission of two radio loud magnetars, PSR J1622$-$4950 and XTE J1810$-$197.
Each pulsar observation typically lasts between 5-20\,minutes, depending on the source flux density and declination. After RFI excision, the observations are then summed in frequency and time to produce averaged pulse profiles. These are then cross-correlated with a standard profile, a template generated from a smoothed, high signal-to-noise profile obtained after many hours of integration, to measure the pulse time of arrival (ToA) at the telescope~\citep{Taylor1992}. This `topocentric' ToA is then converted to the ToA at the Solar System Barycentre via the Jet Propulsion Laboratory's DE430 planetary ephemeris~\citep{Folkner2014}. Due to sensitivity limitations of the telescope, most of these pulsars are bright, isolated southern pulsars with relatively long rotation periods. Their basic observational parameters are drawn from the Australia Telescope National Facility (ATNF) pulsar catalogue ({\sc psrcat};~\citet{Manchester2005})\footnote{\href{http://www.atnf.csiro.au/research/pulsar/psrcat/}{www.atnf.csiro.au/research/pulsar/psrcat/}} and~\citet{Jankowski2019}, where the spin period, position and DM determination epoch is MJD 57600.
Fig.~\ref{fig:ppdot} shows the spin period/period-derivative ($P$-$\dot{P}$) diagram for the full set of pulsars regularly monitored by UTMOST.

The cadence with which we observe individual pulsars varies from days to months, depending on their physical properties, and whether they exhibit interesting behaviour such as nulling, glitches or mode-changing.
Precise observation cadences and lengths are defined by a pulsar's position in the sky, apparent brightness and the scientific benefit of performing observations with daily to monthly cadences. We provide this information to SMIRF, which autonomously schedules and performs the observations. Limiting the amount of mechanical wear on the telescope due to slewing is also factored into how often particular pulsars are observed.

\section{Pulsar-timing strategy}\label{sec:methods}

To determine the timing properties of a pulsar, we must first establish a phase-connected timing solution. 
Pulsars that have undergone glitches or exhibit excess structures due to timing noise are difficult to solve coherently over long timing baselines, often requiring the addition of discrete phase jumps before updating the timing model. 
Once we have a phase coherent solution, we use {\sc tempo2}~\citep{Hobbs2006} to assign relative pulse numbers to each ToA, which we then track to avoid phase wraps when attempting to update the timing model.
The effects of timing noise can be mitigated by including fits for higher order polynomials, corresponding to higher order spin-derivatives, into the timing model, or by subtracting a series of harmonically related sinusoids (e.g. {\sc fitwaves} in {\sc tempo2};~\citet{Hobbs2006}).
While these methods are useful for pre-whitening the timing residuals, they assume the measured pulsar properties and timing noise are uncorrelated.
Attempts to avoid biases induced by correlations in the timing residuals include using a transform of the covariance matrix based on Cholesky decomposition to whiten the timing residuals~\citep{Coles2011}, which enables the timing residual power spectrum to be fit by a steep red power-law. 
However, this method requires some \textit{a priori} knowledge of the covariance matrix, and that the correlated timing noise process is stationary in the post-fit timing residuals. 
\citet{vanHaasteren2013} showed that the assumption of stationarity breaks down during the fitting process, leading to incorrect uncertainties on the spectrum estimates, and an incorrect covariance matrix for the pulsar timing model. 
They instead proposed the use of a Bayesian analysis, in which the stochastic timing noise and pulsar properties are modelled simultaneously, avoiding the possibility of biases in the final posterior distributions. 
This method, in particular the ability to perform model selection, was improved by~\citet{Lentati2013} and~\citet{Lentati2014}, as the~\citet{vanHaasteren2013} method is hampered by large matrix inversions and a growing parameter space as the timing model is linearized.

\subsection{Phenomenological timing noise model}

To model the effects of red timing noise, we utilized the phenomenology outlined in~\cite{Lentati2014} and~\citet{Lentati2016}, where the power spectral density of the red noise process is described in the Fourier domain by a power law\footnote{The division by $12\pi^2$ comes from this power-law originally being derived from the one-sided power spectrum a stochastic gravitational-wave background would induce in pulsar timing residuals~\citep[e.g][]{Jenet2006}} with amplitude $A$ (in units of yr$^{3/2}$) and spectral index $\beta$
\begin{equation}\label{eqn:spec_pl}
    P_{r}(f) = \frac{A^{2}}{12\pi^{2}} \Big(\frac{f}{f_{\mathrm{yr}}}\Big)^{-\beta}.
\end{equation}
In addition to the standard power-law model, we also consider a variation of the spectral turnover model of~\citet{Coles2011}
\begin{equation}\label{eqn:spec_turn}
    P_{r}(f) = \frac{A^{2}}{12\pi^{2}}\frac{(f_{c}/f_{\mathrm{yr}})^{-\beta}}{[1 + (f/f_{c})^{-\beta/2}]^{2}},
\end{equation}
where $f_{c}$ is the frequency at which the spectrum turns over. While the models we test are phenomenological rather than drawn from physical theory, models of superfluid turbulence within neutron star interiors~\citep{Melatos2014}, or the presence of circum-pulsar asteroid belts~\citep{Cordes2008} predict spectral flattening or turnovers in power spectra of pulsar timing residuals.

Excess white noise in the residuals due to pulse jitter or radio interference can be accounted for by modifying the uncertainties of individual ToAs as
\begin{equation}
    \mu_{i} =  \sigma_{q}^{2} + F \sigma_{i}.
\end{equation}
Here $\sigma_{i}$ is the ToA uncertainty on the $i$-th observation derived from the cross-correlation procedure for generating ToAs, $F$ is a fitting factor (commonly referred to as EFAC) that encodes the contribution of unaccounted instrumental effects and imperfect estimates of ToA uncertainties, while $\sigma_{q}$ (error in quadrature: EQUAD) accounts for any additional sources of time-independent uncertainties (e.g pulse jitter).

\subsection{Bayesian framework}\label{sec:param_est}

To characterise timing noise and obtain accurate timing models, we used Bayesian parameter estimation to construct posterior probability distributions for the deterministic and stochastic pulsar properties $\theta$ from the timing residuals $r$. 
Prior to performing parameter estimation we first obtained an initial timing solution from previously computed models. In most cases, the initial timing solutions are re-fitted versions of those found in~\citet{Jankowski2019} or {\sc psrcat}~\citep{Manchester2005}. We fit pulsar parameters that are poorly constrained using a $\chi^{2}$ minimisation scheme with {\sc tempo2}~\citep{Hobbs2006, Edwards2006}, resulting in a phase-connected timing solution.

\begin{table*}
    \centering
    \caption{Prior ranges on pulsar and timing noise parameters. $\Delta_{\mathrm{param}}$ is the uncertainy returned by {\sc tempo2}, $T$ is is length of each pulsar's data set.}
    \label{tab:priors}
    \begin{tabular}{llll}
        \hline
        \hline
        Parameter & Symbol [units] & Prior range & Prior type \\
        \hline
        Spin-frequency and derivatives & $\nu$, $\dot{\nu}$, $\ddot{\nu}$ [Hz, s$^{-2}$, s$^{-3}$]     & $\pm x^{\star}\times\Delta_{\mathrm{param}}$ & Uniform \\
        White noise fitting factor & EFAC                              & ($-$1, 2) & Uniform  \\
        White noise error in quadrature & EQUAD [s]                         & ($-$10, 1)  & log-Uniform  \\
        Red noise amplitude & $A$ [yr$^{3/2}$]                  & ($-$20, $-3$) & log-Uniform  \\
        Red noise spectral index & $\beta$                           & (0, 20)   & Uniform \\
        Red noise turn-over frequency & $f_{c}$ [yr$^{-1}$]               & ($0.01/T$, $10/T$) & log-Uniform \\
        Glitch phase jump & $\Delta\phi$ [rotations]          & ($-$10, 10) & Uniform \\
        Permanent change in spin-frequency & $\Delta\nu_{p}$ [Hz]              & ($-$12, $-5$) & log-Uniform  \\
        Change in spin-down & $\Delta\dot{\nu}_{p}$ [Hz$^{-2}$] & ($-10^{-20}$, $-10^{-9}$) & Uniform \\
        Decaying change in spin-frequency & $\Delta\nu_{d}$ [Hz]              & ($-12$, $-5$) & log-Uniform  \\ 
        Glitch recovery timescale & $\tau_{d}$ [days]                 & (0, 3000) & Uniform \\
        \hline
    \end{tabular}

    {$^{\star}x$ lies between $100 - 100000$ depending on the pulsar.}
\end{table*}

We conducted parameter estimation on these timing models using the {\sc TempoNest}\footnote{\href{https://github.com/LindleyLentati/TempoNest}{github.com/LindleyLentati/TempoNest}} Bayesian pulsar timing software developed by~\cite{Lentati2014}. {\sc TempoNest} utilizes the nested sampling algorithm {\sc MultiNest}~\citep{Skilling2004, Feroz2008, Feroz2009} to sample the posterior distributions of the parameters $\theta$, given timing residuals $r$ and a timing model $\mathcal{M}$, while analytically marginalizing over nuisance parameters. The general form of the posterior probability distribution is given by Bayes' theorem as
\begin{equation}
    p(\theta | r, \mathcal{M}) = \frac{\mathcal{L}(r | \theta, \mathcal{M}) \pi(\theta, \mathcal{M})}{\mathcal{Z}(r | \mathcal{M})},
\end{equation}
where $\mathcal{L}(r | \theta, \mathcal{M})$ is the likelihood function for the residuals given a timing model and model parameters (equation 21 of~\cite{Lentati2014}), $\pi(\theta, \mathcal{M})$ is our prior knowledge, the ranges of which are listed in Table~\ref{tab:priors} and $\mathcal{Z}(r | \mathcal{M})$ is the Bayesian evidence, which is a single number representing the completely marginalized likelihood defined by
\begin{equation}
    \mathcal{Z}(r | \mathcal{M}) = \int d\theta \mathcal{L}(r | \theta, \mathcal{M}) \pi(\theta, \mathcal{M}).
\end{equation}

To account for potential covariances when fitting for the parameters of interest, we included the sky-position of the pulsar as free parameters. However, any improvements in the sky-position uncertainty over the values output by {\sc tempo2} would be marginal at best, as all pulsars in our sample have been timed for more than 1\,yr. Hence, we treat the sky-position as a set of nuisance parameters, $\theta_{n} = \{ \alpha, \delta\}$, that are analytically marginalized over to obtain the marginalized posterior distribution for the parameters of interest ($\theta_{i}$), defined by
\begin{equation}
    p(\theta_{i} | r, \mathcal{M}) = \int \prod_{n \neq i } d\theta_{n} \pi(\theta_{n}, \mathcal{M}) \mathcal{L}(r| \theta_{i} , \theta_{n} , \mathcal{M}).
\end{equation}
Any glitch parameters in the timing model are also marginalized over, unless we are explicitly attempting to measure them. Neglecting to do so biases the recovered spectral index toward larger values (a steeper red spectrum).
For computing posterior distribution confidence intervals we use the maximum likelihood statistics from {\sc ChainConsumer}\footnote{\href{https://github.com/samreay/ChainConsumer/}{github.com/samreay/ChainConsumer/}}~\citep{chainconsumer}.

After conducting parameter estimation, we can use the resulting Bayesian evidences to compare two or more competing hypotheses ($\mathcal{M}_{1}$, $\mathcal{M}_{2}$) by calculating the odds ratio
\begin{equation}\label{eqn:odds}
    \mathcal{O}_{12} = \frac{\mathcal{Z}(r | \mathcal{M}_{1})}{\mathcal{Z}(r | \mathcal{M}_{2})}\frac{\Pi_{1}}{\Pi_{2}},
\end{equation}
where $\Pi_{1}/\Pi_{2}$ is the \textit{a-priori} odds of the two hypotheses. In our case the prior odds are unity as we assume uninformative priors throughout our analysis. This leaves us with an alternative model comparison metric known as the Bayes factor, which can be calculated as
\begin{equation}
    \mathcal{B}_{12} = \frac{\mathcal{Z}(r| \mathcal{M}_{1})}{\mathcal{Z}(r | \mathcal{M}_{2})} = \frac{\int d\theta_{1} \mathcal{L}(r | \theta_{1}, \mathcal{M}_{1}) \pi(\theta_{1}, \mathcal{M}_{1})}{\int d\theta_{2} \mathcal{L}(r | \theta_{2}, \mathcal{M}_{2}) \pi(\theta_{2}, \mathcal{M}_{2})},
\end{equation}
where $\theta_{1}$, $\theta_{2}$ are the parameters associated with models $\mathcal{M}_{1}$ and $\mathcal{M}_{2}$ respectively.
In our analysis, the specific models we compared include:
\begin{itemize}
    \item White timing noise (WTN): fitting for deterministic pulsar paramters, EFAC and EQUAD only.
    \item Power-law red noise (PLRN): fitting for a power-law red noise model (equation~\ref{eqn:spec_pl}) in addition to the WTN parameters.
    \item Power-law red noise with frequency turnover (PL+FC): includes a turnover in the red power spectrum (equation~\ref{eqn:spec_turn}) plus WTN parameters.
    \item Second spin-frequency derivative (PLRN+F2): same as PLRN, but also fitting a cubic term to measure $\ddot{\nu}$.
\end{itemize}
The specific choice of a Bayes factor threshold when performing model selection is largely dependent on what one considers to be an acceptable false positive rate.
For instance, a conservative Bayes factor threshold of $|\ln(\mathcal{B}_{12})| > 8$ (corresponding to a false positive rate of $\sim$\,$1/3000$) is generally used in gravitational-wave astronomy~\citep{Thrane2019}. 
A more common interpretation is outlined in~\citet{Kass1995}, where a Bayes factor of $\ln(\mathcal{B}_{12}) > 5$ (false positive rate $\sim$\,$1/150$) is considered to be `very strong' evidence for one hypothesis over the other. In this work we use the latter interpretation, as it has previously been used in pulsar model selection studies~\citep[e.g.,][]{Lentati2015, Reardon2019, Parthasarathy2019}. In cases where neither model is significantly preferred over the other, i.e. for $|\ln(\mathcal{B}_{12})| < 1$, Occam's razor tells us the least complicated model is preferred.

\subsection{Braking indices}

Over long timescales, the spin-down of a pulsar is often approximated by a power law of the form
\begin{equation}\label{eqn:spin_down}
    \dot{\nu} = -K\nu^{n},
\end{equation}
where $K$ is a scaling constant related to the pulsar moment of inertia and magnetic field structure~\citep{Gunn1969} and $n$ is the `braking index'. The value of the braking index is potentially an indicator of the physical process that dominates the torque acting to slow the rotation of the neutron star. For instance, a braking index of $n = 1$ arises if the spin-down is dominated by an out-flowing particle wind from the pulsar surface~\citep{Harding1999}, $n = 3$ corresponds to magnetic-dipole radiation~\citep[e.g.][]{Shapiro1983}, and $n = 5$ would indicate the pulsar is spinning down due to some form of quadrupole radiation, such as gravitational waves~\citep{Bonazzola1996, Yue2007}. Magnetic field evolution or a varying misalignment between the spin and magnetic axes are also predicted to result in $n < 3$~\citep{Blandford1988, Lyne2013}.
By taking the derivative of equation~\ref{eqn:spin_down} and solving for $n$, we can infer the braking index of a pulsar by measuring its \textit{second} spin-frequency derivative ($\ddot{\nu}$), giving
\begin{equation}
    n = \frac{\nu\ddot{\nu}}{\dot{\nu}^{2}}.
\end{equation}
Obtaining accurate measurements of $\ddot{\nu}$ is difficult, as measured values of $\ddot{\nu}$ in `old' pulsars are not significantly different from zero. 
As with measuring $\nu$ and $\dot{\nu}$, not accounting for timing noise in the pulsar residuals when attempting to measure $\ddot{\nu}$ will lead to biased measurements, as $\ddot{\nu}$ is often highly correlated with timing noise. 

\subsection{Glitch parameter estimation}

While the low frequency structures resulting from red noise affect the long term timing precision of pulsars, pulsar glitches result in neutron stars spinning-up on timescales of seconds~\citep{Ashton2019}. This causes a near-instantaneous difference between the observed ToAs and what is expected from the timing model. Some pulsars take days to months to recover toward their original pre-glitch spin frequency (and sometimes do not fully recover or over-recover). In general the change in rotational phase from a glitch can be expressed in terms of instantaneous, permanent changes in the pulsar spin ($\nu_{p}$) and spin-down ($\dot{\nu}_{p}$), as well as the exponential spin recovery ($\nu_{d}$) over time ($\tau_{d}$)
\begin{equation}\label{eqn:glitch}
\begin{aligned}
    \phi_{\mathrm{g}}(t) = \Delta\phi + \Delta\nu_{p}(t - t_{\mathrm{g}}) + \frac{1}{2}\Delta\dot{\nu}_{p}(t - t_{\mathrm{g}})^{2} \\
     - \Delta\nu_{d}\tau_{d}e^{-(t - t_{\mathrm{g}})/\tau_{d}}.
\end{aligned}
\end{equation}
If the precise epoch at which a glitch occurred ($t_{\mathrm{g}}$) is poorly constrained then there can be some uncertainty in the precise number of pulsar rotations between the last pre-glitch and first post-glitch observations. Hence an unphysical phase jump ($\Delta\phi$) is frequently implemented to maintain a phase-connected timing solution. If the glitch epoch were known precisely, a phase jump would not be required. Glitch recovery is often associated with the re-pinning of superfluid vortices~\citep{Melatos2008}. The degree to which a glitch recovers can be quantified by the recovery parameter $Q = \Delta\nu_{d}/\Delta\nu_{\mathrm{g}}$, where $\Delta\nu_{\mathrm{g}} = \Delta\nu_{p} + \Delta\nu_{d}$.

When fitting for pulsar glitches we include five parameters drawn from equation~\ref{eqn:glitch} that describe the change in pulsar spin and post-glitch recovery,  $\{ \Delta\phi, \Delta\nu_{p}, \Delta\dot{\nu}_{p}, \Delta\nu_{d}, \tau_{d} \}$, in addition to the red noise and spin parameters. 
For pulsars found to have undergone multiple glitches within our timing data, we fitted all of the relevant glitch parameters simultaneously in order to avoid introducing biases from incomplete glitch models when attempting to model them individually.
We then marginalize over the instantaneous phase jump to account for uncertainties on the glitch epoch.
\begin{figure*}
    \centering
    \includegraphics[width=\linewidth]{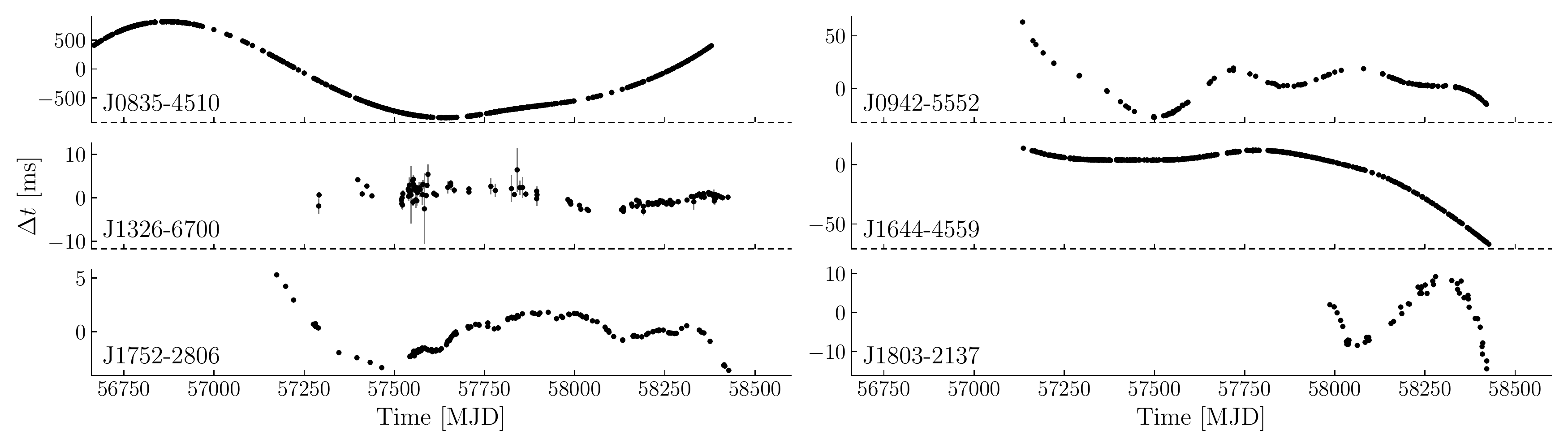}
    \caption{Phase connected timing residuals for 6 pulsars which exhibit various levels of red noise after fitting for $\nu$ and $\dot{\nu}$. Plots showing the timing residuals for all 300 pulsars can be found in the supplementary material.}
    \label{fig:residuals}
\end{figure*}
We employed Bayesian model selection in cases where it is difficult to tell by eye whether a small glitch or glitch-like event is real, or is the result of a cusp in the residuals due to timing noise.

\section{Results}\label{sec:results}

\begin{table*}
    \centering
    \caption{Maximum likelihood posterior measurements of $\ddot{\nu}$ and $n$ compared with reported values in the literature. Values in parentheses indicate the $1\sigma$ uncertainties in the last digits, while the errors in our measurements indicate the 95 percent confidence intervals. Only PSR J0738$-$4042 strongly favours the PLRN+F2 model over the standard PLRN model.}
    \label{tab:F2}
    \renewcommand{\arraystretch}{1.5}
    \begin{tabular}{lccccccc}
    \hline
    \hline
    PSR & $\ln(\mathcal{B})$ & $\ddot{\nu}$ & $\ddot{\nu}_{\mathrm{lit}}$ & $n$ & $n_{\mathrm{lit}}$ & $\tau_{c}$ & Known to glitch? \\
         & & ($10^{-24}$s$^{-3}$) & ($10^{-24}$s$^{-3}$) & & & Myr & \\
    \hline
    J0659$+$1414 & $1.5$ & $1.1^{+1.1}_{-0.5}$ & $0.764(4)$ & $21.1^{+2.7}_{-1.2}$ & $14.44(8)$ & 0.111 & Y \\
    J0729$-$1836 & $3.6$ & $-2.4^{+2.3}_{-1.9}$ & $0.376(15)$ & $-897.1^{+10.3}_{-5.2}$ & $139(5)$ & 0.426 & Y \\
    J0738$-$4042 & $5.4$ & $-3.5 \pm 1.2$ & $-$ & $-96227.0^{+1.8}_{-2.2}$ & $-$ & 4.32 & N \\   
    J0942$-$5552 & $1.6$ & $8.1^{+4.2}_{-4.3}$ & $-$ & $4591.4^{+3.1}_{-3.5}$ & $-$ & 0.461 & N \\
    J1001$-$5507 & $4.9$ & $1.8^{+0.8}_{-1.5}$ & $-$ & $1960.1^{+1.4}_{-4.1}$ & $-$ & 0.441 & N \\   
    J1359$-$6038 & $4.1$ & $-3.4^{+3.3}_{-1.1}$ & $-$ & $-176.9^{+4.6}_{-1.4}$ & $-$ & 0.319 & N \\
    J1413$-$6307 & $1.1$ & $-4.2^{+3.5}_{-11.2}$ & $-$ & $-4606.1^{+2.4}_{-16.5}$ & $-$ & 0.842 & N \\
    J1709$-$4429 & $1.0$ & $106.4^{+45.1}_{-47.3}$ & $173.1(7)$ & $13.3^{+0.8}_{-0.3}$ & $21.35(8)$ & 0.0175 & Y \\
    J1909$+$1102 & $3.2$ & $1.1^{+0.7}_{-1.0}$ & $-2.02(4)$ & $3466.9^{+3.9}_{-5.1}$ & $-6615(131)$ & 1.7 & Y \\
    \hline
    \end{tabular}
    \renewcommand{\arraystretch}{1.5}
\end{table*}

\subsection{Updated timing models}\label{sec:pulsar_models}

Many pulsars in our sample have improved timing measurements over those from version 1.54 of the ATNF pulsar catalogue\footnote{This is the psrcat version from which many of initial ephemerides were drawn from.}, including a number that are not present in~\citet{Jankowski2019}. The updated astrometric and spin parameters are presented in Table~\ref{tab:params}. The resulting timing residuals for all pulsars analysed in this work are presented in Fig.~\ref{fig:residuals}.
Ephemeris files in {\sc tempo2} format, ToAs, plots showing the one- and two-dimensional posterior distributions and a clock correction file are available to download from our online repository\footnote{\href{https://github.com/Molonglo/TimingDataRelease1/}{github.com/Molonglo/TimingDataRelease1/}}.

\subsection{Spin frequency second derivatives and braking indices}

By simultaneously modelling timing noise as a power-law process, we are able to search for unbiased values of $\ddot{\nu}$ by comparing Bayesian evidences for the power-law red noise (PLRN) and red noise with $\ddot{\nu}$ (PLRN+F2) models.
There are 8 pulsars in our sample that marginally prefer the PLRN+F2 model ($1 < \ln(\mathcal{B}) < 5$) for which we recover well constrained $\ddot{\nu}$ posteriors that are inconsistent with zero. We find only {\numfddot} pulsar, PSR J0738$-$4042, significantly favours the PLRN+F2 model with a log Bayes factor $> 5$, while PSR J1001$-$5507 has a marginally sub-threshold preference ($\ln(\mathcal{B} = 4.9)$). These measurements of $\ddot{\nu}$ along with the inferred braking index for each pulsar are compared with those from {\sc psrcat} in Table~\ref{tab:F2}. 
The sign of the braking indices depends on whether the inferred value of $\ddot{\nu}$ is positive or negative.
None of our measurements of $\ddot{\nu}$ and $n$ are consistent with previously published values. Given each of the pulsars with previous $\ddot{\nu}$ measurements have undergone glitches, the difference in results may be due to the accumulated changes in pulsar rotation between measurements or yet to be reported glitches. No glitches have been reported to date in the pulsars that do not have $\ddot{\nu}$ values listed in {\sc psrcat}. While the inferred braking indices are all much larger than the canonical $n = 3$ expected from magnetic dipole radiation, they are consistent with values reported for other young pulsars. It has been speculated these large braking indices may be due to the effects of unmodelled recovery from glitches prior to the start of timing observations~\citep{Johnston1999}. Alternatively, our timing noise model may be incomplete, giving rise to the preference for the PLRN+F2 model. For instance, PSR J0738$-$4042 may be affected by torque variations initially induced after a profile change in 2005, proposed to be evidence for an interaction with an asteroid~\citep{Brook2014}, while PSR J1001$-$5507 is known to exhibit discrete spin-down state switching~\citep{Chukwude2012}. Neither phenomenon is included in the timing models for these pulsars, and are therefore likely causes of these pulsars' strong preference for the PLRN+F2 model.

\subsection{Red timing noise properties}

We assessed the presence of red noise in our pulsars using the Bayes factor found from comparing the PLRN model against the WTN model as a detection statistic, where $\ln(\mathcal{B}_{R/W}) > 5$ is a strong detection. Pulsars for which we obtain Bayes factors of $3 < \ln(\mathcal{B}_{R/W}) < 5$ are categorized as `probable detections' since the PLRN model is favoured, but is subject to an increased false positive rate. Those that have Bayes factors in the range $1 < \ln(\mathcal{B}_{R/W}) < 3$ marginally favour a red noise model, but lack the statistical confidence to be distinguishable from the WTN model.

Out of the $\numpsr$ pulsars analysed, we find $\numred$ strongly favour the PLRN model and 6 that fall into the probable detection category. None of the pulsars in our sample favour the PL+FC model. We find the magnetar PSR J1622$-$4950 to have the largest red noise amplitude at 1\,yr of $\log_{10}(A) = -4.9^{+0.6}_{-0.4}$, with a spectral index $\beta = 7.3^{+3.4}_{-3.6}$. This result should be taken with caution as we do not account for changes in the pulse profile or variations in $\dot{\nu}$ due to short-term changes in magnetic torque that are observed in magnetars~\citep[e.g.][]{Camilo2018}. Excluding magnetars, PSR J0835$-$4510 (the Vela pulsar) has the largest red noise amplitude of any non-recycled pulsar in our data set, with $\log_{10}(A) = -8.2 \pm 0.2$ and a steep spectral index of $\beta = 8.6 \pm 0.9$. This spectral index is different to the value measured by~\citet{Shannon2016} using 21\,years of Vela timing, which may be caused by the occurrence of additional glitches that have recovered since the end of their data set.

We find that three millisecond pulsars favour the PLRN model: PSR J0437$-$4715, PSR J2145$-$0750 and PSR J2241$-$5236. While red noise due to rotational instabilities is known to be present in millisecond pulsars, high precision timing has shown that variations in pulsar DM can mimic timing noise in observations at single frequencies~\citep[e.g.,][]{Lentati2016}. Accounting for DM variations requires observing systems that use either wide-band receivers or are capable of observing at multiple frequencies. Due to the limited bandwidth of UTMOST, there is a covariance between DM variations and achromatic timing noise. Hence it is not possible for us to attribute the red noise we observe in millisecond pulsars to rotational irregularities. Unaccounted instrumental artefacts may also contribute to the red noise in these pulsars~\citep{Jankowski2019}. 

The lack of multi-band observations also means we cannot infer the contribution of DM variations to the red noise in the non-recycled pulsars. However,~\citet{Petroff2013} found only 11 pulsars out of a sample of 160 non-recycled pulsars showed significant changes in DM with time (only setting upper-limits on the remaining 149), while~\citet{Shannon2016} showed the Vela pulsar's DM variations have a sub-dominant contribution to its overall red noise. Hence, any extra red noise induced by DM variations in our non-recycled pulsar sample would be negligible. Additionally, we find no correlation between red noise parameters and DM.

The full list of the maximum likelihood posterior values and associated 95 percent confidence intervals on the red noise parameters are presented in Appendix~\ref{apdx:red_params}.

\subsection{Pulsar glitch reanalysis}\label{sec:glitches}

\begin{figure}
    \centering
    \includegraphics[width=\linewidth]{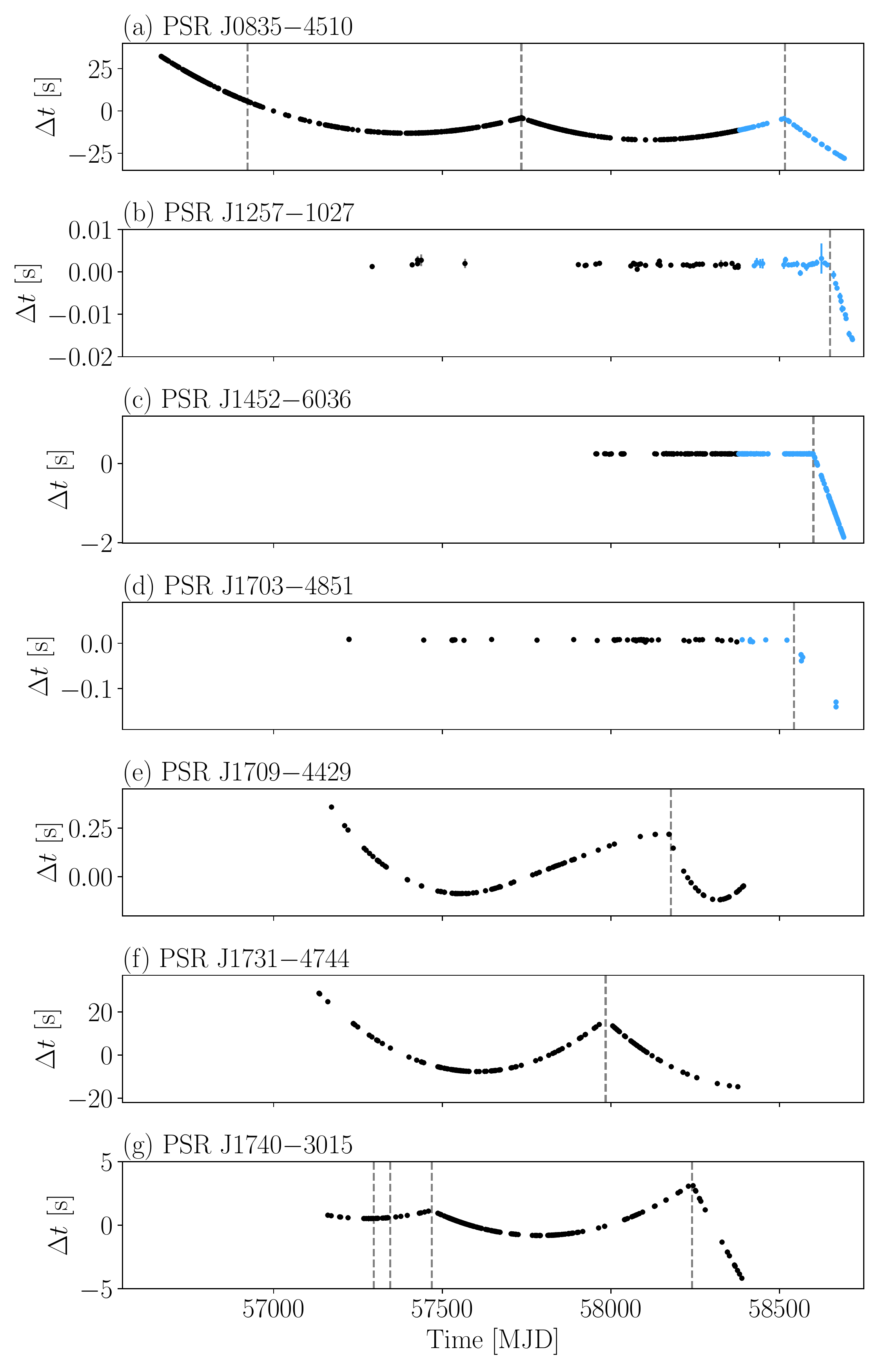}
    \caption{Timing residuals for the four glitched pulsars after fitting for $\nu$ and $\dot{\nu}$. The epochs of the reported glitches are indicated by the dashed vertical lines. Light blue points in (a), (b), (c) and (d) indicate ToAs that are not used in the red noise analysis.}
    \label{fig:glitch_resids}
\end{figure}

So far we have observed twelve glitches in eight pulsars, nine of which have been previously reported.\footnote{Seven of these glitches have been added to the Jodrell Bank glitch catalogue~\citep{Espinoza2011}:~\href{http://www.jb.man.ac.uk/pulsar/glitches.html}{www.jb.man.ac.uk/pulsar/glitches.html}} The timing residuals for the six pulsars prior to adding glitch corrections are depicted in Fig.~\ref{fig:glitch_resids}.
Cusp-like features in the residuals are the result of large glitches. Note that separate ephemeris and ToA files for PSRs J0835$-$4510, J1257$-$1027, J1452$-$6036 and J1703$-$4851 that include post-glitch observations and corrections can be found in the online repository\footnote{\href{https://github.com/Molonglo/TimingDataRelease1/}{github.com/Molonglo/TimingDataRelease1/}}. The extended data sets for these pulsars are used only for the glitch analyses, and are not included in our red noise study.
Currently pulsars that have undergone a glitch are manually identified in the UTMOST data by searching for glitch-like events in the timing residuals `by eye'. This method can be prone to error, with small glitches being glossed over when investigating pulsars that exhibit strong red noise. An automated glitch detection pipeline would be a useful development to search for previously unnoticed glitches in past observations and for near-real time glitch detection. 

\subsubsection*{PSR J0835$-$4510}

There are three reported glitches in PSR J0835$-$4510 that we have observed with UTMOST~\citep{Jankowski2015aATel, Palfreyman2016, Palfreyman2018, Sarkissian2019ATel, Kerr2019ATel}. The first is reported to have occurred on MJD $56922 \pm 3$, with a small glitch amplitude of $\Delta\nu_{\mathrm{g}}/\nu = 0.4 \times 10^{-9}$. Our reanalysis returns only upper limits on the change in $\nu$, but does recover $\Delta\dot{\nu}_{g} = 21.5^{+0.6}_{-1.4} \times 10^{-3}$. 
However, performing model comparison returns a log Bayes factor of $\ln(\mathcal{B}) = -48.0$, indicating a red noise model without a glitch at this epoch is strongly preferred for this event.

For the second glitch, which was observed in real-time by~\citet{Palfreyman2018} at MJD $57734.484991(29)$, we obtain a glitch amplitude that is similar in magnitude to the published value, while our inferred change in the spin-down of the pulsar is $\sim$30\,percent smaller than the previously reported value. This is likely due to our analysis incorporating simultaneous modelling of the glitch and timing noise parameters. Including the short-term recovery found by~\citet{Sarkissian2017}, we find evidence for additional recovery of $\Delta\nu_{d} = 4.2^{+0.6}_{-0.3}$\,nHz over $12.7^{+3.0}_{-1.2}$ days.

The third glitch occurred during observations by the Hawksbury radio observatory~\citep{Sarkissian2019ATel} and the \textit{Fermi} gamma-ray observatory at MJD 58515.5929(5)~\citep{Kerr2019ATel}. Our recovered glitch amplitude is similar in size to the previously reported values, and is typical of other large Vela glitches ($\Delta\nu_{g}/\nu \sim 1000$). We also find a small exponential recovery ($Q = 0.005$) occurred over $11.0 \pm 1.2$\,days. We are unable to test whether this glitch underwent any short-term recovery similar to the previous one as our first post-glitch observation was $\sim$9\,days after the glitch occurred.

In addition to the three glitches we analyse here, a large glitch occurred on MJD 56555.871~\citep{Buchner2013ATel, Shannon2016}, prior to the start of our data set. While we cannot constrain the permanent changes in $\nu$ and $\dot{\nu}$, we can perform parameter estimation on the recovery parameters in the presence of red noise. We obtain a change in spin frequency of $\Delta\nu_{d}/\nu = 1591^{+170}_{-322}$, which decays over $\tau_{d} = 923^{+88}_{-152}$ days. Our measured $\Delta\nu_{d}/\nu$ is inconsistent with the value inferred by~\citet{Shannon2016}, but they only have observations up to 116\,days after the glitch occurred. Hence any further decay in $\Delta\nu_{g}$ beyond the end of their data set will have been missed. When compared against the pure PLRN model, the resulting $\ln(\mathcal{B}) = 65.5$ strongly favours the red noise plus glitch recovery model. This confirms that long-term recovery from glitches prior to the start of timing observations can affect the recovered timing parameters.

\begin{table*}
    \centering
    \begin{threeparttable}
    \caption{Maximum likelihood posterior values from the glitch parameter posterior distributions and associated 95 percent confidence intervals compared with previously reported measurements. Values in parentheses represent the 1-$\sigma$ uncertainties on the last digit.}
    \label{tab:glitches}
    \renewcommand{\arraystretch}{1.5}
    \begin{tabular}{ccccccccc}
        \hline 
        \hline
        PSR & $t_{\mathrm{g}}$ & $\Delta\nu_{\mathrm{g}}/\nu$ & $\Delta\dot{\nu_{\mathrm{g}}}/\dot{\nu}$ & $\tau_{d}$ & $Q$ & $(\Delta\nu_{\mathrm{g}}/\nu)_{\mathrm{lit}}$ & $(\Delta\dot{\nu_{\mathrm{g}}}/\dot{\nu})_{\mathrm{lit}}$ & Ref \\
         & MJD & $\times 10^{-9}$ & $\times 10^{-3}$ & days &  &  $\times 10^{-9}$ & $\times 10^{-3}$ & \\
        \hline
        J0835$-$4510 & $56922(3)$ & $\lesssim 0.2$ & $21.5^{+0.6}_{-1.4}$ & $-$ & $-$ & $0.4$ & $0.1$ & 1, 2 \\
        J0835$-$4510 & $57734.484991(29)$ & $^{\star}1448.8^{+0.9}_{-0.8}$& $7.33^{+0.13}_{-0.11}$ & $12.7^{+3.0}_{-1.2}$ & $^{\star}0.011$ & $1431.24(7)$ & $9.20(83)$ & 3, 4 \\
        J0835$-$4510 & $58515.5929(5)$ & $2501.2^{+2.6}_{-3.2}$& $8.69^{+0.28}_{-0.25}$ & $11.0 \pm 1.2$ & $0.005$ & $2491.1(5)$ & $-$ & 5, 6 \\
        J1257$-$1027 & $58649.3(6)$ & $3.20^{+0.16}_{-0.57}$ & $\lesssim 286$ & $-$ & $-$ & $-$ & $-$ & This work \\
        J1452$-$6036 & $58600.29(5)$ & $270.7^{+0.3}_{-0.4}$ & $\lesssim 16$ & $-$ & $-$ & $-$ & $-$ & This work \\
        J1703$-$4851 & $58543.1(3)$ & $19.0^{+1.0}_{-0.7}$ & $292^{+38}_{-53}$ & $-$ & $-$ & $-$ & $-$ & This work \\
        J1709$-$4429 & $58178(6)$ & $54.6 \pm 1.0$ & $1.06^{+0.36}_{-0.43}$ & $99.1^{+11.3}_{-9.6}$ & $0.995$ & $52.4(1)$ & $7.30(12)$ & 7 \\
        J1731$-$4744 & $57984(20)$ & $3149.5^{+0.5}_{-0.4}$ & $1.2^{+0.7}_{-1.1}$ & $-$ & $-$ & $3147.7(1)$ & $-$ & 8 \\
        J1740$-$3015 & $57296.5(9)$ & $0.122^{+0.086}_{-0.081}$ & $82.2^{+8.8}_{-8.5}$ & $-$ & $-$ & $1.30(4)$ & $<0.66$ & 9, 10 \\
        J1740$-$3015 & $57346.0(6)$ & $\lesssim 0.019$ & $111.1^{+13.6}_{-8.4}$ & $-$ & $-$ & $1.94(2)$ & $<0.07$ & 11 \\
        J1740$-$3015 & $57468.59(40)$ & $237.7^{+13.2}_{-9.3}$ & $1.71^{+3.24}_{-1.54}$ & $430^{+91}_{-101}$ & $0.025$ & $229(2)$ & $2.19(4)$ & 11, 10 \\
        J1740$-$3015 & $58240.781(5)$ & $842.3^{+7.1}_{-5.6}$ & $74.0^{+10.0}_{-13.2}$ & $-$ & $-$ & $837.88(28)$ & $1.63(14)$ & 12, 10 \\
        \hline
    \end{tabular}
    \begin{tablenotes}
      \item References indicated in the last column are (1)~\citet{Jankowski2015aATel}; (2)~\citet{Palfreyman2016}; (3)~\citet{Sarkissian2017}; (4)~\citet{Palfreyman2018}; (5)~\citet{Sarkissian2019ATel}; (6)~\citet{Kerr2019ATel}; (7)~\citet{Lower2018}; (8)~\citet{Jankowski2017ATel}; (9)~\citet{Jankowski2015bATel}; (10)~\citet{Espinoza2011}; (11)~\citet{Jankowski2016ATel}; (12)~\citet{Liu2018}. $^{\star}$Includes a short-term $\Delta\nu_{d} = 129(8)$\,nHz recovery over $0.96(17)$\,days~\citep{Sarkissian2017}. 
    \end{tablenotes}
    \end{threeparttable}
    \renewcommand{\arraystretch}{}
\end{table*}
\subsubsection*{PSR J1257$-$1027}
This glitch is the first to ever be reported in this pulsar. It is well described by a small permanent change in the pulsar spin ($\Delta\nu_{g}/\nu = 3.20^{+0.16}_{-0.57} \times 10^{-9}$) with no evidence for recovery. Including a change in the pulsar's spin-down frequency in our parameter estimation returned only an upper limit of $\Delta\dot{\nu} \lesssim 268 \times 10^{-3}$. Additional observations over longer post-glitch timescales are required for further constraints to be placed on changes in $\dot{\nu}$.

\subsubsection*{PSR J1452$-$6036}

We discovered a glitch with an amplitude of $\Delta \nu_{g}/\nu = 270.7^{+0.3}_{-0.4} \times 10^{-9}$ that occurred in PSR J1452$-$6036 on MJD $58600.29(5)$. This is the second glitch seen in this pulsar to date and is almost a factor of 10 larger than the glitch observed on MJD 55055.22(4) by~\citet{Yu2013}. Performing model selection we find a change in spin-down is weakly disfavoured ($\ln(\mathcal{B}) = 4.2$), hence we can only set an upper limit on $\Delta \dot{\nu}_{g}/\dot{\nu}$ of $\lesssim 16 \times 10^{-3}$ at the 95 percent confidence level. In addition, we find no evidence for an exponential recovery after this glitch. This could be due to a lack of vortex re-pinning following this glitch, or the recovery having occurred on a timescale too short to be resolved with our current observation cadence ($\sim 3.3$\, days between $t_{g}$ and the first post-glitch observation). Alternatively, the recovery timescale may be significantly longer than our current post-glitch data span.

\subsubsection*{PSR J1703$-$4851}

The glitch we observed on MJD $58543.1(3)$ is the first to ever be reported in this pulsar. We recover a moderate change in the pulsar spin of $\Delta \nu_{g}/\nu = 19.0^{+1.0}_{-0.7} \times 10^{-9}$ and a relatively large change in the spin-down of $\Delta \dot{\nu}_{g}/\dot{\nu} = 292^{+38}_{-53} \times 10^{-3}$. We find a recovery model is disfavoured for this glitch. While this pulsar is known to undergo emission state switching~\citep{Wang2007} we have only four post-glitch observations to date, hence we are currently unable to provide any link between the glitch and state switching. The lack of post-glitch observations may also explain why the recovery model is disfavoured, as long-term glitch recoveries require extended observations to detect.

\subsubsection*{PSR J1709$-$4429}

This glitch is the fourth and smallest glitch observed to date in PSR J1709$-$4429. The glitch amplitude we recover ($\Delta\nu_{\mathrm{g}}/\nu = 54.6 \pm 1.0 \times 10^{-9}$) is consistent with the previously reported value~\citep{Lower2018}, but the change in spin-down frequency was overestimated by a factor of $\sim$7. This is likely due to $\Delta\dot{\nu}$ being covariant with the glitch recovery, which was not fit for in~\citet{Lower2018}. We find the change-in-spin period due to this glitch almost completely recovers ($Q = 0.995$) in $99.1^{+11.2}_{-9.6}$,days.

\subsubsection*{PSR J1731$-$4744}

With an amplitude of $\Delta\nu_g/\nu = 3148 \pm 3 \times 10^{-9}$, this is the largest glitch contained within our presented data set, and the largest observed in this pulsar to date~\citep{Espinoza2011}. Previous glitches have shown evidence for linear recoveries~\citep{Yu2013}, but we find no evidence for any spin recovery from this glitch.

\subsubsection*{PSR J1740$-$3015} 

Four previously reported glitches have occurred within our timing measurements of PSR J1740$-$3015~\citep{Jankowski2015bATel, Jankowski2016ATel, Espinoza2011, Liu2018}. We obtain a small amplitude of $\Delta\nu_{g}/\nu = 0.122^{+0.086}_{-0.081} \times 10^{-9}$ and comparatively large change in spin-down ($\Delta\dot{\nu}_{g} = 82.2^{+8.8}_{-8.5} \times 10^{-3}$) associated with the first glitch.
For the second glitch (MJD $57346.0(0.6)$), we are only able to set an upper-limit on the instantaneous change in pulsar spin-frequency. Performing model comparison, we find all models that include the second glitch are strongly disfavoured, suggesting the properties of this glitch are covariant with our red timing noise model.
Our analysis of the third glitch recovers a change in spin-frequency that is largely consistent with previously reported values, with a small recovery ($Q = 0.035$) over $430^{+91.1}_{-100.9}$\,days.
The fourth glitch was discovered in observations of the pulsar at Jodrell Bank~\citep[Shaw et al. 2018, private communication, ][]{Espinoza2011}. It was also seen by the Shanghai Tian Ma Radio Telescope~\citep{Liu2018}. We find no evidence for spin recovery after this latest glitch. However, the large change in spin-down we recover may be evidence of longer-term recovery, as these two effects are strongly covariant while the pulsar remains in the recovery phase.

\begin{figure}
    \centering
    \includegraphics[width=\linewidth]{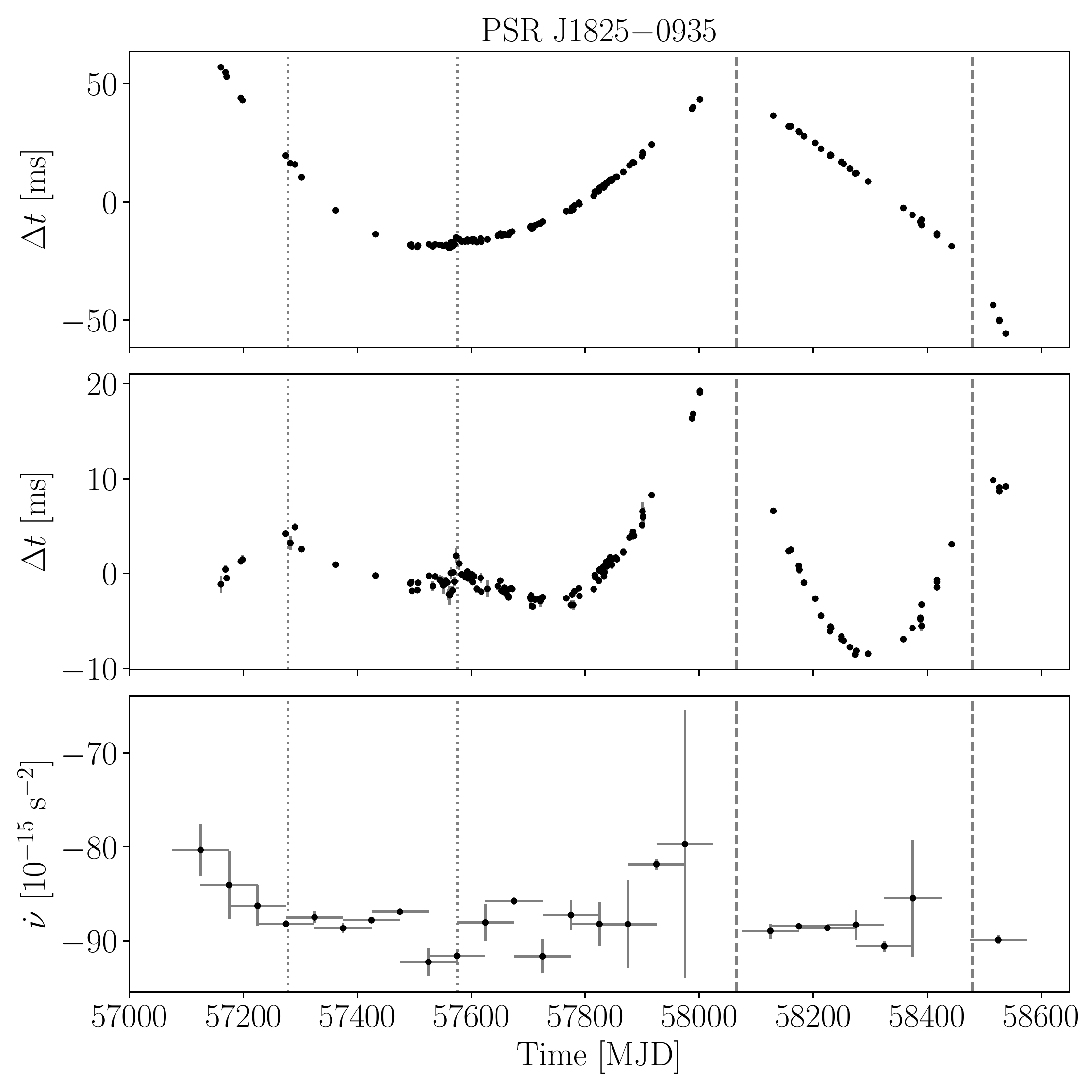}
    \caption{Post-fit timing residuals for PSR J1825$-$0935. Top plot shows the residuals after removing a fit for only $\nu$ and $\dot{\nu}$, while the middle plot includes a fit for $\ddot{\nu}$. Bottom plot shows changes in the spin-down frequency over time as determined via fitting $\dot{\nu}$ over $~\sim50$\,day segments (error bars indicate the 1-sigma error). Vertical dashed lines in all three panels correspond to the approximate epochs of the largest events; dotted lines indicate the two smaller events.}
    \label{fig:J1825_glitch}
\end{figure}

\subsubsection{Glitch-like events in PSR J1825$-$0935} 

Also known as PSR B1822$-$09, this pulsar has been reported to exhibit timing events, sometimes referred to as `slow glitches'~\citep{Zou2004,Shabanova2007}. These events are predominantly characterized by a sharp change in the spin-down of the pulsar, which leads to the pulsar spinning up over the course of a few days. This change in spin-down decays exponentially over timescales of days to months. PSR J1825$-$0935 is also known to switch between two emission states: a `B-mode' where an extra precursor component adjacent to the main pulse is visible, and a `Q-mode', where precursor emission is suppressed and emission from an interpulse component is brightest~\citep{Fowler1981, Morris1981, Gil1994}. \citet{Lyne2010} showed this switching between these two emission modes is correlated with changes in the spin-down rate, concluding the apparent `slow glitches' are not related to the glitch phenomena, but are instead a result of the pulsar spending more time in one emission/spin-down state versus the other. 

The upper panel of Fig.~\ref{fig:J1825_glitch} shows the timing residuals for PSR J1825$-$0935 after subtracting a fit for $\nu$ and $\dot{\nu}$. 
Two glitch-like events are found to have occurred during observing gaps centred at MJD $58065(64)$ and MJD $58484.8(9)$\footnote{The last event is listed as a glitch in the Jodrell Bank glitch catalogue at MJD $58486.2(9)$.}. 
Including a fit for $\ddot{\nu}$ in our timing model reveals two additional events with lower amplitudes that occurred at MJD $57278.5(41)$ and MJD $57576.1(26)$.
Modelling these four events as standard pulsar glitches, we perform parameter estimation using {\sc TempoNest} to fit for step changes in $\nu$ and $\dot{\nu}$. The recovered glitch parameters are presented in Table~\ref{tab:slow_glitches}.
The first two events are only consistent with upper limits on an instantaneous changes in $\nu$, while the changes in $\dot{\nu}$ both peak at negative values.
Changes in $\Delta\dot{\nu}_{g}$ for all four events are consistent with zero at the 95 percent confidence interval. 
However, the standard glitch model we employ does not sufficiently describe the true nature of these events.
By performing model selection, we find a PLRN model with no glitches is preferred over any glitch+PLRN model, with a $\ln(\mathcal{B}) = 14$ when comparing a PLRN-only model to PLRN+4 glitches and $\ln(\mathcal{B}) = 7.8$ in favour of the PLRN-only model versus a fit to only the two large events.
Subtracting off the purely red noise model the post-fit residuals are still dominated by the two larger glitch-like events, implying that at least these two events are not related to the pulsar's red noise.

\begin{table}
    \centering
    \caption{Recovered glitch parameters for the events in PSR J1825$-$0935.}
    \label{tab:slow_glitches}
    \renewcommand{\arraystretch}{1.5}
    \begin{tabular}{lccc}
        \hline
        \hline
        No. & $t_{\mathrm{g}}$ & $\Delta\nu_{\mathrm{g}}/\nu$ & $\Delta\dot{\nu_{\mathrm{g}}}/\dot{\nu}$ \\
        & (MJD) & ($\times 10^{-9}$) & ($\times 10^{-3}$) \\
        \hline
        1 & $57278.5(41)$ & $\lesssim 0.9$      & $-1.3^{+2.7}_{-2.0}$  \\
        2 & $57576.1(26)$ & $\lesssim 0.3$      & $-0.9^{+1.1}_{-2.0}$  \\
        3 & $58065.7(3)$  & $5.2^{+1.7}_{-0.5}$ & $-1.1^{+1.7}_{-2.4}$ \\
        4 & $58486.2(9)$  & $7.6^{+4.5}_{-3.3}$ & $14.6^{+37.3}_{-20.0}$ \\
        \hline
    \end{tabular}
    \renewcommand{\arraystretch}{}
\end{table}
The small variations in $\dot{\nu}$ in the bottom panel of Fig.~\ref{fig:J1825_glitch} at the time of each event are more in line with conventional pulsar glitches as opposed to slow glitches, although the lack of coverage around the more recent large amplitude events means we may have insufficient resolution to detect any rapid changes in spin-down. The mode-changing behaviour and glitch-like events of PSR J1825$-$0935 demand further investigation, as high-cadence coverage of these events, and any that are discovered in other mode-changing pulsars, may allow us to probe the internal dynamics of these neutron stars.


\section{Discussion}\label{sec:discussion}

\subsection{Quantifying timing noise strength}

While a complete characterisation of pulsar timing noise is yet to be achieved, previous work usually followed one of two approaches. The first involves applying a cubic polynomial to fit for $\ddot{\nu}$ to assess the effects of timing noise. The second uses the root-mean-square (RMS) of the residuals after subtracting a quadratic polynomial, which corresponds to a fit for only $\nu$ and $\dot{\nu}$. 

Studies undertaken by~\citet{Urama2006} and~\citet{Chukwude2007} use measurements of $\ddot{\nu}$ to directly infer the strength of the timing noise in their data sets, in addition to searching for correlations with other pulsar parameters. Other users of $\ddot{\nu}$ measurements include work by~\citet{Arzoumanian1994} through the use of a model-dependent parameter
\begin{equation}
    \Delta_{8} = \log\Big(\frac{|\ddot{\nu}|}{6\nu}T_{8}^{3}\Big),
\end{equation}
where $\ddot{\nu}$ is measured over a total (but arbitrary) observation time span of $T_{8} = 10^{8}$ s. A two-sample variance parameter $\sigma_{z}$ is used in~\citet{Matsakis1997} to describe pulsar rotational stability
\begin{equation}\label{eqn:sigma_z}
    \sigma_{z} = \frac{1}{2\sqrt{5}}\Big[\frac{\sigma_{\ddot{\nu}(T)}}{\nu}\Big]T^{2},
\end{equation}
where $\sigma_{\ddot{\nu}}(T)$ is the RMS of the $\ddot{\nu}$ fit over the observing span $T$.
\citet{Shannon2010} note that the $\Delta_{8}$ method is highly model-dependent since the measured $\ddot{\nu}$ will usually increase on longer timescales, requiring additional time dependent scaling to properly compare values of $\ddot{\nu}$ and $\Delta_{8}$. They also state methods based around measurements of $\ddot{\nu}$ (such as $\sigma_{z}$) will often underestimate the amount of timing noise as they neglect contributions from higher-order frequency derivatives.

A method proposed by~\citet{Cordes1980} assesses the RMS of the total timing noise after conducting a second order fit
\begin{equation}
    \sigma_{\mathscr{R},2}^{2}(T) = \frac{1}{N} \sum_{i}^{N}\mathscr{R}(t_{i})^{2},
\end{equation}
where $\mathscr{R}$ refers to the timing residuals and $N$ is the number of ToAs. This can be further broken down into red and white components
\begin{equation}
    \sigma_{\mathscr{R},2}^{2}(T) = \sigma_{\mathrm{TN},2}^{2}(T) + \sigma_{\mathrm{W}}^{2}(T).
\end{equation}
It is assumed the RMS is usually dominated by $\sigma_{\mathrm{TN},2}^{2}$ in slow ($P \sim 1$\,s) pulsars.
The timing noise strength is then estimated via an activity parameter that describes the scaling of $\sigma_{\mathrm{TN},2}$ with respect to PSR J0534+2200 (the Crab pulsar) by
\begin{equation}
    A = \log\Big[\frac{\sigma_{\mathrm{TN},2}(T)}{\sigma_{\mathrm{TN},2}(T)_{\mathrm{Crab}}}\Big].
\end{equation}
This method assumes the timing noise scales in the same way as the Crab pulsar.
\citet{Dewey1989}, and later~\citet{Shannon2010} built upon this method by assessing how timing noise varies across the population by examining a scaling relationship between timing noise strength and $\nu$, $\dot{\nu}$ and the observation time span $T$ (in years) as
\begin{equation}\label{eqn:shannon_metric}
    \hat{\sigma}_{\mathrm{TN},2} = \mathrm{C_{2}}\nu^{a}|\dot{\nu}|^{b} \,T^{\gamma}.
\end{equation}
Here, the fitting factors $\mathrm{C}_{2}$, $a$, $b$ and $\gamma$ are measured from the total pulsar population using fits based on maximum likelihood statistics. 

We note these methods do not attempt to model the timing noise directly. Instead, they assume the RMS of the residuals accurately describes the timing noise strength. This neglects covariances between intrinsic pulsar properties and red noise, which can result in contaminated residuals as some pulsar properties may be over- or under-fit. This is not an issue for modern Bayesian methods that model both deterministic and stochastic properties simultaneously.

The method we use for assessing timing noise strength was developed in parallel with~\citet{Parthasarathy2019}, in which timing noise strength is inferred from the red noise amplitude and spectral index, obtained via parameter estimation with {\sc TempoNest}, and the observation span as
\begin{equation}\label{eqn:rn_strength}
    \sigmaRN^{2} = A^{2} \,T^{\beta-1}.
\end{equation}
Using this metric, we find the magnetar PSR J1622$-$4950 has the strongest timing noise in our sample. 
However, as stated earlier the torque variations due to the magnetar's decaying magnetic field, rather than spin noise, are expected to dominate the observed red noise. In addition,~\citet{Shannon2010} argued that timing noise in magnetars is statistically different to that in millisecond and non-recycled pulsars. Given the red noise we observe in both the millisecond pulsars and PSR J1622$-$4950 can be explained via processes other than rotational irregularities, we restrict our analysis to the {\numcp} non-recycled pulsars in our sample, {\numredcanon} of which strongly favour the PLRN model.

\subsection{Correlations with individual pulsar properties}\label{sec:correlations}

A number of pulsar properties can be inferred from their spin and spin-down. These include the characteristic age ($\tau_{c}$), surface dipole magnetic field strength ($B_{\mathrm{surf}}$) and the rotational kinetic energy loss over time ($\dot{E}$). Simplified approximations to these properties, along with the pulse period derivative ($\dot{P}$), can be expressed in terms of the spin and spin-down frequencies, as follows
\begin{equation}
\begin{aligned}
    & \dot{P} \propto \nu^{-2} |\dot{\nu}|\\
    & \tau_{c} \propto \nu |\dot{\nu}|^{-1}\\
    & B_{\mathrm{surf}} \propto \nu^{-3/2}|\dot{\nu}|^{1/2}\\
    & \dot{E} \propto \nu |\dot{\nu}|.
\end{aligned}
\end{equation}
Similar to previous work on pulsar timing noise, we examine whether correlations exist between the measured red noise strength and these pulsar properties, in addition to the spin and spin-down on their own.
Assuming timing noise strength scales with spin and spin-down frequencies in a similar fashion to Equation~\ref{eqn:shannon_metric}, we compared the inferred strength against a predictive metric
\begin{equation}\label{eqn:tn_propto}
    \chiRN \propto \nu^{a} |\dot{\nu}|^{b},
\end{equation}
where the values of $a$ and $b$ can be set to the approximate pulsar properties we are comparing.

From a frequentist perspective, the amount of correlation between $\sigmaRN$ and $\chiRN$ can be quantified via the Pearson correlation coefficient
\begin{equation}\label{eqn:pearson_corr}
    r_{p} = \frac{\sum_{i=1}^{N} (\sigmaRNi - \mu_{\sigma})(\chiRNi - \mu_{\chi})}{[\sum_{i=1}^{N} (\sigmaRNi - \mu_{\sigma})^{2} \sum_{i=1}^{N} (\chiRNi - \mu_{\chi})^{2}]^{1/2}},
\end{equation}
where $\mu_{\sigma} = \frac{1}{N}\sum_{i=1}^{N} \sigmaRNi$ is the mean of the $\sigma_{\mathrm{RN}}$ values, and $\mu_{\chi}$ is the mean of $\chiRN$. 
However, this approach does not take into account potential covariances between the means of $\sigmaRNi$ and $\nu^{a}|\dot{\nu}|^{b}$, or scatter in the measurements. It is also not robust against the influence of outliers in the data set. 

\begin{table}
    \centering
    \caption{Frequentist ($r_{p}$) and Bayesian ($\rho$) correlations between pulsar properties and red noise strength, using only strong red noise detections (D) and including lower confidence detections (D + PD). Errors represent the 95 percent confidence intervals.}
    \label{tab:correlations}
    \renewcommand{\arraystretch}{1.4}
    \begin{tabular}{lcccc}
        \hline
        \hline
         &  \multicolumn{2}{c}{D} & \multicolumn{2}{c}{D + PD} \\
        \cmidrule(lr){2-3}\cmidrule(lr){4-5}
         & $r_{p}$ & $\rho$ & $r_{p}$ & $\rho$  \\
        \hline
        $\nu$ & $0.13$ & $0.18^{+0.29}_{-0.26}$       & $0.16$ & $-0.18^{+0.33}_{-0.25}$ \\
        $\dot{\nu}$ & $0.45$ & $0.47^{+0.29}_{-0.20}$ & $0.46$ & $0.47^{+0.27}_{-0.20}$ \\
        $\dot{P}$ & $0.49$ & $0.51^{+0.28}_{-0.18}$   & $0.49$ & $0.48^{+0.24}_{-0.21}$ \\
        $\tau$ & $-0.51$ & $-0.51^{+0.19}_{-0.25}$    & $-0.51$ & $-0.53^{+0.17}_{-0.26}$ \\
        $B_{\mathrm{surf}}$ & $0.37$ & $0.39^{+0.30}_{-0.22}$ & $0.37$ & $0.38^{+0.28}_{-0.22}$ \\
        $\dot{E}$ & $0.39$ & $0.41^{+0.30}_{-0.21}$   & $0.41$ & $0.42^{+0.29}_{-0.20}$. \\
        \hline
    \end{tabular}
    \renewcommand{\arraystretch}{}
\end{table}

An alternative approach involves assuming the red noise measurements and values generated from equation~\ref{eqn:tn_propto} are correlated samples drawn from an underlying bivariate Gaussian distribution, the shape of which is best described by the set of hyper-parameters $\{ \mu_{\sigma}, \mu_{\chi}, \sigma_{\sigma}, \sigma_{\chi}, \rho \}$ as
\begin{equation}
\begin{aligned}
    \mathcal{N}_{2}(\sigmaRN, \chiRN) = \frac{1}{2\pi \sigma_{\sigma} \sigma_{\chi}\sqrt{1 - \rho^{2}}} \exp\Big[\frac{-1}{2(1 - \rho^{2})} \\
    \times  \Big(\frac{\sigmaRN^{2}}{\sigma_{\sigma}} + \frac{\chiRN^{2}}{\sigma_{\chi}} - \frac{2\rho\sigmaRN\chiRN}{\sigma_{\sigma}\sigma_{\chi}}\Big)\Big].
\end{aligned}
\end{equation}
Here, $\mu_{\sigma}$, $\sigma_{\sigma}$ are the mean and variance of the distribution in the $\sigmaRN$ direction, and $\mu_{\chi}$, $\sigma_{\chi}$ represent the mean and width in the $\chiRN$ direction. The parameter $\rho$ indicates the direction in which the bivariate Gaussian is rotated and provides an estimate for the level of correlation between $\sigmaRN$ and $\chiRN$. For simplicity, we express the bivariate Gaussian as $\mathcal{N}_{2}(\sigmaRN, \chiRN) = \theta^{\mathrm{T}}\mathbf{C}^{-1}\theta$, where $\mathbf{C}$ is the covariance matrix
\begin{equation}\label{eqn:cov_matrix}
    \mathbf{C} = 
    \begin{bmatrix}
     \sigma_{\sigma}^{2} & \rho\sigma_{\sigma}\sigma_{\chi} \\ \rho\sigma_{\sigma}\sigma_{\chi} &\sigma_{\chi}^{2}
    \end{bmatrix},
\end{equation} 
and $\theta = (\sigmaRN - \mu_{\sigma},\, \chiRN - \mu_{\nu})$.
We can then write the likelihood function from which our samples are drawn from as
\begin{equation}
    \mathcal{L}(\theta | \mathbf{C}) = \frac{1}{2\pi\sqrt{|\mathbf{C}|}} \prod_{i=1}^{N} \exp \Big[ \frac{-1}{2} \theta^{\mathrm{T}}_{i} \mathbf{C}^{-1} \theta_{i} \Big],
\end{equation}
where $|\mathbf{C}| = \sigma_{\sigma}^{2} \sigma_{\chi}^{2} (1 - \rho^{2})$. 
We use the {\sc Bilby} software library~\citep{bilby} and {\sc PyMultiNest}~\citep{Buchner2014}, a Python wrapper for the {\sc MultiNest} algorithm, to sample the hyper-parameter posterior distributions using the {\numredcanon} non-recycled pulsars that strongly favour the PLRN model, ignoring those that are consistent with the WTN model. The resulting Frequentist and Bayesian correlation coefficients for pulsars with strong evidence for red noise (D; $\ln(\mathcal{B}) > 5$) and when including those with less confident evidence (D + PD; $3 < \ln(\mathcal{B}) < 5$) are presented in Table~\ref{tab:correlations}.

We find the strongest correlations exist with $\dot{P}$ and $\dot{\nu}$, in addition to a similar anti-correlation with characteristic age. These correlations are smaller than those from the $\sigma_{z}$ analysis performed by~\citet{Hobbs2010}, who presented an analysis of the ongoing timing campaign of a large sample of pulsars at the Jodrell Bank Observatory ($N = 366$, $T_{\mathrm{mean}} \sim 19$\,yr), but are similar to those from~\citet{Namkham2019} who assessed the timing noise of 129 `middle-aged' ($\tau_{c} \sim 1$\,Myr) pulsars observed by the Parkes radio telescope over $\sim$4\,yr using the $\sigma_{z}$ metric.
We find pulsar spin-frequency has effectively no correlation with timing noise, but the weak correlation of $0.3$ from~\citet{Hobbs2010} does overlap with the 95 percent confidence region of our Bayesian correlation parameter for the pulsars that strongly prefer the PLRN model.
These differences are to be expected as~\citet{Hobbs2010} included both millisecond and partially-recycled pulsars when calculating their correlation coefficients, while we are limited to non-recycled pulsars.

\subsection{Scaling relation fitting and hyper-parameter estimation}\label{sec:model_indep}

\begin{figure*}
    \centering
    \includegraphics[width=0.54\linewidth]{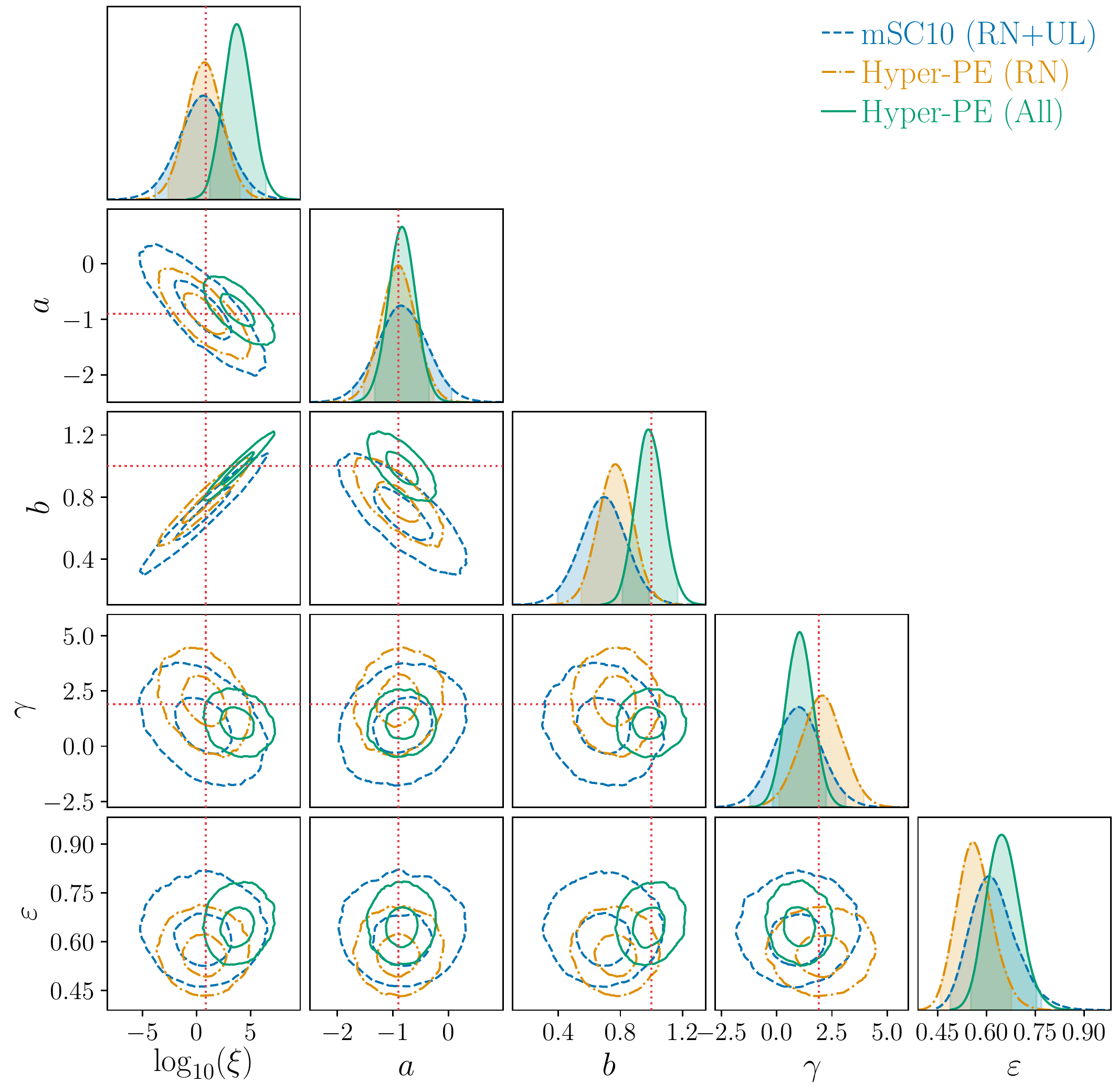}
    \caption{One- and two-dimensional posterior distributions for the scaling parameters across the non-recycled pulsar population. Contours in the two-dimensional posteriors indicate the 50 and 95 percent confidence regions. Shading in the one-dimensional posteriors covers the 95 percent confidence intervals. The red dotted lines indicate the non-recycled pulsar maximum likelihood values from~\citet{Shannon2010}.}
    \label{fig:hyper_posteriors}
\end{figure*}

To see how the timing noise strength varies independently of specific pulsar properties, we build upon previous work by~\citet{Dewey1989} and~\citep{Shannon2010} for finding a scaling relation that maps timing noise strength to a scaling of pulsar $\nu$ and $\dot{\nu}$, and observing timescale $T$. We rewrite their scaling relation (Equation~\ref{eqn:tn_propto}) as
\begin{equation}\label{eqn:improv_rn_metric}
    \chiRN = \xi\,\nu^{a} |\dot{\nu}|^{b} \,T^{\gamma},
\end{equation} 
where~\citet{Shannon2010} use the symbol $\mathrm{C_{2}}$ in place of $\xi$ to represent the linear scaling factor.
Unlike in Section~\ref{sec:correlations}, $a$ and $b$ are not set to fixed values to approximate certain pulsar properties. Instead we allow them to be free parameters with uniformly sampled priors ($-10 < \pi(a,b) < 10$).

\citet{Shannon2010} define the joint likelihood function
\begin{equation}\label{eqn:joint_like}
    \mathcal{L}(\sigmaRN, \sigma_{\mathrm{UL}} | \chiRN, \varepsilon) = \mathcal{L}(\sigmaRN | \chiRN, \varepsilon) \times  \mathcal{L}(\sigma_{\mathrm{UL}} | \chiRN, \varepsilon),
\end{equation}
which is comprised of a standard Gaussian likelihood with an additional hyper-parameter $\varepsilon^{2}$ to describe the scatter in the measured red noise strength in the pulsar sample, which may be attributed to variations in the amount of turbulence in their superfluid interiors~\citep{Melatos2014}, and an upper-limit likelihood. The likelihood functions contributing to Equation~\ref{eqn:joint_like} are given by 
\begin{equation}\label{eqn:like_rn_det}
    \mathcal{L}(\sigmaRN | \chiRN, \varepsilon) = \prod_{i}^{N} \frac{1}{\sqrt{2\pi\varepsilon^{2}}} \exp\Big[ -\frac{(\sigma_{\mathrm{RN}, i} - \chiRNi)^{2}}{2\varepsilon^{2}} \Big],
\end{equation}
and
\begin{equation}\label{eqn:like_rn_upp}
    \mathcal{L}(\sigma_{\mathrm{UL}} | \chiRN, \varepsilon) = \prod_{j}^{N} 1 -  \frac{1}{2} \mathrm{erfc}\Big[ -\frac{(\sigma_{\mathrm{UL}, j} - \mu_{\mathrm{RN}, j})}{\varepsilon\sqrt{2}} \Big],
\end{equation}
where $\sigma_{\mathrm{UL}}$ is the `upper limit' on the red noise strength and $\mathrm{erfc}$ is the complementary error function. We use this likelihood to calculate posterior distributions for the scaling hyper-parameters as follows
\begin{equation}\label{eqn:post_SC10}
    p(\chiRN, \varepsilon | \sigmaRN, \sigma_{\mathrm{UL}}) \propto \mathcal{L}(\sigmaRN, \sigma_{\mathrm{UL}} | \chiRN, \varepsilon) \pi(\chiRN, \varepsilon).
\end{equation}
This modified version of the~\citet{Shannon2010} formalism (mSC10 hereafter) requires making two key assumptions: scatter in the maximum likelihood posterior values of $\sigmaRN$ due to measurement uncertainties are either negligible or absorbed by $\varepsilon$, and the upper limit likelihood holds true for the pulsars with only a marginal preference (i.e. $ 1 < \ln(\mathcal{B}) < 3$) for the PLRN model. It also does not take into account the information that can be gained by including the full posterior distribution for $\sigmaRN$ during the fitting.

We can overcome these shortcomings by assuming our measurements of $\sigmaRN$ for a given pulsar is drawn from a Gaussian distribution, the mean of which depends on the aforementioned scaling of the pulsars spin and spin-down frequencies (equation~\ref{eqn:improv_rn_metric}), and a variance $\varepsilon^{2}$ defined by
\begin{equation}
    \pi(\sigmaRN | \chiRN, \varepsilon) = \frac{1}{\sqrt{2\pi\varepsilon^{2}}} \exp \Big[ -\frac{(\sigmaRN - \chiRN)^{2}}{2\varepsilon^{2}} \Big].
\end{equation}
This distribution represents an approximation to the `true' probability distribution of $\sigmaRN$ across the population.
To compute the posterior distributions for our scaling hyper-parameters ($\{\xi, a, b, \gamma\}$), we use the marginalized likelihood
\begin{equation}
    \mathcal{L}(\mathbf{r} | \chiRN, \varepsilon) = \int d\sigmaRN \mathcal{L}(r | \sigmaRN) \pi(\sigmaRN | \chiRN, \varepsilon).
\end{equation}
As we are using an ensemble of $N$ individual pulsars with residuals $\mathbf{r} = \{ r_{1}, ... r_{N} \}$, we can take the product of the individual likelihoods to obtain the total likelihood of the timing data given the red noise strength for the population
\begin{equation}
    \mathcal{L}_{\mathrm{tot}}(\mathbf{r} | \sigmaRN) = \prod_{i}^{N} \mathcal{L}(r_{i} | \sigmaRNi),
\end{equation}
hence the total marginalized likelihood can be rewritten as
\begin{equation}\label{eqn:hype_like_inc}
    \mathcal{L}_{\mathrm{tot}}(\mathbf{r} | \chiRN, \varepsilon) = \prod_{i}^{N} \int d\sigmaRNi \mathcal{L}(r_{i} | \sigmaRNi) \pi(\sigmaRNi | \chiRNi, \varepsilon).
\end{equation}
From Bayes theorem, we can find $\mathcal{L}(r_{i} | \sigmaRNi)$ as
\begin{equation}\label{eqn:rn_like}
    \mathcal{L}(r_{i} | \sigmaRNi) = \mathcal{Z}(r_{i}) \frac{p(\sigmaRNi | r_{i})}{\pi(\sigmaRNi)}, 
\end{equation}
where the prior on $\sigmaRNi$ is the product of the log-uniform prior on $A$ and the uniform prior on $\beta$
\begin{equation}
    \pi(\sigma_{\mathrm{RN}}) = \pi(A, \beta) = \pi(A) \pi(\beta) = \frac{1}{A},
\end{equation}
for $A \in \{ 10^{-20}, 10^{-3} \}$.
We then substitute equation~\ref{eqn:rn_like} into equation~\ref{eqn:hype_like_inc} to obtain
\begin{equation}
    \mathcal{L}(\mathbf{r} | \chiRN, \varepsilon) = \int  d\sigmaRNi \mathcal{Z}(r_{i}) p(\sigmaRNi | r_{i}) \frac{\pi(\sigmaRNi | \chiRNi, \varepsilon)}{\pi(\sigmaRNi)}.
\end{equation}
Converting the integral over $\sigmaRNi$ to a sum over the posterior samples, as $\int dx\,p(x) f(x) \approx 1/n_{s} \sum_{i}^{n_{s}} f(x_{i})$~\citep{Hogg2018}, the final likelihood function is 
\begin{equation}
    \mathcal{L}(\mathbf{r} | \chiRN, \varepsilon) = \prod_{i}^{N} \frac{\mathcal{Z}(r_{i})}{n_{i}} \sum_{k}^{n_{i}} \frac{\pi(\sigma_{\mathrm{RN},i,k} | \chiRNi, \varepsilon)}{\pi(\sigma_{\mathrm{RN},i,k})},
\end{equation}
where $k$ is the number of posterior samples for the $i$-th pulsar.
We can then combine this likelihood with the prior for the hyper-parameters, $\pi(\chiRN, \varepsilon)$, and the Bayesian evidence for the timing data, to obtain the posterior distributions for the hyper-parameters
\begin{equation}
    p(\chiRN, \varepsilon | \mathbf{r}) = \frac{\mathcal{L}(\mathbf{r} | \chiRN, \varepsilon)\pi(\chiRN, \varepsilon)}{\mathcal{Z}(\mathbf{r})}.
\end{equation}
As with equation~\ref{eqn:post_SC10}, the posterior distributions are sampled using {\sc PyMultiNest}. 
We present the resulting one- and two-dimensional posterior distributions in Fig.~\ref{fig:hyper_posteriors}, comparing results from the mSC10 method, hyper-parameter estimation (Hyper-PE) using only the {\numredcanon} non-recycled pulsars that favour the PLRN model, and the resulting improvement when all 280 non-recycled pulsars are included in the Hyper-PE method regardless of the preferred model.

\begin{table}
    \centering
    \caption{Comparison between recovered maximum likelihood posterior values for the scaling hyper-parameters and their associated 95 percent confidence intervals.}
    \label{tab:hyper_results}
    \renewcommand{\arraystretch}{1.5}
    \begin{tabular}{cccc}
        \hline
        \hline
		Parameter & mSC10 (RN+UL) & Hyper-PE (RN) & Hyper-PE (All) \\
		\hline
		$\log_{10}(\xi)$ & $0.6^{+4.3}_{-4.4}$     & $1.0^{+3.4}_{-3.0}$     & $3.7^{+2.4}_{-2.7}$     \\
		$a$              & $-0.87^{+0.83}_{-0.91}$ & $-0.88^{+0.63}_{-0.60}$ & $-0.84^{+0.47}_{-0.49}$ \\
		$b$              & $0.69 \pm 0.29$         & $0.77^{+0.21}_{-0.22}$  & $0.97^{+0.16}_{-0.19}$  \\
		$\gamma$         & $1.00 \pm 2.17$         & $2.1^{+2.0}_{-1.8}$     & $1.0 \pm 1.2$           \\
		$\varepsilon$    & $0.60^{+0.11}_{-0.16}$  & $0.56^{+0.10}_{-0.12}$  & $0.64^{+0.11}_{-0.16}$  \\
		\hline
    \end{tabular}
    \renewcommand{\arraystretch}{}
\end{table}
Recovered values for each scaling hyper-parameter from both methods are listed in Table~\ref{tab:hyper_results}. It is clear the Hyper-PE method returns improved estimates over the mSC10 method (with the exception of $\varepsilon$, which is consistent between all three methods), as indicated by the smaller confidence regions. Including the additional {\numwtn} white noise dominated pulsars provides additional improvements, as the Hyper-PE method takes into account additional information by summing over the entire posterior distribution of $\sigmaRN$, rather than only using the maximum likelihood posterior value. Our recovered value of $\varepsilon$ differs from the value of $\varepsilon = 1.6 \pm 0.1$ reported by~\citet{Shannon2010}. This inconsistency could be due to the use of two different methods of modelling timing noise in pulsars, resulting in a different amount of measurement scatter.

We can compare our results to those in the literature by looking at the specific scaling relation from equation~\ref{eqn:tn_propto}. From our Hyper-PE method, we find the timing noise strength of the non-recycled pulsars in our sample follow the scaling relation \begin{equation}\label{eqn:scaling}
    \chiRN \propto \nu^{-0.84^{+0.47}_{-0.49}} |\dot{\nu}|^{0.97^{+0.16}_{-0.19}}.
\end{equation}
\citet{Shannon2010} computed a scaling relation of $\sigma_{\mathrm{TN},2} \propto \nu^{-0.9\pm0.2}|\dot{\nu}|^{1.0\pm0.05}$, or $\sigma_{\mathscr{R},2} \propto \nu^{-0.7\pm0.1}|\dot{\nu}|^{0.76\pm0.02}$ when including the effects of additional white noise, while the analysis by \citet{Hobbs2010} found the relation $\sigma_{z}(10\,\mathrm{yr}) = 10^{-11.5}\nu^{-0.4}|\dot{\nu}_{-15}|^{0.8}$, where $|\dot{\nu}_{-15}|$ is the spin-down rate in units of $10^{-15}$\,s$^{-2}$. 
More recently,~\citet{Parthasarathy2019} made use of {\sc TempoNest}, and the same timing noise strength metric we used, to analyse 85 `young' ($\tau_{c} \lesssim 1$\,Myr), high-$\dot{E}$ pulsars with $\sim$10\,years of timing observations. Using a grid search to find the maximally correlated $\nu$ scaling index -- at a fixed scaling parameter of 1 for $\dot{\nu}$ -- they found a scaling relation of $\sigma_{P} \propto \nu^{-0.9 \pm 0.1} |\dot{\nu}|^{1}$. This same grid search method was also used by~\citet{Namkham2019} to infer their scaling of the $\sigma_{z}$ parameter, obtaining the relation $\sigma_{P} \propto \nu^{-1.7}|\dot{\nu}|^{1.0}$.

\begin{figure}
    \centering
    \includegraphics[width=\linewidth]{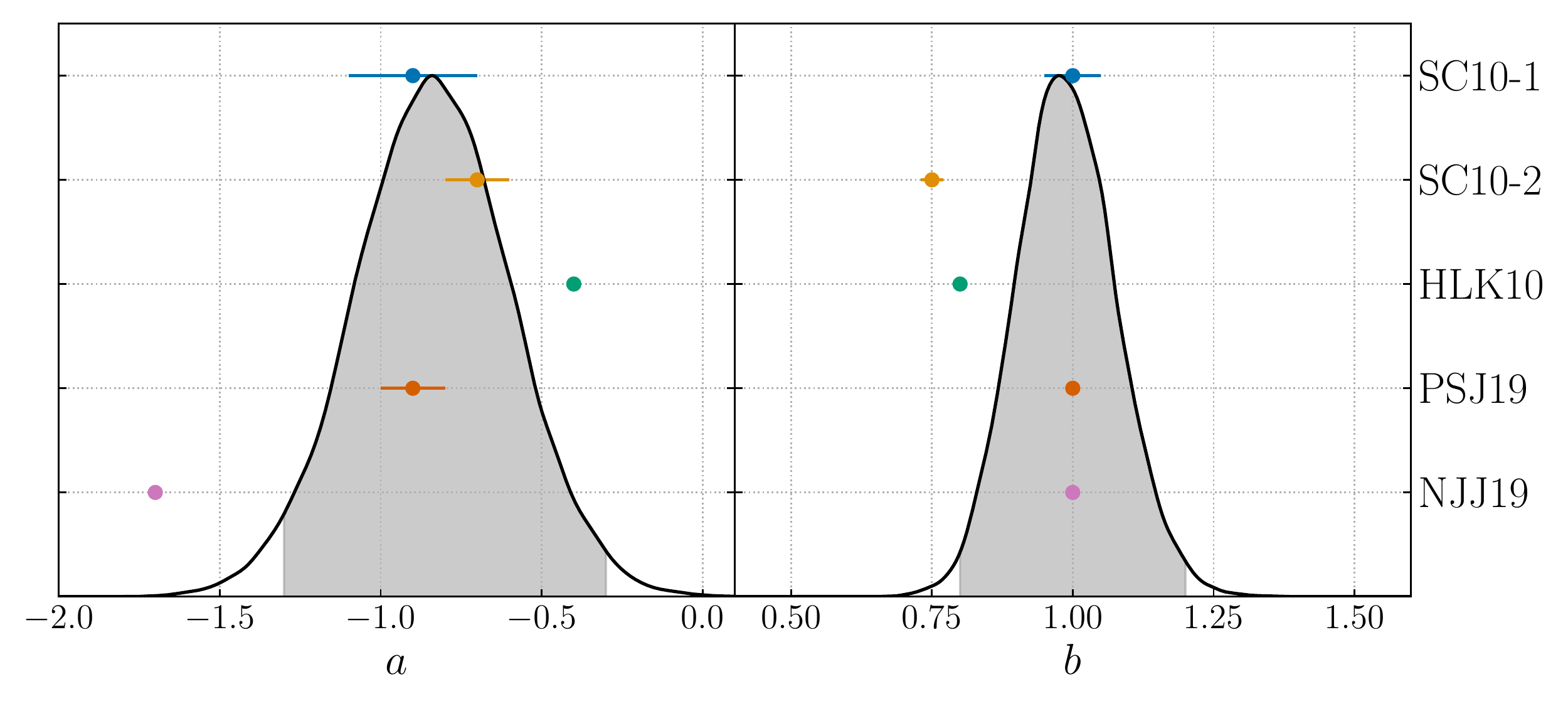}
    \caption{Comparison between our Hyper-PE (All) posteriors for $a$ and $b$ (black curves, shading represents the 95 percent confidence regions), and values (with 1-$\sigma$ errors) from~\citet{Shannon2010} (SC10-1: $\sigma_{\mathrm{TN},2}$ and SC10-2: $\sigma_{\mathscr{R},2}$),~\citet{Hobbs2010} (HLK10),~\citet{Parthasarathy2019} (PSJ19) and~\citet{Namkham2019} (NJJ19).}
    \label{fig:scale_comp}
\end{figure}

The values of $a$ and $b$ from each of these relations are compared with our results in Fig.~\ref{fig:scale_comp}. Our relation is entirely consistent with~\citet{Shannon2010}'s $\sigma_{\mathrm{TN,2}}$ scaling, while both relations from~\citet{Hobbs2010} and~\citet{Parthasarathy2019} fall within our 95-percent confidence regions. Improving our measurements of $a$ and $b$ can be achieved by adding additional pulsars to our sample and/or by extending the lengths of our timing baselines. The improvement made by adding more pulsars is illustrated by the $\sim$22 percent reduction in the Hyper-PE confidence regions in Fig.~\ref{fig:hyper_posteriors} after including the {\numwtn} pulsars that favour the WTN model in our analysis. Additional observations over longer timing baselines may allow us to obtain improved red noise amplitude and spectral index measurements, and detect low amplitude red noise in pulsars that currently favour the WTN model.

Applying a consistent approach to measuring timing noise strength in various data sets is of particular importance when it comes to comparing observations with theoretical models of timing noise processes. Our method of performing parameter estimation on the stochastic properties of individual pulsars with {\sc TempoNest} followed by using hyper-parameter estimation to infer the scaling across the population can be easily extended to other large pulsar timing programmes, or even modified to accommodate astrophysically motivated distributions on the expected spectral properties of timing noise~\citep[see, e.g][]{Melatos2014}. Model selection studies could also allow for different physical timing noise models to be compared, along with their implications for our understanding of the dynamic processes and internal structure of neutron stars.

\subsection{Two noteworthy pulsars}

Here we discuss results for two pulsars of particular interest: PSR J0737$-$3039A, for which we constrain the decay of its orbital period due to gravitational-wave emission, and PSR J1402$-$5124, whose celestial coordinates we find to be different to published values.

\subsubsection*{PSR J0737$-$3039A}

\begin{figure}
    \centering
    \includegraphics[width=\linewidth]{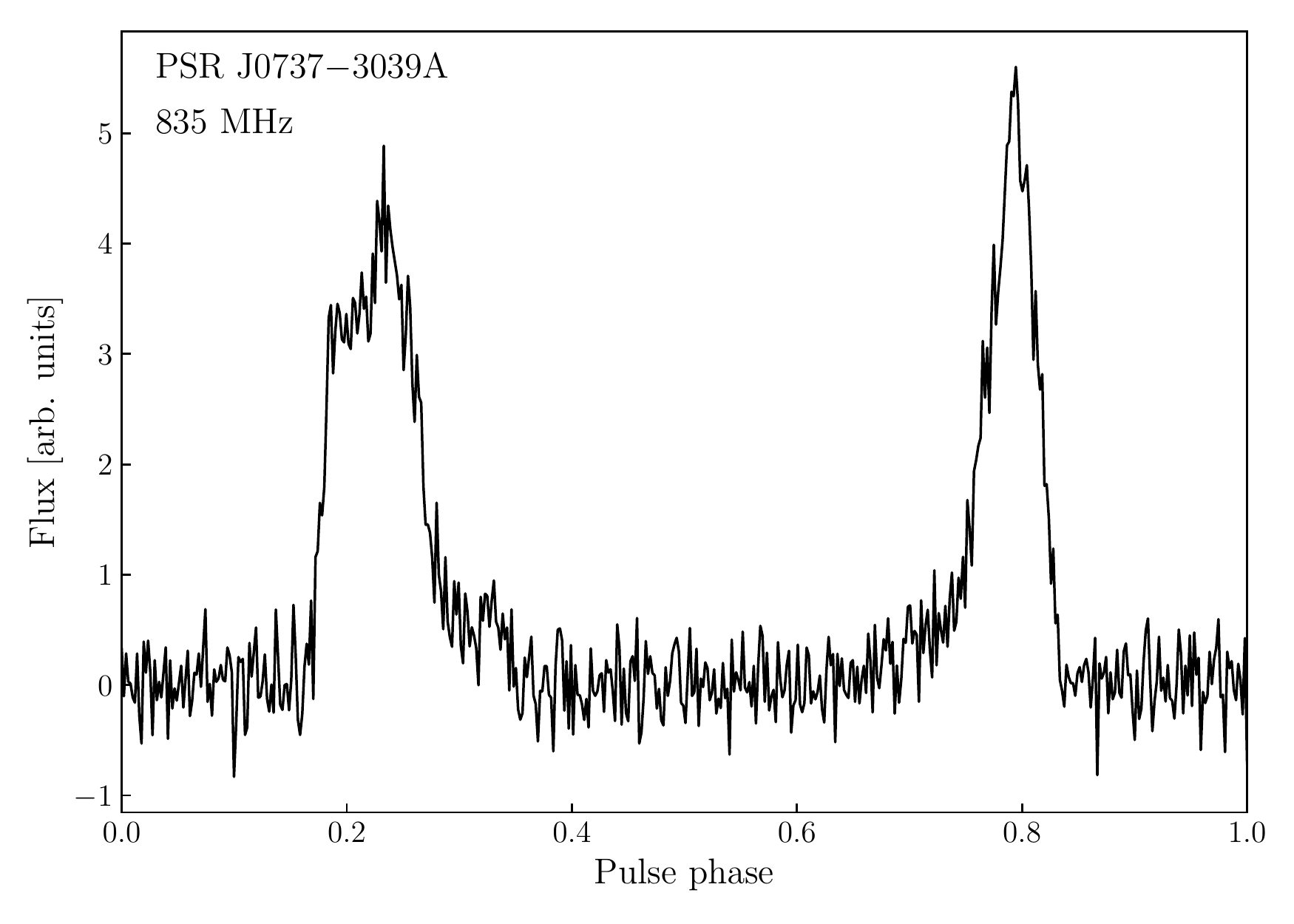}
    \caption{Average pulse profile of PSR J0737$-$3039A at 835\,MHz after summing 71.1 hours worth of observations taken over four years.}
    \label{fig:0737_profile}
\end{figure}

J0737$-$3039A is the `A' pulsar of the renowned double pulsar system discovered by~\cite{Burgay2003}. The `B' pulsar~\citep{Lyne2004} is currently not visible due to its magnetic-axis precessing out of our line-of-sight~\citep{Perera2010}. 
As its name suggests, the double pulsar allowed a determination of the mass ratio $R$ by measuring the two semi-major axes of the pulsars. 
When combined with the sum of the masses derived from the advance of periastron, this completely determines the constituent masses to high precision, and predicts the rate of orbital decay due to the emission of gravitational waves. 
Using {\sc TempoNest} to conduct parameter estimation on the pulsar's rotational and binary parameters, we find the relativistic properties of the system to be consistent with the masses and GR parameters measured by~\citet{Kramer2006b}. This produces the integrated profile seen in Fig.~\ref{fig:0737_profile}. 
The (albeit limited) timing  precision is good enough for us to spot any potential glitches in the pulsar's rotation, assist in dispersion measure variation monitoring, and to be used in undergraduate projects to demonstrate post-Keplerian effects such as advance of periastron and orbital decay to better than 1\,percent accuracy.

\subsubsection*{PSR J1402$-$5124}

\begin{figure}
    \centering
    \includegraphics[width=\linewidth]{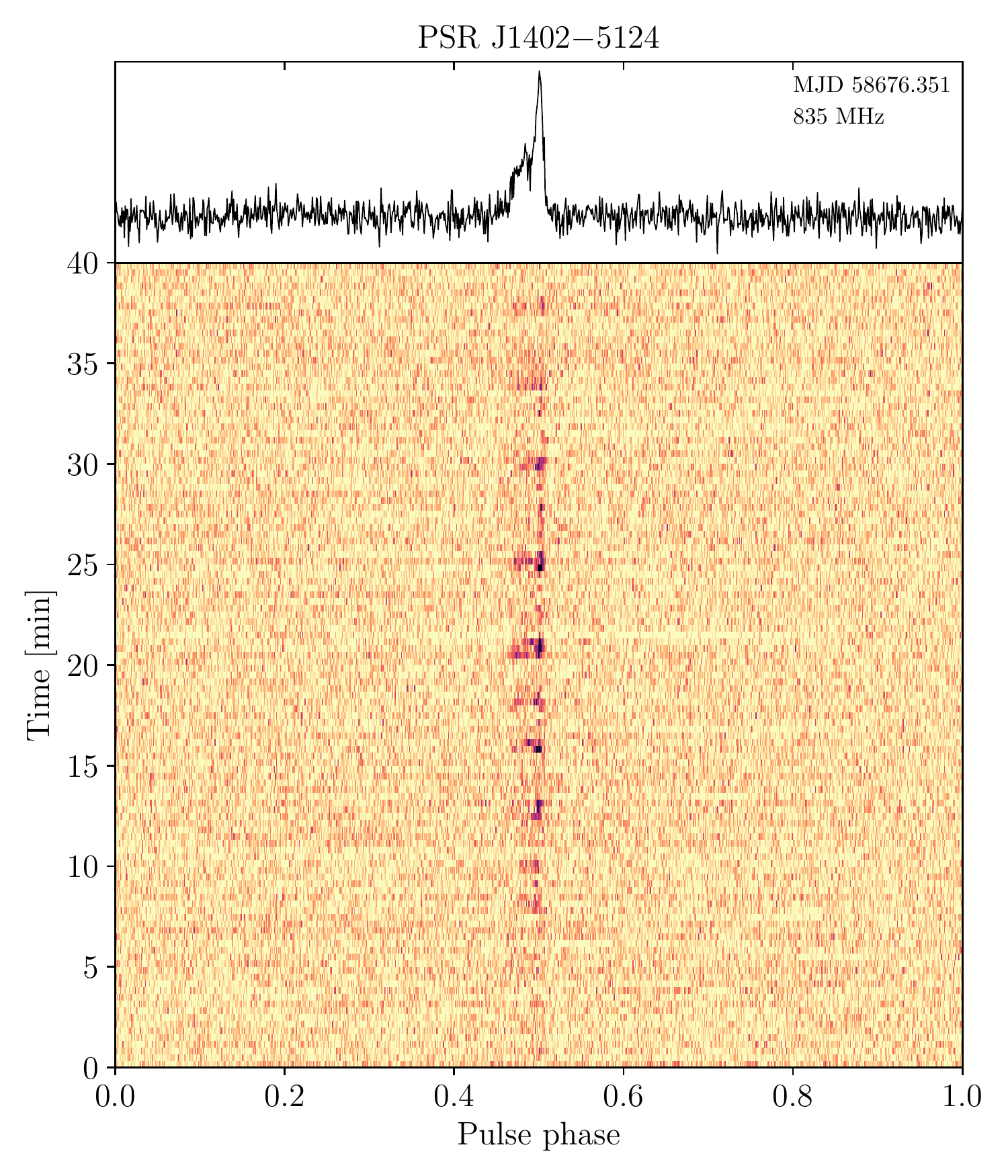}
    \caption{Stack of $20$\,s sub-integrations over 40 minutes (bottom) and the integrated pulse profile (top) of PSR J1402$-$5124 at $835$\,MHz. Brightening (dimming) of the pulsar toward the lower (upper) edge of the figure is due to the telescope's beam response. Dynamic range has been reduced to highlight profile changes between sub-integrations.}
    \label{fig:J1402}
\end{figure}

During our regular FRB-search transit observations, the real-time detection pipeline reported a candidate pulse from an unknown source with a DM of 53\,pc\,cm$^{-3}$ and a S/N of 10.2 at MJD 58657.40992245. Upon inspection of the data, we detected many faint single pulses with similar morphology drifting through our fan-beams. A periodicity search on the data `stitched' according to the sky-drift-rate revealed a high S/N pulsar candidate with a period of 1.38 s and DM $= 51(9)$\,pc\,cm$^{-3}$, closely matching the properties of pulsar PSR J1402$-$5124 reported by~\citet{Manchester1978}. A first-order localisation of the source, however, yielded a sky position that was inconsistent with the coordinates reported in the pulsar catalogue. Tracking the source using finely-spaced fan-beams over the next few days, we optimized the coordinates of the pulsar to: RA $= 14$:$02$:$56.0(2)$, DEC $= -50$:$21$:$43(49)$. The improved declination measurement is consistent with the value of DEC $= -50$:$20(5)$ reported by~\citet{Edwards2001}. In Fig.~\ref{fig:J1402} we highlight the variability of the pulse profile by plotting the phase vs time of the pulsar throughout a 40\,minute observation after placing a tied-array-beam on the updated coordinates.
Using 5 epochs of timing observations, we are able to constrain the spin-period of the pulsar to $P = 1.380182295(4)$\,s. Subtracting this new period measurement from the value reported in~\citet{Manchester1978} we derive an estimated spin-down of $\dot{P} = -5.413(4) \times 10^{-15}$, placing it in the population of `middle-aged' pulsars ($\tau_{c} \approx$ 4\,Myr). The astrometric and rotational properties will be further constrained as we continue to time the pulsar.

\section{Conclusions}\label{sec:conclusion}
We have performed an initial study of the rotational properties of $\numpsr$ bright, southern-sky radio pulsars observed by UTMOST using the Bayesian pulsar timing software {\sc TempoNest} to characterise the stochastic properties of our pulsar sample and to obtain unbiased measurements of $\nu$ and $\dot{\nu}$. Three millisecond pulsars in our sample favour the Power-Law Red Noise (PLRN) model, but this excess noise is due to a combination of instrumental artefacts and extrinsic astrophysical sources as opposed to rotational irregularities. 

We also used {\sc TempoNest} to reanalyse nine previously reported pulsar glitches. While the posterior distributions we recover for the change in spin-frequency are generally consistent with previously published values, we are able to place tighter constraints on the change in spin-down and spin-recovery.
Performing model selection, we find a PLRN-only model is preferred for two previously reported glitches, one in PSR J0835$-$4510 on MJD $56922(3)$ and PSR J1740$-$3015 on MJD $57346.0(6)$. 
This highlights the importance of accounting for timing noise of a pulsar when estimating glitch properties, and the potential use of model selection as a means of verifying glitch detections any additional, undiscovered glitch/micro-glitch candidates that may be present in the data. It also indicates conducting reliable parameter estimation on micro-glitch candidates in the presence of strong red noise is difficult.
We also present the discovery of a new glitch in PSR J1452$-$6036 and the first ever glitch observed in PSR J1703$-$4851. Additionally, we observed four unusual glitch-like events in PSR J1825$-$0935, the larger of which cannot be adequately explained by timing noise alone. While it is possible the two large events were due to `slow glitches', we have insufficient observations around their epochs to obtain high resolution measurements of $\dot{\nu}$. As a result, we do not observe a sharp increase in the spin-down typically associated with slow glitches.

Limiting ourselves to only the non-recycled pulsars in our sample, we find the strongest correlation between various pulsar properties and the relative red noise strength exists with pulsar spin-down ($\dot{P}$) and $\dot{\nu}$, with a similar anti-correlation with characteristic age. These correlations agree with recent work by~\citet{Namkham2019}, but are generally weaker than those found by~\citet{Hobbs2010}. We conclude this difference is likely caused by~\citet{Hobbs2010} including millisecond pulsars in their analysis, while we are limited to non-recycled pulsars. Building on work by~\citet{Shannon2010}, we developed a new Bayesian hyper-parameter estimation (Hyper-PE) framework for inferring the scaling between red noise strength and pulsar spin-frequency and spin-down across the population. This relation can be used to estimate the expected red noise strength of a pulsar based on its measured spin and spin-down. Our inferred scaling relation of $\chi_{\mathrm{RN}} \propto \nu^{a} |\dot{\nu}|^{b}$, where $a = -0.84^{+0.47}_{-0.49}$ and $b = 0.97^{+0.16}_{-0.19}$, is consistent with those found in previous studies by~\citet{Shannon2010},~\citet{Hobbs2010} and~\citet{Parthasarathy2019}.
As timing noise strength scales with the length of each pulsar data span, applying our Hyper-PE method to a much larger data set with longer timing baselines will enable more stringent constraints to be placed on the scaling between timing noise strength and pulsar rotational properties. These data could be obtained by UTMOST in the future, or other large, long-term timing programmes such as those undertaken at Jodrell Bank and CHIME/Pulsar~\citep{Ng2018}. A natural extension of our study would be to include measurements of red noise in a large sample of millisecond pulsars and magnetars. In addition, the ability to perform model selection studies using astrophysically motivated noise models could allow us to place constraints on the precise mechanism behind spin noise. 

Finally, we discussed the capability of UTMOST to contribute to the monitoring of relativistic binary systems such as the double pulsar PSR J0737$-$3039. 
We also used the interferometer nature of the instrument to measure an updated position for the bright, mode-changing pulsar PSR J1402$-$5124 in addition to providing the first estimate of this pulsar's spin-down rate.

\section*{Acknowledgements}

The Molonglo Observatory is owned and operated by the University of Sydney. Major support for the UTMOST project has been provided by Swinburne University of Technology. We acknowledge the Australian Research Council grants CE110001020 (CAASTRO) and the Laureate Fellowship FL150100148. MEL acknowledges support from the Australian Government Research Training Program and CSIRO Astronomy and Space Science. FJ acknowledges funding from the European Research Council (ERC) under the European Union's Horizon 2020 research and innovation programme (grant agreement No. 694745). This work made use of the gSTAR and OzSTAR national HPC facilities. gSTAR is funded by Swinburne and the Australian Government's Education Investment Fund. OzSTAR is funded by Swinburne and the National Collaborative Research Infrastructure Strategy (NCRIS). We made use of the standard scientific Python packages~\citep{numpy, scipy, pandas}, {\sc matplotlib}~\citep{matplotlib}, {\sc psrqpy}~\citep{psrqpy}, {\sc Bilby}~\citep{bilby}, {\sc ChainConsumer}~\citep{chainconsumer} and {\sc astropy}~\citep{astropy1, astropy2}. We also acknowledge use of the Astronomer's Telegram (ATel), NASA Astrophysics Data Service (ADS) and NASA/IPAC Extragalactic Database (NED). We give thanks to the anonymous referee for their helpful comments and suggestions.


\bibliographystyle{mnras}
\bibliography{utmost2}


\appendix

\section{Pulsar astrometric and spin parameters}\label{apdx:ephem}

We list the measured astrometric and rotational parameters for each pulsar in our sample in Table~\ref{tab:params}.

\begin{landscape}
\begin{table}
  \centering
  \caption{Astrometric and rotational parameters of all pulsars analysed in this work, including the sky-locations in equatorial coordinates, spin frequencies, spin-down and second spin-frequency derivative. The period, position and DM epoch is MJD 57600 for all pulsars. Errors for RAJ and DEC represent the one-sigma uncertainty on the last digit from {\sc tempo2}. Uncertainties on $\nu$ and $\dot{\nu}$ represent the 95 percent confidence intervals scaled to the last digit. Flags indicate: PPTA $-$ pulsar is observed as part of the Parkes Pulsar Timing Array project~\citep{Manchester2013}, B $-$ pulsar is in a binary. The full table contains {\numpsr} pulsars and is available in the supplementary material.}
    \label{tab:params}
    \renewcommand{\arraystretch}{1.5}
    \begin{tabular}{lllcccccc}
        \hline
        \hline
        PSRJ & RAJ & DECJ & $\nu$ & $\dot{\nu}$ & $\ddot{\nu}$ & $N_{\mathrm{ToA}}$ & $T$ & Flags \\
            & (hh:mm:ss) & ($\degr$:$\arcmin$:$\arcsec$) & (Hz) & ($10^{-15}$ s$^{-2}$) & ($10^{-24}$ s$^{-3}$) &  & (yr) &  \\
        \hline
J0030$+$0451 & $00$:$30$:$27.423(9)$ & $+04$:$51$:$39.7(3)$ & $205.530699027(+7,-8)$ & $-0.39(\pm1)$ & $-$ & 58 & 1.04 & $-$ \\
J0134$-$2937 & $01$:$34$:$18.6939(2)$ & $-29$:$37$:$17.157(5)$ & $7.30131486798(\pm8)$ & $-4.17767(\pm2)$ & $(\pm0.1)$ & 198 & 3.42 & $-$ \\
J0151$-$0635 & $01$:$51$:$22.718(4)$ & $-06$:$35$:$02.98(1)$ & $0.682750032508(+6,-5)$ & $-0.20585(\pm2)$ & $(+0.038,-0.2)$ & 196 & 3.74 & $-$ \\
J0152$-$1637 & $01$:$52$:$10.854(1)$ & $-16$:$37$:$53.63(3)$ & $1.20085114658(+7,-9)$ & $-1.87464(+2,-3)$ & $(+0.064,-0.098)$ & 119 & 3.29 & $-$ \\
J0206$-$4028 & $02$:$06$:$01.2931(1)$ & $-40$:$28$:$03.616(1)$ & $1.58591388856(\pm5)$ & $-3.01013(\pm2)$ & $(+0.1,-0.08)$ & 131 & 3.35 & $-$ \\
J0255$-$5304 & $02$:$55$:$56.2939(4)$ & $-53$:$04$:$21.250(4)$ & $2.23359632409(\pm2)$ & $-0.155852(+7,-8)$ & $(+0.038,-0.054)$ & 270 & 3.75 & $-$ \\
J0348$+$0432 & $03$:$48$:$43.639(3)$ & $+04$:$32$:$11.45(2)$ & $25.5606365903(+4,-5)$ & $-0.158(\pm2)$ & $(+26.2,-14.7)$ & 41 & 2.85 & B \\
J0401$-$7608 & $04$:$01$:$51.75(1)$ & $-76$:$08$:$12.95(5)$ & $1.83400700914(+9,-7)$ & $-5.1927(+3,-4)$ & $(+0.9,-0.4)$ & 110 & 3.15 & $-$ \\
J0418$-$4154 & $04$:$18$:$03.7748(4)$ & $-41$:$54$:$14.42(6)$ & $1.32079643042(\pm2)$ & $-2.30176(\pm6)$ & $(+0.13,-0.58)$ & 99 & 3.38 & $-$ \\
J0437$-$4715 & $04$:$37$:$15.8961(6)$ & $-47$:$15$:$09.1107(3)$ & $173.68794581(+2,-1)$ & $-1.72(+3,-4)$ & $-$ & 230 & 1.41 & PPTA, B \\
J0450$-$1248 & $04$:$50$:$08.7903(2)$ & $-12$:$48$:$07.088(8)$ & $2.28303085312(+6,-4)$ & $-0.5358(\pm1)$ & $(+1.7,-0.9)$ & 81 & 3.14 & $-$ \\
J0452$-$1759 & $04$:$52$:$34.119(1)$ & $-17$:$59$:$23.15(3)$ & $1.82168155657(+5,-4)$ & $-19.0941(+1,-3)$ & $(+0.08,-0.45)$ & 121 & 3.73 & $-$ \\
J0525$+$1115 & $05$:$25$:$56.498(1)$ & $+11$:$15$:$18.8(1)$ & $2.82137062404(\pm2)$ & $-0.58697(+8,-7)$ & $(+0.21,-0.58)$ & 59 & 3.15 & $-$ \\
J0529$-$6652 & $05$:$29$:$50.90(3)$ & $-66$:$52$:$39.9(3)$ & $1.02486651379(+2,-1)$ & $-16.2526(+3,-4)$ & $(+20.7,-2.7)$ & 59 & 2.24 & $-$ \\
J0533$+$0402 & $05$:$33$:$25.828(5)$ & $+04$:$01$:$59.7(2)$ & $1.03840237842(\pm2)$ & $-0.17255(+8,-1)$ & $(+0.46,-0.16)$ & 54 & 3.14 & $-$ \\
J0536$-$7543 & $05$:$36$:$30.829(4)$ & $-75$:$43$:$54.63(2)$ & $0.802660896061(+1,-2)$ & $-0.37076(\pm5)$ & $(+0.06,-0.066)$ & 189 & 3.63 & $-$ \\
J0601$-$0527 & $06$:$01$:$58.9731(8)$ & $-05$:$27$:$50.92(2)$ & $2.52544324442(+6,-7)$ & $-8.30641(+3,-2)$ & $(+0.085,-0.053)$ & 208 & 3.72 & $-$ \\
J0624$-$0424 & $06$:$24$:$20.025(1)$ & $-04$:$24$:$50.56(4)$ & $0.962392542206(\pm1)$ & $-0.769(\pm2)$ & $(+0.12,-0.12)$ & 120 & 3.14 & $-$ \\
J0627$+$0706 & $06$:$27$:$44.172(4)$ & $+07$:$06$:$33.0(2)$ & $2.10134828979(\pm1)$ & $-131.6248(\pm5)$ & $(+0.3,-1.5)$ & 111 & 3.15 & $-$ \\
J0630$-$2834 & $06$:$30$:$49.35(1)$ & $-28$:$34$:$42.1(2)$ & $0.803583722117(+3,-2)$ & $-4.6323(+1,-9)$ & $(+0.16,-0.13)$ & 87 & 3.65 & $-$ \\
J0646$+$0905 & $06$:$46$:$31.025(5)$ & $+09$:$05$:$49.6(3)$ & $1.10630072318(+7,-9)$ & $-0.9009(\pm2)$ & $(+1.3,-1.1)$ & 160 & 3.01 & $-$ \\
J0659$+$1414 & $06$:$59$:$48.188(5)$ & $+14$:$14$:$19.2(4)$ & $2.59788422925(\pm2)$ & $-370.7966(\pm9)$ & $1^{+1.1}_{-0.5}$ & 162 & 3.15 & $-$ \\
J0711$-$6830 & $07$:$11$:$54.1654(1)$ & $-68$:$30$:$47.296(1)$ & $182.117234537(+9,-1)$ & $-0.4928(\pm3)$ & $(+4.0,-2.5)$ & 43 & 2.89 & PPTA \\
J0729$-$1836 & $07$:$29$:$32.30(1)$ & $-18$:$36$:$42.1(2)$ & $1.96011842607(+2,-1)$ & $-72.8289(+6,-7)$ & $-2^{+2.4}_{-1.8}$ & 169 & 3.16 & $-$ \\
J0737$-$3039A & $07$:$37$:$51.24669(2)$ & $-30$:$39$:$40.6895(3)$ & $44.0540680812(+7,-6)$ & $-3.4149(\pm3)$ & $(+0.82,-0.63)$ & 144 & 3.49 & B \\
J0738$-$4042 & $07$:$38$:$32.244(3)$ & $-40$:$42$:$39.43(4)$ & $2.66723044109(\pm5)$ & $-9.805(\pm2)$ & $-3.5\pm1.2$ & 243 & 3.66 & $-$ \\
J0742$-$2822 & $07$:$42$:$48.91(4)$ & $-28$:$22$:$44.0(7)$ & $5.996127853(\pm2)$ & $-604.187(\pm1)$ & $(+1.0,-35.8)$ & 180 & 3.46 & $-$ \\
J0758$-$1528 & $07$:$58$:$29.061(2)$ & $-15$:$28$:$08.333(4)$ & $1.46570344504(+3,-4)$ & $-3.4786(+2,-1)$ & $(+0.21,-0.28)$ & 144 & 3.12 & $-$ \\
        \hline
    \end{tabular}
    \renewcommand{\arraystretch}{}
\end{table}
\end{landscape}


 \begin{landscape}
 \begin{table}
   \centering
   \contcaption{}
     \renewcommand{\arraystretch}{1.5}
     \begin{tabular}{lllcccccc}
         \hline
         \hline
         PSRJ & RAJ & DECJ & $\nu$ & $\dot{\nu}$ & $\ddot{\nu}$ & $N_{\mathrm{ToA}}$ & $T$ & Flags \\
             & (hh:mm:ss) & ($\degr$:$\arcmin$:$\arcsec$) & (Hz) & ($10^{-15}$ s$^{-2}$) & ($10^{-24}$ s$^{-3}$) &  & (yr) &  \\
         \hline
 J0809$-$4753 & $08$:$09$:$43.834(3)$ & $-47$:$53$:$54.85(2)$ & $1.82747830432(+6,-4)$ & $-10.2748(\pm2)$ & $(+0.07,-0.32)$ & 79 & 3.12 & $-$ \\
 J0820$-$1350 & $08$:$20$:$26.407(1)$ & $-13$:$50$:$56.32(4)$ & $0.807668884066(\pm4)$ & $-1.37174(\pm2)$ & $(+0.084,-0.09)$ & 51 & 3.57 & $-$ \\
 J0820$-$4114 & $08$:$20$:$15.46(1)$ & $-41$:$14$:$35.2(1)$ & $1.8333635346(\pm7)$ & $-0.0664(\pm2)$ & $(+0.48,-0.78)$ & 88 & 3.39 & $-$ \\
 J0835$-$4510 & $08$:$35$:$20.6(2)$ & $-45$:$10$:$33(1)$ & $11.18677868(\pm2)$ & $-13161(+5.4,-5.3)$ & $(+19949.0,-7024.5)$ & 1420 & 4.81 & $-$ \\
 J0837$+$0610 & $08$:$37$:$05.6462(1)$ & $+06$:$10$:$15.87(6)$ & $0.785068914181(+2,-3)$ & $-4.19046(+1,-8)$ & $(+0.033,-0.016)$ & 78 & 3.75 & $-$ \\
 J0837$-$4135 & $08$:$37$:$21.1922(4)$ & $-41$:$35$:$14.589(4)$ & $1.33044994223(\pm5)$ & $-6.26572(\pm3)$ & $(+0.038,-0.09)$ & 140 & 3.64 & $-$ \\
 J0840$-$5332 & $08$:$40$:$33.726(9)$ & $-53$:$32$:$35.95(6)$ & $1.38770592261(\pm3)$ & $-3.154(\pm1)$ & $(+0.09,-0.23)$ & 59 & 3.14 & $-$ \\
 J0842$-$4851 & $08$:$42$:$05.4443(9)$ & $-48$:$51$:$20.6(1)$ & $1.5519428595(\pm9)$ & $-23.0246(+7,-8)$ & $(+2.7,-0.1)$ & 53 & 3.13 & $-$ \\
 J0846$-$3533 & $08$:$46$:$06.0712(4)$ & $-35$:$33$:$40.91(6)$ & $0.895978340332(\pm1)$ & $-1.28499(\pm4)$ & $(+0.13,-0.3)$ & 65 & 3.3 & $-$ \\
 J0855$-$3331 & $08$:$55$:$38.421(3)$ & $-33$:$31$:$38.99(4)$ & $0.788929778283(+1,-2)$ & $-3.93327(+4,-3)$ & $(+0.2,-0.28)$ & 76 & 3.12 & $-$ \\
 J0856$-$6137 & $08$:$56$:$59.27(1)$ & $-61$:$37$:$52.71(8)$ & $1.03894958562(+2,-3)$ & $-1.813(+1,-8)$ & $(+0.33,-0.11)$ & 48 & 3.39 & $-$ \\
 J0904$-$4246 & $09$:$04$:$59.083(9)$ & $-42$:$46$:$13.4(1)$ & $1.03608336991(+3,-2)$ & $-2.0148(+9,-1)$ & $(+0.24,-0.4)$ & 50 & 3.37 & $-$ \\
 J0904$-$7459 & $09$:$04$:$10.47(3)$ & $-74$:$59$:$41.7(1)$ & $1.81965845843(+1,-7)$ & $-1.5278(+4,-3)$ & $(+1.1,-1.0)$ & 44 & 2.98 & $-$ \\
 J0907$-$5157 & $09$:$07$:$15.901(3)$ & $-51$:$57$:$59.36(2)$ & $3.9438751026(+7,-5)$ & $-28.5412(\pm3)$ & $(+0.87,-0.12)$ & 121 & 3.5 & $-$ \\
 J0908$-$1739 & $09$:$08$:$38.227(4)$ & $-17$:$39$:$39.9(1)$ & $2.48987780015(\pm4)$ & $-4.1492(\pm2)$ & $(+0.27,-0.98)$ & 37 & 3.16 & $-$ \\
 J0908$-$4913 & $09$:$08$:$35.46(1)$ & $-49$:$13$:$05.00(1)$ & $9.36601123448(\pm6)$ & $-1324.821(+3,-4)$ & $4^{+1.3}_{-1.5}$ & 173 & 3.48 & $-$ \\
 J0909$-$7212 & $09$:$09$:$35.81(3)$ & $-72$:$12$:$08.94(1)$ & $0.733734742115(\pm4)$ & $-0.1762(\pm2)$ & $(+0.15,-0.41)$ & 31 & 3.15 & $-$ \\
 J0922$+$0638 & $09$:$22$:$13.85(3)$ & $+06$:$38$:$19(1)$ & $2.32217901642(\pm3)$ & $-73.985(\pm1)$ & $(+2.5,-9.6)$ & 68 & 3.02 & $-$ \\
 J0924$-$5302 & $09$:$24$:$08.722(4)$ & $-53$:$02$:$42.6(3)$ & $1.33987540461(\pm2)$ & $-63.431(+1,-9)$ & $(+3.9,-4.1)$ & 137 & 3.2 & $-$ \\
 J0924$-$5814 & $09$:$24$:$30.82(1)$ & $-58$:$14$:$05.10(1)$ & $1.35225504146(+4,-3)$ & $-9.0013(\pm2)$ & $(+0.48,-0.28)$ & 72 & 3.39 & $-$ \\
 J0934$-$5249 & $09$:$34$:$28.237(5)$ & $-52$:$49$:$27.30(5)$ & $0.692148271001(+1,-8)$ & $-2.22898(\pm3)$ & $(+0.07,-0.14)$ & 152 & 3.16 & $-$ \\
 J0942$-$5552 & $09$:$42$:$14.88(6)$ & $-55$:$52$:$55.1(5)$ & $1.50514304056(\pm3)$ & $-51.376(\pm2)$ & $8^{+4.1}_{-4.3}$ & 150 & 3.69 & $-$ \\
 J0942$-$5657 & $09$:$42$:$54.422(5)$ & $-56$:$57$:$43.21(3)$ & $1.23737204376(\pm4)$ & $-60.6357(\pm2)$ & $(+0.37,-0.09)$ & 90 & 3.18 & $-$ \\
 J0944$-$1354 & $09$:$44$:$28.967(1)$ & $-13$:$54$:$41.88(2)$ & $1.75357327379(+5,-7)$ & $-0.13929(\pm2)$ & $(+0.069,-0.118)$ & 61 & 3.11 & $-$ \\
 J0953$+$0755 & $09$:$53$:$09.3121(2)$ & $+07$:$55$:$36.9(1)$ & $3.95154788907(+9,-8)$ & $-3.58768(\pm3)$ & $(+0.021,-0.07)$ & 73 & 3.71 & $-$ \\
 J0955$-$5304 & $09$:$55$:$29.461(1)$ & $-53$:$04$:$16.64(1)$ & $1.15992862694(\pm4)$ & $-4.74318(\pm1)$ & $(+0.16,-0.078)$ & 155 & 3.26 & $-$ \\
 J0959$-$4809 & $09$:$59$:$26.212(6)$ & $-48$:$09$:$47.47(7)$ & $1.49234604(+5,-4)$ & $-0.1887(+1,-9)$ & $(+0.49,-0.27)$ & 78 & 3.37 & $-$ \\
 J1001$-$5507 & $10$:$01$:$37.85(5)$ & $-55$:$07$:$07.8(5)$ & $0.696073036122(+9,-1)$ & $-24.9925(+7,-6)$ & $1^{+0.8}_{-1.6}$ & 138 & 3.7 & $-$ \\
 J1003$-$4747 & $10$:$03$:$21.529(1)$ & $-47$:$47$:$01.190(2)$ & $3.25654170022(\pm3)$ & $-21.96544(+9,-6)$ & $(+0.29,-0.36)$ & 69 & 3.13 & $-$ \\
 J1012$-$5857 & $10$:$12$:$48.470(5)$ & $-58$:$57$:$48.50(3)$ & $1.21962189426(\pm1)$ & $-26.47184(\pm4)$ & $(+0.05,-0.23)$ & 193 & 3.71 & $-$ \\
         \hline
     \end{tabular}
     \renewcommand{\arraystretch}{}
 \end{table}
 \end{landscape}

 \begin{landscape}
 \begin{table}
   \centering
   \contcaption{}
     \renewcommand{\arraystretch}{1.5}
     \begin{tabular}{lllcccccc}
         \hline
         \hline
         PSRJ & RAJ & DECJ & $\nu$ & $\dot{\nu}$ & $\ddot{\nu}$ & $N_{\mathrm{ToA}}$ & $T$ & Flags \\
             & (hh:mm:ss) & ($\degr$:$\arcmin$:$\arcsec$) & (Hz) & ($10^{-15}$ s$^{-2}$) & ($10^{-24}$ s$^{-3}$) &  & (yr) &  \\
         \hline
 J1013$-$5934 & $10$:$13$:$31.848(2)$ & $-59$:$34$:$26.63(1)$ & $2.25784124141(\pm1)$ & $-2.83682(+3,-4)$ & $(+0.12,-0.18)$ & 132 & 3.43 & $-$ \\
 J1016$-$5345 & $10$:$16$:$31.135(4)$ & $-53$:$45$:$14.26(3)$ & $1.29940016176(\pm2)$ & $-3.25316(+6,-4)$ & $(+0.24,-0.8)$ & 102 & 3.14 & $-$ \\
 J1017$-$5621 & $10$:$17$:$12.831(1)$ & $-56$:$21$:$30.517(7)$ & $1.98624477638(\pm5)$ & $-12.38704(\pm2)$ & $(+0.2,-0.052)$ & 127 & 3.17 & B \\
 J1017$-$7156 & $10$:$17$:$51.3172(5)$ & $-71$:$56$:$41.596(2)$ & $427.621905026(\pm2)$ & $-0.415(+1,-8)$ & $(+10.3,-5.6)$ & 51 & 3.38 & B \\
 J1022$+$1001 & $10$:$22$:$58.3(1)$ & $+10$:$01$:$58(5)$ & $60.7794479207(+6,-9)$ & $-0.1564(\pm3)$ & $(+0.36,-0.52)$ & 39 & 3.43 & PPTA \\
 J1032$-$5911 & $10$:$32$:$04.876(1)$ & $-59$:$11$:$54.8(1)$ & $2.15418526489(+2,-3)$ & $-8.3414(\pm5)$ & $(+4.9,-0.4)$ & 152 & 2.51 & $-$ \\
 J1034$-$3224 & $10$:$34$:$19.46(1)$ & $-32$:$24$:$26.2(2)$ & $0.869118880561(+3,-2)$ & $-0.17372(\pm1)$ & $(+0.14,-0.14)$ & 41 & 3.74 & $-$ \\
 J1036$-$4926 & $10$:$36$:$13.121(7)$ & $-49$:$26$:$21.2(1)$ & $1.95936510762(\pm1)$ & $-6.3385(\pm3)$ & $(+6.1,-7.5)$ & 30 & 2.36 & $-$ \\
 J1041$-$1942 & $10$:$41$:$36.191(9)$ & $-19$:$42$:$13.7(2)$ & $0.721308789858(+2,-3)$ & $-0.4925(\pm1)$ & $(+0.12,-0.13)$ & 34 & 3.41 & $-$ \\
 J1042$-$5521 & $10$:$42$:$00.4853(9)$ & $-55$:$21$:$05.793(6)$ & $0.854067790376(\pm2)$ & $-4.89999(+6,-5)$ & $(+0.24,-0.14)$ & 101 & 3.36 & $-$ \\
 J1043$-$6116 & $10$:$43$:$55.222(8)$ & $-61$:$16$:$51.29(8)$ & $3.46491578971(+8,-7)$ & $-124.947(\pm2)$ & $(+77.6,-108.0)$ & 78 & 1.47 & $-$ \\
 J1045$-$4509 & $10$:$45$:$50.1794(5)$ & $-45$:$09$:$54.106(6)$ & $133.79314947(\pm2)$ & $-0.3153(+8,-6)$ & $(+2.0,-1.0)$ & 38 & 3.31 & PPTA, B \\
 J1046$-$5813 & $10$:$46$:$18.815(2)$ & $-58$:$13$:$51.89(2)$ & $2.70688676045(\pm1)$ & $-8.4(+4,-3)$ & $(+0.48,-0.5)$ & 193 & 3.25 & $-$ \\
 J1047$-$6709 & $10$:$47$:$28.285(5)$ & $-67$:$09$:$51.61(4)$ & $5.0389844311(\pm1)$ & $-42.8418(+4,-3)$ & $(+0.9,-1.2)$ & 39 & 2.43 & $-$ \\
 J1048$-$5832 & $10$:$48$:$13.1(1)$ & $-58$:$32$:$03(1)$ & $8.0824185121(+4,-5)$ & $-6273.49(+2,-3)$ & $90^{+41.8}_{-51.2}$ & 232 & 3.44 & $-$ \\
 J1056$-$6258 & $10$:$56$:$25.53(1)$ & $-62$:$58$:$47.7(1)$ & $2.36714106203(\pm1)$ & $-20.057(+8,-7)$ & $(+0.9,-1.0)$ & 198 & 3.69 & $-$ \\
 J1057$-$5226 & $10$:$57$:$59.068(8)$ & $-52$:$26$:$56.10(8)$ & $5.0731886204(+2,-3)$ & $-150.205(+2,-1)$ & $(+0.1,-5.6)$ & 126 & 2.62 & $-$ \\
 J1057$-$7914 & $10$:$57$:$27.7(1)$ & $-79$:$14$:$23.6(3)$ & $0.74216802567(+8,-7)$ & $-0.7321(+4,-3)$ & $(+1.4,-0.8)$ & 41 & 2.84 & $-$ \\
 J1059$-$5742 & $10$:$59$:$00.8886(4)$ & $-57$:$42$:$14.55(3)$ & $0.843879990895(+1,-8)$ & $-3.0668(\pm3)$ & $(+0.08,-0.31)$ & 178 & 3.18 & $-$ \\
 J1105$-$6107 & $11$:$05$:$26.2(1)$ & $-61$:$07$:$48.0(8)$ & $15.8222513283(+4,-3)$ & $-3966.97(\pm1)$ & $(+6.0,-31.4)$ & 145 & 2.85 & $-$ \\
 J1110$-$5637 & $11$:$10$:$00.3712(6)$ & $-56$:$37$:$32.57(4)$ & $1.79129810299(+1,-2)$ & $-6.6125(+8,-4)$ & $(+1.8,-0.2)$ & 130 & 3.17 & $-$ \\
 J1112$-$6613 & $11$:$12$:$38.414(4)$ & $-66$:$13$:$04.663(2)$ & $2.9920963178(+5,-3)$ & $-7.385(+1,-2)$ & $(+0.1,-1.6)$ & 96 & 2.93 & $-$ \\
 J1112$-$6926 & $11$:$12$:$50.78(1)$ & $-69$:$26$:$32.33(6)$ & $1.21878739947(+4,-3)$ & $-4.1912(\pm1)$ & $(+0.52,-0.6)$ & 97 & 3.22 & $-$ \\
 J1114$-$6100 & $11$:$14$:$22.69(5)$ & $-61$:$00$:$32.1(3)$ & $1.13525643826(\pm2)$ & $-59.3019(+6,-5)$ & $(+4.8,-4.2)$ & 137 & 2.45 & $-$ \\
 J1116$-$4122 & $11$:$16$:$43.083(4)$ & $-41$:$22$:$44.86(8)$ & $1.06026074416(+2,-3)$ & $-8.955(\pm1)$ & $(+0.04,-0.27)$ & 47 & 3.53 & $-$ \\
 J1121$-$5444 & $11$:$21$:$19.23(1)$ & $-54$:$44$:$04.90(1)$ & $1.86641502454(\pm2)$ & $-9.7309(+7,-8)$ & $(+2.2,-3.8)$ & 117 & 2.98 & $-$ \\
 J1123$-$6259 & $11$:$23$:$55.53(1)$ & $-62$:$59$:$10.92(8)$ & $3.68409189328(\pm1)$ & $-71.2863(+3,-4)$ & $(+5.3,-6.1)$ & 70 & 2.99 & $-$ \\
 J1126$-$6942 & $11$:$26$:$21.66(4)$ & $-69$:$42$:$15.8(1)$ & $1.72586751278(+3,-2)$ & $-9.8111(+9,-1)$ & $-0.69^{+6}_{-2}$ & 34 & 2.06 & $-$ \\
 J1133$-$6250 & $11$:$33$:$51.3(1)$ & $-62$:$50$:$51(1)$ & $0.9776360471(+6,-4)$ & $-0.448(+9,-1)$ & $(+132.0,-20.1)$ & 128 & 1.22 & $-$ \\
 J1136$+$1551 & $11$:$36$:$03.0946(5)$ & $+15$:$51$:$15.9(1)$ & $0.841809871701(+6,-2)$ & $-2.64185(+9,-3)$ & $(+0.34,-0.09)$ & 36 & 3.63 & $-$ \\
         \hline
     \end{tabular}
     \renewcommand{\arraystretch}{}
 \end{table}
 \end{landscape}

 \begin{landscape}
 \begin{table}
   \centering
   \contcaption{}
     \renewcommand{\arraystretch}{1.5}
     \begin{tabular}{lllcccccc}
         \hline
         \hline
         PSRJ & RAJ & DECJ & $\nu$ & $\dot{\nu}$ & $\ddot{\nu}$ & $N_{\mathrm{ToA}}$ & $T$ & Flags \\
             & (hh:mm:ss) & ($\degr$:$\arcmin$:$\arcsec$) & (Hz) & ($10^{-15}$ s$^{-2}$) & ($10^{-24}$ s$^{-3}$) &  & (yr) &  \\
         \hline
 J1136$-$5525 & $11$:$36$:$02.2354(5)$ & $-55$:$25$:$06.843(5)$ & $2.74188009799(+1,-2)$ & $-61.8(+1,-8)$ & $(+0.6,-3.7)$ & 147 & 3.53 & $-$ \\
 J1141$-$3322 & $11$:$41$:$42.756(2)$ & $-33$:$22$:$37.31(5)$ & $3.43091071248(\pm3)$ & $-5.4774(\pm1)$ & $(+0.24,-0.14)$ & 51 & 3.13 & $-$ \\
 J1141$-$6545 & $11$:$41$:$07.0006(6)$ & $-65$:$45$:$19.05(3)$ & $2.53871590079(+6,-7)$ & $-27.7621(\pm4)$ & $0.48^{+7}_{-3}$ & 273 & 3.55 & B \\
 J1146$-$6030 & $11$:$46$:$07.7152(1)$ & $-60$:$30$:$59.622(9)$ & $3.65798554138(\pm1)$ & $-23.93026(\pm4)$ & $(+0.53,-0.3)$ & 169 & 3.35 & $-$ \\
 J1157$-$6224 & $11$:$57$:$15.208(7)$ & $-62$:$24$:$50.90(5)$ & $2.49671858326(+9,-7)$ & $-24.5068(+4,-5)$ & $(+0.9,-0.63)$ & 229 & 3.55 & $-$ \\
 J1202$-$5820 & $12$:$02$:$28.358(6)$ & $-58$:$20$:$33.41(5)$ & $2.20846545731(+7,-6)$ & $-10.3828(\pm3)$ & $(+0.95,-0.6)$ & 134 & 3.51 & $-$ \\
 J1210$-$5559 & $12$:$10$:$05.98706(2)$ & $-55$:$59$:$03.8501(2)$ & $3.57439188715(\pm5)$ & $-9.26691(+2,-3)$ & $(+0.032,-0.003)$ & 158 & 3.51 & $-$ \\
 J1224$-$6407 & $12$:$24$:$22.264(2)$ & $-64$:$07$:$53.79(1)$ & $4.61934676092(+4,-5)$ & $-105.6992(+3,-2)$ & $(+0.09,-0.92)$ & 367 & 3.55 & $-$ \\
 J1231$-$6303 & $12$:$31$:$13.0(1)$ & $-63$:$03$:$18(1)$ & $0.74006295676(\pm3)$ & $-0.0723(+8,-7)$ & $(+2.4,-0.8)$ & 79 & 3.38 & $-$ \\
 J1239$-$6832 & $12$:$39$:$58.96(2)$ & $-68$:$32$:$28.94(9)$ & $0.768094857398(\pm2)$ & $-7.01054(+7,-8)$ & $(+1.1,-0.9)$ & 60 & 3.15 & $-$ \\
 J1243$-$6423 & $12$:$43$:$17.120(6)$ & $-64$:$23$:$23.92(4)$ & $2.57410111798(\pm5)$ & $-29.8026(\pm3)$ & $(+0.0,-0.49)$ & 345 & 3.63 & $-$ \\
 J1253$-$5820 & $12$:$53$:$28.418(2)$ & $-58$:$20$:$40.47(2)$ & $3.91392670035(\pm7)$ & $-32.2492(\pm3)$ & $(+1.24,-0.01)$ & 204 & 3.38 & $-$ \\
 J1257$-$1027 & $12$:$57$:$04.7796(9)$ & $-10$:$27$:$04.77(3)$ & $1.61993710136(+2,-1)$ & $-0.94879(\pm4)$ & $(+0.27,-0.39)$ & 39 & 3.18 & $-$ \\
 J1259$-$6741 & $12$:$59$:$22.64(1)$ & $-67$:$41$:$40.27(6)$ & $1.5075450023(+5,-4)$ & $-1.9434(\pm1)$ & $(+0.8,-1.6)$ & 42 & 2.54 & $-$ \\
 J1305$-$6455 & $13$:$05$:$23.47(2)$ & $-64$:$55$:$28.5(1)$ & $1.74931666048(+1,-2)$ & $-12.3373(+9,-8)$ & $(+1.2,-2.7)$ & 175 & 3.47 & $-$ \\
 J1306$-$6617 & $13$:$06$:$38.12(1)$ & $-66$:$17$:$21.2(1)$ & $2.11404065516(\pm2)$ & $-26.7181(\pm9)$ & $(+1.41,-0.41)$ & 125 & 3.38 & $-$ \\
 J1312$-$5402 & $13$:$12$:$04.708(2)$ & $-54$:$02$:$42.5(2)$ & $1.37333511678(+4,-5)$ & $-0.2765(+2,-1)$ & $(+0.7,-1.03)$ & 36 & 3.18 & $-$ \\
 J1312$-$5516 & $13$:$12$:$53.533(9)$ & $-55$:$16$:$47.3(1)$ & $1.1775198318(\pm3)$ & $-7.9101(\pm1)$ & $(+0.3,-0.99)$ & 98 & 3.07 & $-$ \\
 J1319$-$6056 & $13$:$19$:$20.250(7)$ & $-60$:$56$:$46.79(6)$ & $3.51675963943(+2,-9)$ & $-18.8878(+3,-7)$ & $(+0.87,-0.14)$ & 199 & 3.19 & $-$ \\
 J1320$-$5359 & $13$:$20$:$53.932(2)$ & $-53$:$59$:$04.967(3)$ & $3.57477758551(\pm6)$ & $-118.146(\pm3)$ & $0.36^{+1}_{-3}$ & 129 & 3.5 & $-$ \\
 J1326$-$5859 & $13$:$26$:$58.219(7)$ & $-58$:$59$:$29.29(7)$ & $2.09207813232(+7,-8)$ & $-14.2359(+5,-4)$ & $(+1.4,-1.0)$ & 306 & 3.63 & $-$ \\
 J1326$-$6408 & $13$:$26$:$32.433(2)$ & $-64$:$08$:$43.80(1)$ & $1.26155286486(+9,-6)$ & $-4.91831(\pm2)$ & $(+0.122,-0.15)$ & 161 & 2.85 & $-$ \\
 J1326$-$6700 & $13$:$26$:$02.706(4)$ & $-67$:$00$:$50.1(3)$ & $1.84156958655(+4,-3)$ & $-18.037(\pm2)$ & $(+1.3,-4.2)$ & 132 & 3.54 & $-$ \\
 J1327$-$6222 & $13$:$27$:$17.36(7)$ & $-62$:$22$:$44.7(5)$ & $1.8870445541(+4,-3)$ & $-66.926(\pm2)$ & $2.37^{+7}_{-5}$ & 308 & 3.63 & $-$ \\
 J1327$-$6301 & $13$:$27$:$07.4320(3)$ & $-63$:$01$:$15.51(2)$ & $5.08957797539(\pm4)$ & $-39.6313(\pm1)$ & $(+0.49,-0.29)$ & 238 & 3.41 & $-$ \\
 J1328$-$4357 & $13$:$28$:$06.4198(5)$ & $-43$:$57$:$44.50(8)$ & $1.87722052793(\pm7)$ & $-10.7557(+4,-3)$ & $0.29\pm2$ & 93 & 2.99 & $-$ \\
 J1338$-$6204 & $13$:$38$:$09.247(7)$ & $-62$:$04$:$18.7(5)$ & $0.80710212469(\pm2)$ & $-8.9837(+6,-5)$ & $(+2.9,-4.5)$ & 198 & 2.47 & $-$ \\
 J1350$-$5115 & $13$:$50$:$16.159(2)$ & $-51$:$15$:$24.56(3)$ & $3.38180924275(+8,-6)$ & $-8.6634(+2,-1)$ & $(+4.0,-5.1)$ & 95 & 2.43 & $-$ \\
 J1355$-$5153 & $13$:$55$:$58.692(2)$ & $-51$:$53$:$53.95(2)$ & $1.55206115637(\pm3)$ & $-6.7736(\pm1)$ & $(+0.26,-0.23)$ & 123 & 3.19 & $-$ \\
 J1356$-$5521 & $13$:$56$:$50.49(2)$ & $-55$:$21$:$15.2(2)$ & $1.97090897627(\pm2)$ & $-2.8152(+5,-4)$ & $(+14.3,-9.1)$ & 31 & 2.39 & $-$ \\
         \hline
     \end{tabular}
     \renewcommand{\arraystretch}{}
 \end{table}
 \end{landscape}

 \begin{landscape}
 \begin{table}
   \centering
   \contcaption{}
     \renewcommand{\arraystretch}{1.5}
     \begin{tabular}{lllcccccc}
         \hline
         \hline
         PSRJ & RAJ & DECJ & $\nu$ & $\dot{\nu}$ & $\ddot{\nu}$ & $N_{\mathrm{ToA}}$ & $T$ & Flags \\
             & (hh:mm:ss) & ($\degr$:$\arcmin$:$\arcsec$) & (Hz) & ($10^{-15}$ s$^{-2}$) & ($10^{-24}$ s$^{-3}$) &  & (yr) &  \\
         \hline
 J1359$-$6038 & $13$:$59$:$58.230(9)$ & $-60$:$38$:$07.671(7)$ & $7.84261649234(\pm2)$ & $-389.488(\pm1)$ & $-3^{+3.3}_{-1.1}$ & 429 & 3.54 & $-$ \\
 J1401$-$6357 & $14$:$01$:$52.45(1)$ & $-63$:$57$:$42.0(1)$ & $1.18651362793(+8,-1)$ & $-23.6871(+5,-4)$ & $(+0.4,-1.3)$ & 245 & 3.56 & $-$ \\
 J1413$-$6307 & $14$:$13$:$31.32(4)$ & $-63$:$07$:$34.6(3)$ & $2.5319550816(\pm1)$ & $-47.9(+4,-3)$ & $-4^{+3.3}_{-11.1}$ & 152 & 2.43 & $-$ \\
 J1418$-$3921 & $14$:$18$:$50.28(1)$ & $-39$:$21$:$18.6(2)$ & $0.911737714389(+2,-3)$ & $-0.73841(+8,-6)$ & $(+0.42,-0.39)$ & 57 & 3.63 & $-$ \\
 J1420$-$5416 & $14$:$20$:$29.11(1)$ & $-54$:$16$:$22.7(1)$ & $1.06863614348(\pm2)$ & $-0.26515(+6,-7)$ & $(+0.22,-0.3)$ & 74 & 3.08 & $-$ \\
 J1424$-$5822 & $14$:$24$:$32.130(8)$ & $-58$:$22$:$55.7(1)$ & $2.7267557238(+1,-9)$ & $-29.262(\pm2)$ & $(+120.5,-91.9)$ & 188 & 1.25 & $-$ \\
 J1428$-$5530 & $14$:$28$:$26.240(3)$ & $-55$:$30$:$50.06(4)$ & $1.75348642691(\pm1)$ & $-6.41562(+4,-3)$ & $0.055\pm5$ & 168 & 3.66 & $-$ \\
 J1430$-$6623 & $14$:$30$:$40.732(1)$ & $-66$:$23$:$05.546(1)$ & $1.2731663027(\pm6)$ & $-4.50256(+4,-3)$ & $(+0.032,-0.039)$ & 170 & 3.57 & $-$ \\
 J1435$-$5954 & $14$:$35$:$00.208(1)$ & $-59$:$54$:$49.5(1)$ & $2.11418109778(\pm4)$ & $-6.9189(\pm1)$ & $(+0.81,-0.52)$ & 254 & 3.49 & $-$ \\
 J1452$-$6036 & $14$:$52$:$51.80(1)$ & $-60$:$36$:$30.00(8)$ & $6.4519415824(\pm1)$ & $-60.401(+3,-2)$ & $(+105.1,-110.5)$ & 152 & 1.32 & $-$ \\
 J1453$-$6413 & $14$:$53$:$32.652(1)$ & $-64$:$13$:$16.095(9)$ & $5.571424352(+2,-3)$ & $-85.1854(+2,-1)$ & $(+0.18,-0.24)$ & 234 & 3.56 & $-$ \\
 J1456$-$6843 & $14$:$55$:$59.914(1)$ & $-68$:$43$:$39.49(1)$ & $3.79684009011(+5,-6)$ & $-1.42687(+3,-2)$ & $(+0.044,-0.009)$ & 119 & 4.21 & $-$ \\
 J1457$-$5122 & $14$:$57$:$40.093(8)$ & $-51$:$22$:$54.9(1)$ & $0.57198175779(+1,-2)$ & $-1.73305(+5,-7)$ & $(+0.22,-0.24)$ & 39 & 3.03 & $-$ \\
 J1507$-$4352 & $15$:$07$:$34.175(4)$ & $-43$:$52$:$04.05(1)$ & $3.48725495693(\pm1)$ & $-19.2672(+5,-4)$ & $(+0.8,-1.4)$ & 56 & 3.17 & $-$ \\
 J1507$-$6640 & $15$:$07$:$48.634(1)$ & $-66$:$40$:$57.86(1)$ & $2.81170331276(+8,-9)$ & $-9.1066(\pm2)$ & $(+0.22,-0.14)$ & 111 & 2.97 & $-$ \\
 J1511$-$5414 & $15$:$11$:$51.285(3)$ & $-54$:$14$:$40.32(6)$ & $4.99041973147(+8,-6)$ & $-12.072(+1,-2)$ & $(+144.8,-15.3)$ & 101 & 1.25 & $-$ \\
 J1512$-$5759 & $15$:$12$:$43.13(1)$ & $-58$:$00$:$00.43(1)$ & $7.77001479211(\pm5)$ & $-413.71(\pm2)$ & $(+1.7,-13.0)$ & 177 & 3.19 & $-$ \\
 J1514$-$4834 & $15$:$14$:$14.563(2)$ & $-48$:$34$:$19.97(4)$ & $2.19857371563(\pm3)$ & $-4.47652(+9,-8)$ & $(+2.2,-0.3)$ & 51 & 2.55 & $-$ \\
 J1522$-$5829 & $15$:$22$:$42.244(4)$ & $-58$:$29$:$02.815(3)$ & $2.52937565301(+7,-9)$ & $-13.1588(+3,-4)$ & $(+0.25,-1.07)$ & 187 & 3.21 & $-$ \\
 J1527$-$3931 & $15$:$27$:$58.828(9)$ & $-39$:$31$:$34.2(2)$ & $0.41363243246(\pm1)$ & $-3.26122(+4,-5)$ & $(+0.19,-0.15)$ & 36 & 3.21 & $-$ \\
 J1527$-$5552 & $15$:$27$:$40.734(4)$ & $-55$:$52$:$08.352(6)$ & $0.953544682096(+3,-4)$ & $-10.2459(+2,-1)$ & $(+0.25,-0.0)$ & 134 & 3.21 & $-$ \\
 J1528$-$3146 & $15$:$28$:$34.952(1)$ & $-31$:$46$:$06.944(6)$ & $16.4413569253(+2,-1)$ & $-0.068(+5,-6)$ & $(+1.36,-0.57)$ & 25 & 3.05 & B \\
 J1534$-$5334 & $15$:$34$:$08.2790(1)$ & $-53$:$34$:$19.57(2)$ & $0.730523027415(+2,-3)$ & $-0.76251(\pm7)$ & $(+0.013,-0.016)$ & 231 & 3.63 & $-$ \\
 J1534$-$5405 & $15$:$34$:$33.59(1)$ & $-54$:$05$:$40.5(2)$ & $3.4519643055(+1,-8)$ & $-18.528(+2,-3)$ & $4^{+5.5}_{-4.5}$ & 100 & 2.43 & $-$ \\
 J1539$-$5626 & $15$:$39$:$14.07(1)$ & $-56$:$26$:$26.2(1)$ & $4.10852985562(+3,-6)$ & $-81.894(+3,-1)$ & $(+1.5,-7.0)$ & 179 & 3.13 & $-$ \\
 J1542$-$5034 & $15$:$42$:$45.32(2)$ & $-50$:$34$:$03.66(3)$ & $1.6687581689(+9,-1)$ & $-11.208(+4,-3)$ & $(+5.2,-7.3)$ & 59 & 2.39 & $-$ \\
 J1543$+$0929 & $15$:$43$:$38.82(2)$ & $+09$:$29$:$16.4(5)$ & $1.33609682985(\pm2)$ & $-0.7773(\pm7)$ & $(+1.7,-1.6)$ & 28 & 2.75 & $-$ \\
 J1544$-$5308 & $15$:$44$:$59.8294(6)$ & $-53$:$08$:$46.953(9)$ & $5.60055059845(+1,-9)$ & $-1.88991(+3,-4)$ & $(+0.17,-0.26)$ & 164 & 3.58 & $-$ \\
 J1549$-$4848 & $15$:$49$:$21.027(6)$ & $-48$:$48$:$36.1(1)$ & $3.46791653545(+9,-1)$ & $-169.702(\pm2)$ & $(+63.2,-56.4)$ & 44 & 1.46 & $-$ \\
 J1553$-$5456 & $15$:$53$:$59.61(1)$ & $-54$:$56$:$06.25(1)$ & $0.9247724034(+6,-5)$ & $-13.4399(\pm1)$ & $(+36.6,-44.9)$ & 73 & 1.21 & $-$ \\
         \hline
     \end{tabular}
     \renewcommand{\arraystretch}{}
 \end{table}
 \end{landscape}

 \begin{landscape}
 \begin{table}
   \centering
   \contcaption{}
     \renewcommand{\arraystretch}{1.5}
     \begin{tabular}{lllcccccc}
         \hline
         \hline
         PSRJ & RAJ & DECJ & $\nu$ & $\dot{\nu}$ & $\ddot{\nu}$ & $N_{\mathrm{ToA}}$ & $T$ & Flags \\
             & (hh:mm:ss) & ($\degr$:$\arcmin$:$\arcsec$) & (Hz) & ($10^{-15}$ s$^{-2}$) & ($10^{-24}$ s$^{-3}$) &  & (yr) &  \\
         \hline
 J1555$-$3134 & $15$:$55$:$17.947(2)$ & $-31$:$34$:$20.3(1)$ & $1.93009273101(\pm2)$ & $-0.23061(\pm7)$ & $(+0.34,-0.11)$ & 49 & 3.19 & $-$ \\
 J1557$-$4258 & $15$:$57$:$00.25445(6)$ & $-42$:$58$:$12.35(1)$ & $3.0377858242(+1,-2)$ & $-3.04646(+8,-7)$ & $(+0.05,-0.131)$ & 90 & 3.54 & $-$ \\
 J1559$-$4438 & $15$:$59$:$41.525(1)$ & $-44$:$38$:$45.85(3)$ & $3.89018703598(+2,-3)$ & $-15.4484(\pm2)$ & $(+0.53,-0.27)$ & 66 & 3.37 & $-$ \\
 J1559$-$5545 & $15$:$59$:$20.7(1)$ & $-55$:$45$:$47(1)$ & $1.04464080328(\pm5)$ & $-21.733(\pm3)$ & $(+12.5,-3.9)$ & 87 & 3.49 & $-$ \\
 J1600$-$3053 & $16$:$00$:$51.8941(7)$ & $-30$:$53$:$49.70(3)$ & $277.937706823(\pm2)$ & $-0.687(\pm5)$ & $(+47.9,-46.2)$ & 32 & 3.39 & PPTA, B \\
 J1600$-$5044 & $16$:$00$:$53.033(5)$ & $-50$:$44$:$20.93(8)$ & $5.19197591119(\pm2)$ & $-136.452(+1,-9)$ & $(+3.1,-1.2)$ & 175 & 3.63 & $-$ \\
 J1603$-$2531 & $16$:$03$:$04.8253(6)$ & $-25$:$31$:$47.9(4)$ & $3.53267858106(\pm4)$ & $-19.8906(\pm2)$ & $(+1.13,-0.29)$ & 29 & 3.36 & $-$ \\
 J1603$-$2712 & $16$:$03$:$08.036(1)$ & $-27$:$13$:$27.0(7)$ & $1.28482652061(\pm4)$ & $-4.9683(+1,-2)$ & $(+0.11,-0.28)$ & 29 & 3.0 & $-$ \\
 J1603$-$7202 & $16$:$03$:$35.6736(9)$ & $-72$:$02$:$32.795(7)$ & $67.3765811129(+1,-2)$ & $-0.074(+6,-5)$ & $(+0.66,-0.56)$ & 38 & 3.11 & PPTA, B \\
 J1604$-$4909 & $16$:$04$:$22.985(2)$ & $-49$:$09$:$58.33(5)$ & $3.05419496456(\pm5)$ & $-9.4894(\pm3)$ & $(+1.25,-0.25)$ & 113 & 3.47 & $-$ \\
 J1605$-$5257 & $16$:$05$:$16.265(3)$ & $-52$:$57$:$34.80(5)$ & $1.51972586121(\pm1)$ & $-0.59109(+4,-3)$ & $(+0.12,-0.43)$ & 162 & 3.75 & $-$ \\
 J1613$-$4714 & $16$:$13$:$29.018(4)$ & $-47$:$14$:$26.41(8)$ & $2.61522196138(\pm2)$ & $-4.33493(\pm6)$ & $(+0.31,-0.23)$ & 65 & 3.35 & $-$ \\
 J1622$-$4950 & $16$:$22$:$44.80(3)$ & $-49$:$50$:$54.5(5)$ & $0.2311087(\pm3)$ & $-526(+54.3,-56.9)$ & $(+5145.6,-7084.3)$ & 77 & 1.21 & $-$ \\
 J1623$-$0908 & $16$:$23$:$17.658(4)$ & $-09$:$08$:$48.9(3)$ & $0.783424111867(+2,-1)$ & $-1.58401(\pm5)$ & $(+0.36,-0.49)$ & 31 & 3.13 & $-$ \\
 J1623$-$4256 & $16$:$23$:$48.291(6)$ & $-42$:$56$:$52.6(1)$ & $2.74279572052(\pm1)$ & $-7.5624(+8,-7)$ & $(+1.16,-0.06)$ & 57 & 3.47 & $-$ \\
 J1626$-$4537 & $16$:$26$:$48.94(1)$ & $-45$:$37$:$25.8(6)$ & $2.701641249(\pm2)$ & $-60.541(\pm4)$ & $(+190.9,-230.7)$ & 35 & 1.21 & $-$ \\
 J1633$-$4453 & $16$:$33$:$47.03(3)$ & $-44$:$53$:$08.58(7)$ & $2.2908895877(\pm3)$ & $-32.539(+5,-6)$ & $(+327.5,-153.7)$ & 33 & 1.24 & $-$ \\
 J1633$-$5015 & $16$:$33$:$00.0861(1)$ & $-50$:$15$:$08.358(3)$ & $2.83973605453(\pm1)$ & $-30.54746(\pm4)$ & $(+0.128,-0.07)$ & 110 & 3.74 & $-$ \\
 J1639$-$4604 & $16$:$39$:$21.198(3)$ & $-46$:$04$:$33.23(7)$ & $1.88992880367(\pm3)$ & $-20.60947(\pm9)$ & $(+0.69,-0.88)$ & 57 & 2.47 & $-$ \\
 J1644$-$4559 & $16$:$44$:$49.234(6)$ & $-45$:$59$:$10.3(1)$ & $2.19742452445(+4,-3)$ & $-96.9653(+4,-5)$ & $1^{+1.1}_{-1.4}$ & 648 & 4.1 & $-$ \\
 J1646$-$6831 & $16$:$46$:$54.91(3)$ & $-68$:$31$:$51.7(1)$ & $0.560031669373(\pm2)$ & $-0.5337(\pm1)$ & $(+0.074,-0.14)$ & 27 & 3.35 & $-$ \\
 J1651$-$4246 & $16$:$51$:$48.797(6)$ & $-42$:$46$:$09.97(1)$ & $1.18472094037(+4,-5)$ & $-6.662(+3,-2)$ & $(+0.72,-0.76)$ & 148 & 3.46 & $-$ \\
 J1651$-$5222 & $16$:$51$:$42.962(2)$ & $-52$:$22$:$58.38(3)$ & $1.5746588888(\pm1)$ & $-4.48968(\pm3)$ & $(+0.24,-0.017)$ & 95 & 3.38 & $-$ \\
 J1651$-$5255 & $16$:$51$:$41.41(1)$ & $-52$:$55$:$47.7(2)$ & $1.12291858733(\pm1)$ & $-2.6677(\pm6)$ & $(+0.94,-0.04)$ & 71 & 3.14 & $-$ \\
 J1652$-$2404 & $16$:$52$:$58.50(5)$ & $-24$:$03$:$54(7)$ & $0.586943472123(+4,-3)$ & $-1.0877(\pm1)$ & $(+0.29,-0.26)$ & 27 & 2.87 & $-$ \\
 J1700$-$3312 & $17$:$00$:$52.96(2)$ & $-33$:$12$:$45(1)$ & $0.736209097583(+6,-5)$ & $-2.5543(\pm2)$ & $(+0.19,-0.37)$ & 53 & 3.36 & $-$ \\
 J1701$-$3726 & $17$:$01$:$18.45(1)$ & $-37$:$26$:$27.2(5)$ & $0.407395359535(\pm2)$ & $-1.84611(+5,-6)$ & $(+0.23,-0.21)$ & 62 & 3.17 & $-$ \\
 J1703$-$1846 & $17$:$03$:$51.102(9)$ & $-18$:$46$:$13(1)$ & $1.24325189271(\pm3)$ & $-2.67613(\pm1)$ & $(+0.06,-0.21)$ & 34 & 3.18 & $-$ \\
 J1703$-$3241 & $17$:$03$:$22.514(2)$ & $-32$:$41$:$48.5(1)$ & $0.825228539025(\pm5)$ & $-0.44787(\pm2)$ & $(+0.057,-0.067)$ & 81 & 3.54 & $-$ \\
 J1703$-$4851 & $17$:$03$:$54.541(7)$ & $-48$:$52$:$01.04(1)$ & $0.716124628374(\pm2)$ & $-2.60178(+5,-7)$ & $(+0.22,-0.09)$ & 50 & 3.36 & $-$ \\
         \hline
     \end{tabular}
     \renewcommand{\arraystretch}{}
 \end{table}
 \end{landscape}

 \begin{landscape}
 \begin{table}
   \centering
   \contcaption{}
     \renewcommand{\arraystretch}{1.5}
     \begin{tabular}{lllcccccc}
         \hline
         \hline
         PSRJ & RAJ & DECJ & $\nu$ & $\dot{\nu}$ & $\ddot{\nu}$ & $N_{\mathrm{ToA}}$ & $T$ & Flags \\
             & (hh:mm:ss) & ($\degr$:$\arcmin$:$\arcsec$) & (Hz) & ($10^{-15}$ s$^{-2}$) & ($10^{-24}$ s$^{-3}$) &  & (yr) &  \\
         \hline
 J1705$-$1906 & $17$:$05$:$36.093(2)$ & $-19$:$06$:$39.2(3)$ & $3.34458679304(+6,-7)$ & $-46.2498(\pm3)$ & $(+0.27,-0.89)$ & 78 & 3.52 & $-$ \\
 J1705$-$3423 & $17$:$05$:$42.362(3)$ & $-34$:$23$:$43.1(2)$ & $3.91501633777(\pm1)$ & $-16.4861(+6,-5)$ & $(+0.6,-1.6)$ & 112 & 3.43 & $-$ \\
 J1707$-$4053 & $17$:$07$:$21.78(2)$ & $-40$:$53$:$55.1(9)$ & $1.72111797519(\pm2)$ & $-5.6882(+5,-4)$ & $(+2.0,-1.5)$ & 57 & 3.51 & $-$ \\
 J1708$-$3426 & $17$:$08$:$57.79(1)$ & $-34$:$26$:$44.0(6)$ & $1.44484514046(+3,-5)$ & $-8.7827(\pm2)$ & $(+0.42,-0.59)$ & 52 & 3.35 & $-$ \\
 J1709$-$1640 & $17$:$09$:$26.452(4)$ & $-16$:$40$:$59.2(4)$ & $1.53125350203(+1,-7)$ & $-14.8003(+5,-6)$ & $0.61\pm2$ & 38 & 3.63 & $-$ \\
 J1709$-$4429 & $17$:$09$:$42.62(5)$ & $-44$:$29$:$12(1)$ & $9.7542901224(+6,-4)$ & $-8850.16(+6,-8)$ & $196^{+31.4}_{-20.1}$ & 111 & 3.5 & $-$ \\
 J1711$-$5350 & $17$:$11$:$53.13(1)$ & $-53$:$50$:$18.3(2)$ & $1.11205916031(\pm1)$ & $-19.2133(\pm5)$ & $(+0.47,-0.73)$ & 46 & 3.1 & $-$ \\
 J1715$-$4034 & $17$:$15$:$40.92(3)$ & $-40$:$34$:$18(1)$ & $0.482589475307(+7,-6)$ & $-0.7063(\pm2)$ & $(+0.48,-0.52)$ & 76 & 3.5 & $-$ \\
 J1717$-$3425 & $17$:$17$:$20.30(1)$ & $-34$:$25$:$00.31(8)$ & $1.52368077976(+5,-6)$ & $-22.72(\pm2)$ & $2^{+1.7}_{-1.6}$ & 56 & 2.38 & $-$ \\
 J1717$-$4054 & $17$:$17$:$52.31(1)$ & $-41$:$03$:$13.0(4)$ & $1.12648202933(+6,-8)$ & $-4.7161(+8,-5)$ & $-1.54^{+9}_{-2}$ & 31 & 4.09 & $-$ \\
 J1720$-$1633 & $17$:$20$:$25.23(1)$ & $-16$:$33$:$35(1)$ & $0.638730665146(\pm3)$ & $-2.3719(+9,-1)$ & $(+0.23,-0.06)$ & 41 & 3.13 & $-$ \\
 J1720$-$2933 & $17$:$20$:$34.131(5)$ & $-29$:$33$:$14.0(5)$ & $1.61173637049(+3,-2)$ & $-1.9396(\pm1)$ & $(+0.22,-0.22)$ & 43 & 3.29 & $-$ \\
 J1722$-$3207 & $17$:$22$:$02.9641(1)$ & $-32$:$07$:$45.07(6)$ & $2.09574210095(\pm3)$ & $-2.8316(\pm1)$ & $(+0.2,-0.22)$ & 89 & 3.17 & $-$ \\
 J1722$-$3712 & $17$:$22$:$59.17(5)$ & $-37$:$12$:$03.(2)$ & $4.23402576366(\pm4)$ & $-194.486(\pm2)$ & $16^{+13.7}_{-12.7}$ & 116 & 3.1 & $-$ \\
 J1727$-$2739 & $17$:$27$:$30.98(3)$ & $-27$:$38$:$53(4)$ & $0.7733354277(+3,-2)$ & $-0.6399(+5,-6)$ & $3^{+15.8}_{-1.7}$ & 34 & 2.42 & $-$ \\
 J1730$-$2304 & $17$:$30$:$21.682(4)$ & $-23$:$04$:$30(1)$ & $123.110287079(+1,-9)$ & $-0.3023(+2,-4)$ & $(+1.6,-1.2)$ & 42 & 3.3 & PPTA \\
 J1731$-$4744 & $17$:$31$:$42.21(1)$ & $-47$:$44$:$38.7(4)$ & $1.2049313854(+2,-3)$ & $-237.394(\pm5)$ & $-9^{+3.2}_{-0.6}$ & 145 & 3.58 & $-$ \\
 J1733$-$2228 & $17$:$33$:$26.43(3)$ & $-22$:$28$:$37(10)$ & $1.14720621377(\pm5)$ & $-0.0585(+2,-1)$ & $23^{+0.0}_{-14.1}$ & 40 & 3.05 & $-$ \\
 J1736$-$2457 & $17$:$36$:$45.4(1)$ & $-24$:$57$:$50(33)$ & $0.3784689286(+3,-2)$ & $-0.452(+5,-6)$ & $(+174.6,-205.3)$ & 25 & 1.14 & $-$ \\
 J1739$-$2903 & $17$:$39$:$34.285(2)$ & $-29$:$03$:$03.96(3)$ & $3.09704932896(+9,-8)$ & $-75.5355(+3,-4)$ & $(+0.36,-0.6)$ & 88 & 3.02 & $-$ \\
 J1740$-$3015 & $17$:$40$:$33.98(5)$ & $-30$:$15$:$22(5)$ & $1.647450502(+2,-3)$ & $-1263.51(+7,-9)$ & $(+46.0,-51.7)$ & 229 & 3.47 & $-$ \\
 J1741$-$3927 & $17$:$41$:$18.079(1)$ & $-39$:$27$:$38.12(7)$ & $1.95231526583(\pm1)$ & $-6.4645(+5,-6)$ & $2.81^{+3}_{-5}$ & 74 & 3.14 & $-$ \\
 J1743$-$3150 & $17$:$43$:$36.710(8)$ & $-31$:$50$:$22.7(9)$ & $0.414138298084(\pm1)$ & $-20.7152(+4,-5)$ & $(+0.13,-0.09)$ & 84 & 3.16 & $-$ \\
 J1745$-$3040 & $17$:$45$:$56.3081(6)$ & $-30$:$40$:$23.30(6)$ & $2.72156341619(+1,-2)$ & $-79.04005(+8,-9)$ & $(+0.23,-0.1)$ & 110 & 3.5 & $-$ \\
 J1751$-$4657 & $17$:$51$:$42.185(1)$ & $-46$:$57$:$26.72(4)$ & $1.34706694407(+4,-3)$ & $-2.35478(+1,-2)$ & $(+0.09,-0.083)$ & 53 & 3.61 & $-$ \\
 J1752$-$2806 & $17$:$52$:$58.707(8)$ & $-28$:$06$:$36(1)$ & $1.77757096075(\pm6)$ & $-25.6877(+8,-7)$ & $0.51^{+3}_{-4}$ & 145 & 4.1 & $-$ \\
 J1757$-$2421 & $17$:$57$:$29.37(1)$ & $-24$:$19$:$54(10)$ & $4.2715099866(+2,-3)$ & $-236.544(+5,-4)$ & $(+166.8,-147.0)$ & 66 & 1.31 & $-$ \\
 J1759$-$2205 & $17$:$59$:$24.164(4)$ & $-22$:$05$:$33(2)$ & $2.16928428064(+1,-2)$ & $-51.0746(+7,-6)$ & $(+1.52,-0.26)$ & 54 & 3.02 & $-$ \\
 J1759$-$3107 & $17$:$59$:$22.056(4)$ & $-31$:$07$:$21.8(5)$ & $0.926822758345(\pm3)$ & $-3.24135(\pm9)$ & $(+0.81,-0.48)$ & 40 & 2.39 & $-$ \\
 J1801$-$0357 & $18$:$01$:$22.628(3)$ & $-03$:$57$:$55.7(2)$ & $1.08519559579(\pm4)$ & $-3.8928(\pm1)$ & $(+0.78,-0.46)$ & 29 & 2.34 & $-$ \\
         \hline
     \end{tabular}
     \renewcommand{\arraystretch}{}
 \end{table}
 \end{landscape}

 \begin{landscape}
 \begin{table}
   \centering
   \contcaption{}
     \renewcommand{\arraystretch}{1.5}
     \begin{tabular}{lllcccccc}
         \hline
         \hline
         PSRJ & RAJ & DECJ & $\nu$ & $\dot{\nu}$ & $\ddot{\nu}$ & $N_{\mathrm{ToA}}$ & $T$ & Flags \\
             & (hh:mm:ss) & ($\degr$:$\arcmin$:$\arcsec$) & (Hz) & ($10^{-15}$ s$^{-2}$) & ($10^{-24}$ s$^{-3}$) &  & (yr) &  \\
         \hline
 J1801$-$2920 & $18$:$01$:$46.839(3)$ & $-29$:$20$:$38.1(3)$ & $0.924290873961(\pm1)$ & $-2.81266(\pm3)$ & $(+0.24,-0.02)$ & 60 & 3.26 & $-$ \\
 J1803$-$2137 & $18$:$03$:$51.4(1)$ & $-21$:$37$:$07.(27)$ & $7.478883401(+4,-6)$ & $-7488(+1.1,-0.9)$ & $283^{+29.1}_{-39.6}$ & 52 & 1.29 & $-$ \\
 J1805$-$1504 & $18$:$05$:$06.1(2)$ & $-15$:$04$:$36(10)$ & $0.84654711(\pm1)$ & $-0.31(\pm2)$ & $(+378.9,-625.7)$ & 28 & 1.25 & $-$ \\
 J1807$-$0847 & $18$:$07$:$38.0259(2)$ & $-08$:$47$:$43.28(1)$ & $6.10771328217(+6,-5)$ & $-1.06808(+2,-1)$ & $(+0.12,-0.057)$ & 74 & 3.7 & $-$ \\
 J1807$-$2715 & $18$:$07$:$08.4918(3)$ & $-27$:$15$:$02.07(5)$ & $1.20804374592(\pm6)$ & $-17.8128(\pm2)$ & $(+0.65,-0.2)$ & 77 & 3.12 & $-$ \\
 J1808$-$0813 & $18$:$08$:$09.432(1)$ & $-08$:$13$:$01.8(4)$ & $1.14149384538(\pm5)$ & $-1.6108(\pm2)$ & $(+0.13,-0.35)$ & 32 & 3.38 & $-$ \\
 J1809$-$2109 & $18$:$09$:$14.32(3)$ & $-21$:$09$:$02.(5)$ & $1.42365721129(\pm4)$ & $-7.747(\pm1)$ & $(+22.7,-13.0)$ & 29 & 2.4 & $-$ \\
 J1810$-$5338 & $18$:$10$:$44.473(3)$ & $-53$:$38$:$07.631(6)$ & $3.8306868647(\pm4)$ & $-5.6604(\pm1)$ & $(+0.78,-0.49)$ & 32 & 2.96 & $-$ \\
 J1816$-$2650 & $18$:$16$:$35.399(6)$ & $-26$:$49$:$53(1)$ & $1.68666719259(+2,-3)$ & $-0.18919(+9,-1)$ & $(+0.75,-0.83)$ & 51 & 3.35 & $-$ \\
 J1818$-$1422 & $18$:$18$:$23.77(1)$ & $-14$:$22$:$39(1)$ & $3.43064845763(\pm2)$ & $-23.9924(+6,-7)$ & $(+1.2,-3.0)$ & 51 & 3.02 & $-$ \\
 J1820$-$0427 & $18$:$20$:$52.559(2)$ & $-04$:$27$:$37.9(1)$ & $1.67201171071(+6,-9)$ & $-17.6967(+5,-3)$ & $-0.64^{+5}_{-6}$ & 55 & 3.63 & $-$ \\
 J1822$-$2256 & $18$:$22$:$58.95(4)$ & $-22$:$56$:$29(16)$ & $0.53354117731(\pm2)$ & $-0.38531(\pm5)$ & $(+0.31,-0.14)$ & 57 & 3.33 & $-$ \\
 J1823$-$0154 & $18$:$23$:$52.138(3)$ & $-01$:$54$:$04.94(1)$ & $1.31617369972(+3,-2)$ & $-1.95718(\pm9)$ & $(+0.41,-0.34)$ & 35 & 3.14 & $-$ \\
 J1823$-$1115 & $18$:$23$:$40.3(1)$ & $-11$:$15$:$11(1)$ & $3.57360247363(\pm2)$ & $-17.5869(+7,-8)$ & $(+2.8,-2.4)$ & 53 & 3.19 & B \\
 J1823$-$3106 & $18$:$23$:$46.819(4)$ & $-31$:$06$:$48.0(3)$ & $3.52042950493(+1,-9)$ & $-36.3594(+4,-6)$ & $0.69\pm3$ & 35 & 3.11 & $-$ \\
 J1824$-$0127 & $18$:$24$:$53.43(1)$ & $-01$:$27$:$51.4(4)$ & $0.400084842618(\pm3)$ & $-0.62531(+8,-1)$ & $(+0.81,-0.82)$ & 30 & 2.4 & $-$ \\
 J1824$-$1945 & $18$:$24$:$00.4360(4)$ & $-19$:$45$:$44.5(8)$ & $5.28154642765(\pm1)$ & $-146.2029(+8,-7)$ & $-2^{+1.9}_{-1.9}$ & 95 & 3.38 & $-$ \\
 J1825$-$0935 & $18$:$25$:$30.62(6)$ & $-09$:$35$:$22(4)$ & $1.3003801253(\pm1)$ & $-88.397(\pm3)$ & $(+59.4,-52.1)$ & 144 & 3.84 & $-$ \\
 J1827$-$0750 & $18$:$27$:$02.7071(6)$ & $-07$:$50$:$15.4(2)$ & $3.69682174881(+6,-5)$ & $-21.224(\pm2)$ & $(+8.0,-6.7)$ & 49 & 2.36 & $-$ \\
 J1829$-$1751 & $18$:$29$:$43.15(1)$ & $-17$:$51$:$13(1)$ & $3.25587939511(+2,-3)$ & $-58.78(\pm1)$ & $(+0.9,-1.1)$ & 79 & 3.41 & $-$ \\
 J1830$-$1135 & $18$:$30$:$01.787(6)$ & $-11$:$35$:$27(6)$ & $0.160730937421(\pm6)$ & $-1.2319(\pm2)$ & $(+0.51,-0.61)$ & 40 & 2.99 & $-$ \\
 J1832$-$0827 & $18$:$32$:$37.013(2)$ & $-08$:$27$:$03.16(1)$ & $1.54478860938(\pm6)$ & $-152.4959(+3,-2)$ & $-0.39\pm3$ & 93 & 3.21 & $-$ \\
 J1833$-$0338 & $18$:$33$:$42.028(8)$ & $-03$:$39$:$08.00(3)$ & $1.45617035595(+1,-9)$ & $-88.0832(+4,-5)$ & $1^{+1.1}_{-1.7}$ & 102 & 3.19 & $-$ \\
 J1833$-$0827 & $18$:$33$:$40.245(2)$ & $-08$:$27$:$30.9(1)$ & $11.7247184586(\pm5)$ & $-1261.993(\pm2)$ & $(+4.1,-5.2)$ & 54 & 2.96 & $-$ \\
 J1834$-$0426 & $18$:$34$:$25.621(3)$ & $-04$:$26$:$15.7(2)$ & $3.44698922332(\pm3)$ & $-0.8605(\pm1)$ & $(+1.0,-0.8)$ & 53 & 3.22 & $-$ \\
 J1835$-$1020 & $18$:$35$:$57.44(3)$ & $-10$:$19$:$51(2)$ & $3.3063181115(\pm1)$ & $-64.651(\pm5)$ & $(+5.2,-16.0)$ & 53 & 3.23 & $-$ \\
 J1836$-$0436 & $18$:$36$:$51.77(1)$ & $-04$:$36$:$37.3(7)$ & $2.82296346028(\pm4)$ & $-13.232(\pm1)$ & $(+11.7,-48.8)$ & 31 & 2.42 & $-$ \\
 J1836$-$1008 & $18$:$36$:$53.922(3)$ & $-10$:$08$:$09.39(2)$ & $1.77708391524(+8,-1)$ & $-37.1805(+8,-4)$ & $(+1.2,-1.3)$ & 74 & 3.53 & $-$ \\
 J1837$-$0653 & $18$:$37$:$14.53(7)$ & $-06$:$52$:$55(5)$ & $0.52471147608(+4,-5)$ & $-0.194(+2,-1)$ & $(+6.1,-6.1)$ & 50 & 2.24 & $-$ \\
 J1840$-$0809 & $18$:$40$:$33.365(6)$ & $-08$:$09$:$03.62(4)$ & $1.04638272527(\pm3)$ & $-2.57306(+9,-1)$ & $(+0.17,-0.2)$ & 75 & 3.43 & $-$ \\
         \hline
     \end{tabular}
     \renewcommand{\arraystretch}{}
 \end{table}
 \end{landscape}

 \begin{landscape}
 \begin{table}
   \centering
   \contcaption{}
     \renewcommand{\arraystretch}{1.5}
     \begin{tabular}{lllcccccc}
         \hline
         \hline
         PSRJ & RAJ & DECJ & $\nu$ & $\dot{\nu}$ & $\ddot{\nu}$ & $N_{\mathrm{ToA}}$ & $T$ & Flags \\
             & (hh:mm:ss) & ($\degr$:$\arcmin$:$\arcsec$) & (Hz) & ($10^{-15}$ s$^{-2}$) & ($10^{-24}$ s$^{-3}$) &  & (yr) &  \\
         \hline
 J1840$-$0815 & $18$:$40$:$13.756(9)$ & $-08$:$15$:$08.88(4)$ & $0.912041662037(\pm3)$ & $-2.01835(\pm8)$ & $(+0.6,-1.3)$ & 51 & 2.44 & $-$ \\
 J1841$-$0425 & $18$:$41$:$05.663(5)$ & $-04$:$25$:$19.5(2)$ & $5.37198570613(\pm4)$ & $-184.318(\pm1)$ & $(+3.2,-6.3)$ & 31 & 2.24 & $-$ \\
 J1841$+$0912 & $18$:$41$:$55.921(7)$ & $+09$:$12$:$07.29(2)$ & $2.62246808546(+3,-4)$ & $-7.505(+2,-1)$ & $(+0.3,-4.2)$ & 29 & 3.23 & $-$ \\
 J1842$-$0359 & $18$:$42$:$26.49(1)$ & $-04$:$00$:$01.52(7)$ & $0.543494594895(\pm5)$ & $-0.1501(\pm1)$ & $(+0.4,-1.0)$ & 77 & 3.44 & $-$ \\
 J1843$-$0000 & $18$:$43$:$27.965(9)$ & $-00$:$00$:$41.5(2)$ & $1.13593208324(+1,-2)$ & $-10.0305(\pm7)$ & $(+1.0,-0.7)$ & 60 & 3.19 & $-$ \\
 J1844$-$0433 & $18$:$44$:$33.446(3)$ & $-04$:$33$:$12.5(1)$ & $1.00905187281(\pm2)$ & $-3.98545(+4,-5)$ & $(+0.29,-0.24)$ & 46 & 2.68 & $-$ \\
 J1845$-$0743 & $18$:$45$:$57.1763(9)$ & $-07$:$43$:$38.16(5)$ & $9.55157970586(+5,-4)$ & $-33.444(\pm1)$ & $(+1.5,-2.4)$ & 72 & 2.46 & $-$ \\
 J1847$-$0402 & $18$:$47$:$22.850(1)$ & $-04$:$02$:$14.70(7)$ & $1.67277577163(+6,-4)$ & $-144.6391(+2,-3)$ & $(+0.36,-0.0)$ & 127 & 3.53 & $-$ \\
 J1848$-$0123 & $18$:$48$:$23.596(1)$ & $-01$:$23$:$58.48(6)$ & $1.51644857592(+3,-6)$ & $-11.9808(+4,-2)$ & $0.19^{+2}_{-1}$ & 111 & 3.49 & $-$ \\
 J1849$-$0636 & $18$:$49$:$06.4647(2)$ & $-06$:$37$:$06.91(1)$ & $0.689011346415(\pm3)$ & $-21.9601(\pm1)$ & $(+0.02,-0.23)$ & 108 & 3.23 & $-$ \\
 J1852$-$0635 & $18$:$52$:$57.448(5)$ & $-06$:$36$:$00.45(2)$ & $1.90782618111(+5,-4)$ & $-53.2438(\pm1)$ & $(+0.59,-0.23)$ & 109 & 3.25 & $-$ \\
 J1852$-$2610 & $18$:$52$:$59.471(5)$ & $-26$:$10$:$13.6(6)$ & $2.9732067598(\pm6)$ & $-0.7704(+1,-2)$ & $(+1.5,-1.9)$ & 25 & 2.43 & $-$ \\
 J1857$+$0212 & $18$:$57$:$43.654(8)$ & $+02$:$12$:$41.0(3)$ & $2.40470716439(+9,-1)$ & $-232.7342(\pm3)$ & $(+1.4,-0.35)$ & 71 & 2.72 & $-$ \\
 J1900$-$2600 & $19$:$00$:$47.542(5)$ & $-26$:$00$:$44.8(6)$ & $1.63342812459(\pm1)$ & $-0.54862(+4,-5)$ & $(+0.15,-0.167)$ & 40 & 3.44 & $-$ \\
 J1901$+$0331 & $19$:$01$:$31.76(1)$ & $+03$:$31$:$06.73(4)$ & $1.52565744968(\pm2)$ & $-17.3341(+8,-6)$ & $(+2.3,-2.6)$ & 127 & 3.13 & $-$ \\
 J1901$+$0716 & $19$:$01$:$39.02(1)$ & $+07$:$16$:$33.6(5)$ & $1.55279458439(\pm4)$ & $-5.577(\pm1)$ & $(+4.6,-1.3)$ & 28 & 2.61 & $-$ \\
 J1901$-$0906 & $19$:$01$:$53.007(3)$ & $-09$:$06$:$10.9(2)$ & $0.561189668479(+6,-7)$ & $-0.516(\pm2)$ & $(+0.08,-0.19)$ & 52 & 3.12 & $-$ \\
 J1902$+$0556 & $19$:$02$:$42.60(1)$ & $+05$:$56$:$25.8(1)$ & $1.33943019014(\pm1)$ & $-23.0839(\pm4)$ & $(+1.3,-2.6)$ & 32 & 2.44 & $-$ \\
 J1902$+$0615 & $19$:$02$:$50.277(3)$ & $+06$:$16$:$33.41(7)$ & $1.48476895959(\pm4)$ & $-16.9975(\pm1)$ & $(+0.5,-2.3)$ & 45 & 2.62 & $-$ \\
 J1903$+$0135 & $19$:$03$:$29.973(1)$ & $+01$:$35$:$38.73(4)$ & $1.37116475483(\pm2)$ & $-7.57312(\pm8)$ & $(+0.16,-0.28)$ & 94 & 3.18 & $-$ \\
 J1903$-$0632 & $19$:$03$:$37.934(2)$ & $-06$:$32$:$21.52(9)$ & $2.31540809127(+7,-8)$ & $-18.1244(+4,-3)$ & $(+0.16,-0.71)$ & 69 & 3.14 & $-$ \\
 J1905$-$0056 & $19$:$05$:$27.752(6)$ & $-00$:$56$:$40.8(3)$ & $1.55476661905(+7,-6)$ & $-7.3951(\pm1)$ & $(+2.0,-1.7)$ & 29 & 2.45 & $-$ \\
 J1909$+$0007 & $19$:$09$:$35.252(2)$ & $+00$:$07$:$56.84(9)$ & $0.983329997648(+4,-7)$ & $-5.3391(+3,-2)$ & $(+0.37,-0.68)$ & 71 & 3.07 & $-$ \\
 J1909$+$0254 & $19$:$09$:$38.311(2)$ & $+02$:$54$:$50.36(9)$ & $1.01026940483(\pm1)$ & $-5.61185(\pm4)$ & $(+0.24,-0.1)$ & 51 & 3.19 & $-$ \\
 J1909$+$1102 & $19$:$09$:$48.6829(9)$ & $+11$:$02$:$03.044(3)$ & $3.5255695764(+1,-9)$ & $-32.8178(\pm4)$ & $1^{+0.7}_{-1.1}$ & 95 & 3.19 & $-$ \\
 J1909$-$3744 & $19$:$09$:$47.42783(7)$ & $-37$:$44$:$14.767(3)$ & $339.315686856(\pm5)$ & $-1.6153(\pm2)$ & $(+4,-5)$ & 68 & 3.54 & PPTA, B \\
 J1910$-$0309 & $19$:$10$:$29.692(2)$ & $-03$:$09$:$54.1(1)$ & $1.98174395507(+3,-4)$ & $-8.61183(\pm1)$ & $-0.22^{+2}_{-4}$ & 43 & 3.09 & $-$ \\
 J1910$+$0358 & $19$:$10$:$09.016(3)$ & $+03$:$58$:$30(1)$ & $0.429135601854(+1,-8)$ & $-0.8134(\pm3)$ & $(+1.6,-1.9)$ & 44 & 2.39 & $-$ \\
 J1913$-$0440 & $19$:$13$:$54.1624(9)$ & $-04$:$40$:$47.56(4)$ & $1.21074218559(\pm2)$ & $-5.9681(\pm1)$ & $(+0.38,-0.21)$ & 88 & 3.52 & $-$ \\
 J1913$+$1400 & $19$:$13$:$24.352(1)$ & $+14$:$00$:$52.50(3)$ & $1.91764388163(+2,-3)$ & $-2.95953(\pm7)$ & $(+0.29,-0.24)$ & 66 & 2.63 & $-$ \\
         \hline
     \end{tabular}
     \renewcommand{\arraystretch}{}
 \end{table}
 \end{landscape}

 \begin{landscape}
 \begin{table}
   \centering
   \contcaption{}
     \renewcommand{\arraystretch}{1.5}
     \begin{tabular}{lllcccccc}
         \hline
         \hline
         PSRJ & RAJ & DECJ & $\nu$ & $\dot{\nu}$ & $\ddot{\nu}$ & $N_{\mathrm{ToA}}$ & $T$ & Flags \\
             & (hh:mm:ss) & ($\degr$:$\arcmin$:$\arcsec$) & (Hz) & ($10^{-15}$ s$^{-2}$) & ($10^{-24}$ s$^{-3}$) &  & (yr) &  \\
         \hline
 J1915$+$1009 & $19$:$15$:$29.993(1)$ & $+10$:$09$:$43.58(3)$ & $2.47187153893(\pm3)$ & $-93.22115(+9,-8)$ & $(+0.71,-0.2)$ & 105 & 3.13 & $-$ \\
 J1916$+$0951 & $19$:$16$:$32.333(1)$ & $+09$:$51$:$25.97(3)$ & $3.70019376623(\pm8)$ & $-34.52(+4,-3)$ & $(+0.1,-0.51)$ & 86 & 3.17 & $-$ \\
 J1916$+$1312 & $19$:$16$:$58.67(2)$ & $+13$:$12$:$50.0(4)$ & $3.548050079(\pm2)$ & $-46.069(+4,-5)$ & $(+49.9,-9.6)$ & 39 & 161.67 & $-$ \\
 J1917$+$1353 & $19$:$17$:$39.794(3)$ & $+13$:$53$:$57.16(8)$ & $5.13779943103(\pm4)$ & $-189.936(\pm1)$ & $1^{+1.1}_{-1.5}$ & 71 & 2.47 & $-$ \\
 J1919$+$0021 & $19$:$19$:$50.670(2)$ & $+00$:$21$:$39.8(1)$ & $0.78599927(+8,-7)$ & $-4.74138(+2,-3)$ & $(+0.18,-0.15)$ & 97 & 3.26 & $-$ \\
 J1926$+$0431 & $19$:$26$:$24.472(2)$ & $+04$:$31$:$31.54(8)$ & $0.931029279866(+1,-2)$ & $-2.13409(+4,-5)$ & $(+0.18,-0.1)$ & 130 & 3.18 & $-$ \\
 J1932$+$1059 & $19$:$32$:$14.038(2)$ & $+10$:$59$:$33.21(5)$ & $4.41464565156(\pm1)$ & $-22.5369(+7,-6)$ & $(+1.6,-1.4)$ & 116 & 3.64 & $-$ \\
 J1932$-$3655 & $19$:$32$:$06.1280(6)$ & $-36$:$55$:$01.78(3)$ & $1.75002463079(+9,-1)$ & $-0.8767(\pm3)$ & $(+0.6,-3.6)$ & 39 & 2.38 & $-$ \\
 J1935$+$1616 & $19$:$35$:$47.8255(2)$ & $+16$:$16$:$39.723(4)$ & $2.78750145981(+3,-5)$ & $-46.6373(+1,-8)$ & $(+0.51,-0.31)$ & 59 & 2.15 & $-$ \\
 J1941$-$2602 & $19$:$41$:$00.4169(1)$ & $-26$:$02$:$05.884(9)$ & $2.48226091399(+9,-8)$ & $-5.89424(\pm2)$ & $(+0.21,-0.098)$ & 105 & 3.09 & $-$ \\
 J1943$-$1237 & $19$:$43$:$25.461(3)$ & $-12$:$37$:$42.9(2)$ & $1.02835150982(+1,-9)$ & $-1.75624(+4,-3)$ & $(+1.4,-0.2)$ & 54 & 3.22 & $-$ \\
 J1945$-$0040 & $19$:$45$:$28.33(3)$ & $-00$:$40$:$59(1)$ & $0.9563585837(\pm1)$ & $-0.47(\pm2)$ & $(+79.1,-88.5)$ & 64 & 1.3 & $-$ \\
 J1946$-$2913 & $19$:$46$:$51.757(5)$ & $-29$:$13$:$48.1(3)$ & $1.04226478935(\pm1)$ & $-1.61748(+5,-4)$ & $(+0.2,-0.14)$ & 63 & 3.27 & $-$ \\
 J2006$-$0807 & $20$:$06$:$16.365(4)$ & $-08$:$07$:$02.16(2)$ & $1.72155151633(\pm5)$ & $-0.1355(+1,-9)$ & $(+0.52,-0.4)$ & 255 & 3.42 & $-$ \\
 J2033$+$0042 & $20$:$33$:$31.12(2)$ & $+00$:$42$:$24.1(9)$ & $0.199465428208(+3,-2)$ & $-0.38564(+6,-7)$ & $(+0.21,-0.09)$ & 129 & 3.28 & $-$ \\
 J2038$-$3816 & $20$:$38$:$54.36(3)$ & $-38$:$16$:$15.6(9)$ & $0.633999188596(\pm9)$ & $-1.6728(\pm3)$ & $(+2.6,-6.7)$ & 48 & 2.4 & $-$ \\
 J2046$-$0421 & $20$:$46$:$00.1760(2)$ & $-04$:$21$:$26.3(1)$ & $0.646437789195(+1,-9)$ & $-0.61473(+3,-2)$ & $(+0.044,-0.064)$ & 141 & 3.41 & $-$ \\
 J2046$+$1540 & $20$:$46$:$39.336(5)$ & $+15$:$40$:$33.6(1)$ & $0.878513972444(\pm3)$ & $-0.14056(\pm7)$ & $(+0.21,-0.24)$ & 72 & 3.39 & $-$ \\
 J2048$-$1616 & $20$:$48$:$35.74(2)$ & $-16$:$16$:$45(1)$ & $0.509792367545(\pm6)$ & $-2.84929(\pm2)$ & $(+0.071,-0.055)$ & 105 & 4.04 & $-$ \\
 J2051$-$0827 & $20$:$51$:$07.52058(5)$ & $-08$:$27$:$37.61(2)$ & $221.796283548(\pm3)$ & $-0.6248(+9,-7)$ & $(+2.6,-0.5)$ & 193 & 3.23 & B \\
 J2053$-$7200 & $20$:$53$:$47.280(4)$ & $-72$:$00$:$42.48(2)$ & $2.9296611297(\pm3)$ & $-1.69606(+9,-8)$ & $(+0.1,-0.19)$ & 64 & 3.22 & $-$ \\
 J2116$+$1414 & $21$:$16$:$13.761(1)$ & $+14$:$14$:$20.38(4)$ & $2.27193569866(\pm4)$ & $-1.49439(\pm8)$ & $(+0.13,-0.38)$ & 127 & 3.18 & $-$ \\
 J2129$-$5721 & $21$:$29$:$22.77664(9)$ & $-57$:$21$:$14.2954(7)$ & $268.359226956(\pm2)$ & $-1.5024(\pm5)$ & $(+14.1,-15.3)$ & 100 & 2.22 & PPTA, B \\
 J2144$-$3933 & $21$:$44$:$12.01(1)$ & $-39$:$33$:$58.4(3)$ & $0.117511188481(+4,-5)$ & $-0.0064(+1,-2)$ & $(+0.038,-0.089)$ & 95 & 3.2 & $-$ \\
 J2145$-$0750 & $21$:$45$:$50.4552(8)$ & $-07$:$50$:$18.56(3)$ & $62.2958878113(\pm3)$ & $-0.111(\pm1)$ & $(+2.5,-0.1)$ & 162 & 3.12 & PPTA, B \\
 J2155$-$3118 & $21$:$55$:$13.64(1)$ & $-31$:$18$:$53.8(2)$ & $0.97087088287(+3,-2)$ & $-1.16876(+7,-8)$ & $(+0.25,-0.21)$ & 61 & 3.08 & $-$ \\
 J2222$-$0137 & $22$:$22$:$05.96713(1)$ & $-01$:$37$:$15.731(5)$ & $30.4712133291(\pm1)$ & $-0.0535(+3,-5)$ & $(+0.67,-0.17)$ & 216 & 3.2 & B \\
 J2241$-$5236 & $22$:$41$:$42.03154(6)$ & $-52$:$36$:$36.2491(6)$ & $457.310149438(+9,-1)$ & $-1.4408(\pm6)$ & $(+0.65,-0.81)$ & 295 & 3.27 & PPTA, B \\
 J2248$-$0101 & $22$:$48$:$26.884(6)$ & $-01$:$01$:$48.0(2)$ & $2.09541027394(\pm5)$ & $-2.8961(\pm1)$ & $(+0.8,-2.5)$ & 172 & 2.54 & $-$ \\
 J2324$-$6054 & $23$:$24$:$27.14(1)$ & $-60$:$54$:$05.794(9)$ & $0.425987202198(+1,-9)$ & $-0.46843(+3,-4)$ & $(+0.061,-0.12)$ & 87 & 3.07 & $-$ \\
         \hline
     \end{tabular}
     \renewcommand{\arraystretch}{}
 \end{table}
 \end{landscape}

 \begin{landscape}
 \begin{table}
   \centering
   \contcaption{}
     \renewcommand{\arraystretch}{1.5}
     \begin{tabular}{lllcccccc}
         \hline
         \hline
         PSRJ & RAJ & DECJ & $\nu$ & $\dot{\nu}$ & $\ddot{\nu}$ & $N_{\mathrm{ToA}}$ & $T$ & Flags \\
             & (hh:mm:ss) & ($\degr$:$\arcmin$:$\arcsec$) & (Hz) & ($10^{-15}$ s$^{-2}$) & ($10^{-24}$ s$^{-3}$) &  & (yr) &  \\
         \hline
 J2330$-$2005 & $23$:$30$:$26.986(2)$ & $-20$:$05$:$29.75(7)$ & $0.608411174931(\pm4)$ & $-1.71419(\pm1)$ & $(+0.042,-0.097)$ & 172 & 3.52 & $-$ \\
 J2346$-$0609 & $23$:$46$:$50.54(1)$ & $-06$:$10$:$01.04(4)$ & $0.846407381972(\pm3)$ & $-0.9728(\pm1)$ & $(+0.04,-0.46)$ & 236 & 3.23 & $-$ \\
         \hline
     \end{tabular}
     \renewcommand{\arraystretch}{}
 \end{table}
 \end{landscape}

\section{Timing noise parameters}\label{apdx:red_params}

Preferred models indicated are: white timing noise (WTN), power-law red noise (PLRN), power law red noise with low-frequency turnover (PL+FC) and power-law red noise with $\ddot{\nu}$ (PLRN+F2). The listed Bayes factor is taken as being the difference in evidences between the best model and the next simplest model.

\begin{table}
  \centering
  \caption{List of the preferred timing noise model, Bayes factor when compared to the next simplest model and associated red noise parameters (errors indicate the 95 percent confidence intervals) for each pulsar in our data set. MSPs are indicated by a $\star$ and clock reference pulsars by a $\dagger$. The red noise of clock reference pulsars (such as PSR J0437$-$4715) are contaminated by residual clock jumps, so should be used with caution. The full table can be found in the supplementary material.}
    \label{tab:timing_noise}
    \renewcommand{\arraystretch}{1.5}
    \begin{tabular}{llccc}
        \hline
        \hline
        PSRJ & Model & $\ln(\mathcal{B})$ & $\log_{10}(A)$ & $\beta$ \\
             & & & ($\mathrm{yr}^{3/2}$) & \\
        \hline
J0030$+$0451$^{\star}$ & WTN & $-$ & $-$ & $-$ \\
J0134$-$2937 & WTN & $-$ & $-$ & $-$ \\
J0151$-$0635 & WTN & $-$ & $-$ & $-$ \\
J0152$-$1637 & WTN & $-$ & $-$ & $-$ \\
J0206$-$4028 & WTN & $-$ & $-$ & $-$ \\
J0255$-$5304 & WTN & $-$ & $-$ & $-$ \\
J0348$+$0432$^{\star}$ & WTN & $-$ & $-$ & $-$ \\
J0401$-$7608 & PLRN & 10.6 & $-10.2^{+0.2}_{-0.5}$ & $4.0^{+14.4}_{--2.5}$ \\
J0418$-$4154 & WTN & $-$ & $-$ & $-$ \\
J0437$-$4715$^{\star\dagger}$ & PLRN & 4.2 & $-10.8^{+0.7}_{-0.4}$ & $9.5^{+3.8}_{-2.5}$ \\
J0450$-$1248 & WTN & $-$ & $-$ & $-$ \\
J0452$-$1759 & PLRN & 22.0 & $-10.4^{+0.3}_{-0.7}$ & $3.1^{+3.5}_{-2.0}$ \\
J0525$+$1115 & WTN & $-$ & $-$ & $-$ \\
J0529$-$6652 & WTN & $-$ & $-$ & $-$ \\
J0533$+$0402 & WTN & $-$ & $-$ & $-$ \\
J0536$-$7543 & WTN & $-$ & $-$ & $-$ \\
J0601$-$0527 & WTN & $-$ & $-$ & $-$ \\
J0624$-$0424 & WTN & $-$ & $-$ & $-$ \\
J0627$+$0706 & PLRN & 60.2 & $-10.0^{+0.4}_{-0.4}$ & $3.5^{+2.5}_{-1.7}$ \\
J0630$-$2834 & WTN & $-$ & $-$ & $-$ \\
J0646$+$0905 & WTN & $-$ & $-$ & $-$ \\
J0659$+$1414 & PLRN & 26.8 & $-10.1^{+0.4}_{-3.5}$ & $5.6^{+14.3}_{-2.1}$ \\
J0711$-$6830$^{\star}$ & WTN & $-$ & $-$ & $-$ \\
J0729$-$1836 & PLRN & 191.6 & $-9.7^{+0.3}_{-0.3}$ & $6.1^{+3.0}_{-2.0}$ \\
J0737$-$3039A$^{\star}$ & WTN & $-$ & $-$ & $-$ \\
J0738$-$4042 & PLRN+F2 & 5.4 & $-9.8^{+0.3}_{-0.2}$ & $6.5^{+1.8}_{-1.4}$ \\
J0742$-$2822 & PLRN & 512.3 & $-9.0^{+0.2}_{-0.1}$ & $4.8^{+1.4}_{-1.0}$ \\
J0758$-$1528 & PLRN & 3.2 & $-10.7^{+0.6}_{-3.9}$ & $4.1^{+15.8}_{-1.8}$ \\
J0809$-$4753 & PLRN & 3.5 & $-11.1^{+1.2}_{-3.6}$ & $5.7^{+14.2}_{-3.5}$ \\
J0820$-$1350 & WTN & $-$ & $-$ & $-$ \\
J0820$-$4114 & WTN & $-$ & $-$ & $-$ \\
J0835$-$4510 & PLRN & 3173.2 & $-8.2^{+0.2}_{-0.2}$ & $8.6^{+0.9}_{-0.9}$ \\
J0837$+$0610 & WTN & $-$ & $-$ & $-$ \\
J0837$-$4135 & PLRN & 138.6 & $-11.7^{+0.8}_{-1.0}$ & $7.5^{+5.5}_{-2.9}$ \\
J0840$-$5332 & WTN & $-$ & $-$ & $-$ \\
J0842$-$4851 & WTN & $-$ & $-$ & $-$ \\
J0846$-$3533 & WTN & $-$ & $-$ & $-$ \\
        \hline
    \end{tabular}
    \renewcommand{\arraystretch}{}
\end{table}

 \begin{table}
   \centering
   \contcaption{}
     \label{tab:timing_noise}
     \renewcommand{\arraystretch}{1.5}
     \begin{tabular}{llcccc}
         \hline
         \hline
         PSRJ & Model & $\ln(\mathcal{B})$ & $\log_{10}(A)$ & $\beta$ \\
              & & & ($\mathrm{yr}^{3/2}$) & \\
         \hline
 J0855$-$3331 & WTN & $-$ & $-$ & $-$ \\
 J0856$-$6137 & WTN & $-$ & $-$ & $-$ \\
 J0904$-$4246 & WTN & $-$ & $-$ & $-$ \\
 J0904$-$7459 & WTN & $-$ & $-$ & $-$ \\
 J0907$-$5157 & PLRN & 92.7 & $-10.8^{+0.7}_{-3.4}$ & $5.8^{+14.2}_{-2.1}$ \\
 J0908$-$1739 & WTN & $-$ & $-$ & $-$ \\
 J0908$-$4913 & PLRN & 523.7 & $-9.6^{+0.2}_{-0.2}$ & $5.0^{+1.0}_{-0.8}$ \\
 J0909$-$7212 & WTN & $-$ & $-$ & $-$ \\
 J0922$+$0638 & PLRN & 101.1 & $-9.4^{+0.2}_{-0.2}$ & $5.6^{+1.8}_{-2.0}$ \\
 J0924$-$5302 & PLRN & 279.9 & $-9.3^{+0.3}_{-0.2}$ & $4.5^{+1.3}_{-1.1}$ \\
 J0924$-$5814 & WTN & $-$ & $-$ & $-$ \\
 J0934$-$5249 & WTN & $-$ & $-$ & $-$ \\
 J0942$-$5552 & PLRN & 523.1 & $-9.0^{+0.2}_{-0.2}$ & $5.9^{+1.6}_{-1.1}$ \\
 J0942$-$5657 & PLRN & 26.5 & $-13.1^{+2.6}_{-1.0}$ & $17.4^{+2.6}_{-11.6}$ \\
 J0944$-$1354 & WTN & $-$ & $-$ & $-$ \\
 J0953$+$0755 & WTN & $-$ & $-$ & $-$ \\
 J0955$-$5304 & WTN & $-$ & $-$ & $-$ \\
 J0959$-$4809 & WTN & $-$ & $-$ & $-$ \\
 J1001$-$5507 & PLRN & 492.9 & $-9.1^{+0.2}_{-0.1}$ & $4.6^{+1.2}_{-0.8}$ \\
 J1003$-$4747 & WTN & $-$ & $-$ & $-$ \\
 J1012$-$5857 & WTN & $-$ & $-$ & $-$ \\
 J1013$-$5934 & WTN & $-$ & $-$ & $-$ \\
 J1016$-$5345 & WTN & $-$ & $-$ & $-$ \\
 J1017$-$5621 & WTN & $-$ & $-$ & $-$ \\
 J1017$-$7156$^{\star}$ & WTN & $-$ & $-$ & $-$ \\
 J1022$+$1001$^{\star}$ & WTN & $-$ & $-$ & $-$ \\
 J1032$-$5911 & WTN & $-$ & $-$ & $-$ \\
 J1034$-$3224 & WTN & $-$ & $-$ & $-$ \\
 J1036$-$4926 & WTN & $-$ & $-$ & $-$ \\
 J1041$-$1942 & WTN & $-$ & $-$ & $-$ \\
 J1042$-$5521 & WTN & $-$ & $-$ & $-$ \\
 J1043$-$6116 & WTN & $-$ & $-$ & $-$ \\
 J1045$-$4509$^{\star}$ & WTN & $-$ & $-$ & $-$ \\
 J1046$-$5813 & PLRN & 7.0 & $-13.2^{+2.9}_{-1.1}$ & $18.3^{+1.6}_{-14.4}$ \\
 J1047$-$6709 & WTN & $-$ & $-$ & $-$ \\
 J1048$-$5832 & PLRN & 1258.2 & $-8.6^{+0.2}_{-0.1}$ & $6.3^{+1.2}_{-1.0}$ \\
 J1056$-$6258 & PLRN & 297.3 & $-9.7^{+0.2}_{-0.2}$ & $2.9^{+1.1}_{-1.0}$ \\
 J1057$-$5226 & PLRN & 267.5 & $-9.9^{+0.2}_{-0.2}$ & $5.9^{+2.3}_{-1.5}$ \\
 J1057$-$7914 & WTN & $-$ & $-$ & $-$ \\
 J1059$-$5742 & WTN & $-$ & $-$ & $-$ \\
 J1105$-$6107 & PLRN & 347.5 & $-8.9^{+0.3}_{-0.2}$ & $4.1^{+1.7}_{-1.2}$ \\
         \hline
     \end{tabular}
     \renewcommand{\arraystretch}{}
 \end{table}

 \begin{table}
   \centering
   \contcaption{}
     \label{tab:timing_noise}
     \renewcommand{\arraystretch}{1.5}
     \begin{tabular}{llcccc}
         \hline
         \hline
         PSRJ & Model & $\ln(\mathcal{B})$ & $\log_{10}(A)$ & $\beta$ \\
              & & & ($\mathrm{yr}^{3/2}$) & \\
         \hline
 J1110$-$5637 & PLRN & 49.0 & $-10.3^{+0.5}_{-0.9}$ & $6.5^{+5.7}_{-3.1}$ \\
 J1112$-$6613 & PLRN & 35.2 & $-9.6^{+0.3}_{-0.2}$ & $3.5^{+2.8}_{-1.7}$ \\
 J1112$-$6926 & WTN & $-$ & $-$ & $-$ \\
 J1114$-$6100 & WTN & $-$ & $-$ & $-$ \\
 J1116$-$4122 & PLRN & 4.8 & $-13.9^{+3.8}_{-0.8}$ & $17.0^{+2.1}_{-5.2}$ \\
 J1121$-$5444 & PLRN & 111.1 & $-9.8^{+0.3}_{-0.3}$ & $6.0^{+3.6}_{-2.1}$ \\
 J1123$-$6259 & WTN & $-$ & $-$ & $-$ \\
 J1126$-$6942 & WTN & $-$ & $-$ & $-$ \\
 J1133$-$6250 & WTN & $-$ & $-$ & $-$ \\
 J1136$+$1551 & PLRN & 6.5 & $-10.4^{+0.5}_{-3.9}$ & $4.0^{+15.9}_{-2.1}$ \\
 J1136$-$5525 & PLRN & 174.3 & $-10.2^{+0.5}_{-0.6}$ & $7.7^{+4.3}_{-2.5}$ \\
 J1141$-$3322 & WTN & $-$ & $-$ & $-$ \\
 J1141$-$6545 & PLRN & 186.8 & $-10.3^{+0.4}_{-0.5}$ & $4.7^{+3.2}_{-1.9}$ \\
 J1146$-$6030 & WTN & $-$ & $-$ & $-$ \\
 J1157$-$6224 & PLRN & 97.5 & $-10.1^{+0.2}_{-0.2}$ & $3.3^{+1.8}_{-1.4}$ \\
 J1202$-$5820 & PLRN & 69.2 & $-10.3^{+0.4}_{-0.6}$ & $5.3^{+4.0}_{-2.0}$ \\
 J1210$-$5559 & PLRN & 4.5 & $-14.7^{+3.2}_{-1.4}$ & $13.8^{+6.1}_{-10.4}$ \\
 J1224$-$6407 & PLRN & 372.6 & $-10.7^{+0.2}_{-0.2}$ & $5.0^{+1.9}_{-1.5}$ \\
 J1231$-$6303 & WTN & $-$ & $-$ & $-$ \\
 J1239$-$6832 & WTN & $-$ & $-$ & $-$ \\
 J1243$-$6423 & PLRN & 950.4 & $-10.2^{+0.2}_{-0.2}$ & $4.5^{+1.0}_{-0.8}$ \\
 J1253$-$5820 & PLRN & 67.0 & $-10.7^{+0.5}_{-0.7}$ & $5.6^{+4.5}_{-2.7}$ \\
 J1257$-$1027 & WTN & $-$ & $-$ & $-$ \\
 J1259$-$6741 & WTN & $-$ & $-$ & $-$ \\
 J1305$-$6455 & PLRN & 197.0 & $-9.7^{+0.3}_{-0.6}$ & $4.9^{+3.0}_{-1.6}$ \\
 J1306$-$6617 & PLRN & 7.8 & $-11.5^{+1.6}_{-2.6}$ & $16.5^{+3.4}_{-12.3}$ \\
 J1312$-$5402 & WTN & $-$ & $-$ & $-$ \\
 J1312$-$5516 & WTN & $-$ & $-$ & $-$ \\
 J1319$-$6056 & PLRN & 24.8 & $-10.3^{+0.4}_{-3.7}$ & $3.1^{+15.5}_{--3.4}$ \\
 J1320$-$5359 & PLRN & 43.1 & $-13.8^{+3.1}_{-0.9}$ & $19.4^{+0.6}_{-13.4}$ \\
 J1326$-$5859 & PLRN & 718.2 & $-10.1^{+0.3}_{-0.2}$ & $5.4^{+1.3}_{-1.0}$ \\
 J1326$-$6408 & WTN & $-$ & $-$ & $-$ \\
 J1326$-$6700 & PLRN & 107.9 & $-9.3^{+0.2}_{-0.2}$ & $3.5^{+1.8}_{-1.4}$ \\
 J1327$-$6222 & PLRN & 946.9 & $-9.1^{+0.2}_{-0.2}$ & $4.2^{+1.1}_{-1.0}$ \\
 J1327$-$6301 & WTN & $-$ & $-$ & $-$ \\
 J1328$-$4357 & PLRN & 11.0 & $-13.2^{+3.0}_{-0.8}$ & $19.9^{+0.0}_{-15.4}$ \\
 J1338$-$6204 & WTN & $-$ & $-$ & $-$ \\
 J1350$-$5115 & WTN & $-$ & $-$ & $-$ \\
 J1355$-$5153 & PLRN & 4.1 & $-13.4^{+2.9}_{-1.0}$ & $16.4^{+3.5}_{-11.8}$ \\
 J1356$-$5521 & WTN & $-$ & $-$ & $-$ \\
 J1359$-$6038 & PLRN & 1556.6 & $-10.0^{+0.2}_{-0.1}$ & $5.1^{+0.9}_{-0.8}$ \\
         \hline
     \end{tabular}
     \renewcommand{\arraystretch}{}
 \end{table}

 \begin{table}
   \centering
   \contcaption{}
     \label{tab:timing_noise}
     \renewcommand{\arraystretch}{1.5}
     \begin{tabular}{llcccc}
         \hline
         \hline
         PSRJ & Model & $\ln(\mathcal{B})$ & $\log_{10}(A)$ & $\beta$ \\
              & & & ($\mathrm{yr}^{3/2}$) & \\
         \hline
 J1401$-$6357 & PLRN & 693.7 & $-9.8^{+0.3}_{-0.3}$ & $7.5^{+2.9}_{-2.1}$ \\
 J1413$-$6307 & PLRN & 143.4 & $-9.4^{+0.4}_{-0.3}$ & $4.6^{+3.0}_{-2.4}$ \\
 J1418$-$3921 & WTN & $-$ & $-$ & $-$ \\
 J1420$-$5416 & WTN & $-$ & $-$ & $-$ \\
 J1424$-$5822 & WTN & $-$ & $-$ & $-$ \\
 J1428$-$5530 & WTN & $-$ & $-$ & $-$ \\
 J1430$-$6623 & PLRN & 26.4 & $-11.0^{+0.2}_{-0.0}$ & $3.2^{+7.2}_{--0.2}$ \\
 J1435$-$5954 & WTN & $-$ & $-$ & $-$ \\
 J1452$-$6036 & WTN & $-$ & $-$ & $-$ \\
 J1453$-$6413 & PLRN & 156.9 & $-10.9^{+0.2}_{-0.2}$ & $3.2^{+1.5}_{-1.3}$ \\
 J1456$-$6843 & WTN & $-$ & $-$ & $-$ \\
 J1457$-$5122 & WTN & $-$ & $-$ & $-$ \\
 J1507$-$4352 & PLRN & 16.2 & $-10.3^{+0.4}_{-0.9}$ & $3.6^{+5.5}_{-2.5}$ \\
 J1507$-$6640 & WTN & $-$ & $-$ & $-$ \\
 J1511$-$5414 & WTN & $-$ & $-$ & $-$ \\
 J1512$-$5759 & PLRN & 254.7 & $-9.9^{+0.3}_{-0.4}$ & $6.8^{+3.9}_{-2.4}$ \\
 J1514$-$4834 & WTN & $-$ & $-$ & $-$ \\
 J1522$-$5829 & PLRN & 28.9 & $-12.3^{+2.1}_{-1.6}$ & $11.0^{+9.0}_{-6.3}$ \\
 J1527$-$3931 & WTN & $-$ & $-$ & $-$ \\
 J1527$-$5552 & PLRN & 11.8 & $-10.0^{+0.2}_{-4.1}$ & $17.1^{+2.8}_{-15.8}$ \\
 J1528$-$3146$^{\star}$ & WTN & $-$ & $-$ & $-$ \\
 J1534$-$5334 & WTN & $-$ & $-$ & $-$ \\
 J1534$-$5405 & PLRN & 37.8 & $-9.7^{+0.3}_{-0.3}$ & $6.1^{+12.2}_{--1.8}$ \\
 J1539$-$5626 & PLRN & 17.2 & $-9.8^{+0.3}_{-0.2}$ & $3.1^{+2.7}_{-2.4}$ \\
 J1542$-$5034 & PLRN & 10.0 & $-11.5^{+2.4}_{-0.6}$ & $15.3^{+4.6}_{-11.1}$ \\
 J1543$+$0929 & WTN & $-$ & $-$ & $-$ \\
 J1544$-$5308 & WTN & $-$ & $-$ & $-$ \\
 J1549$-$4848 & WTN & $-$ & $-$ & $-$ \\
 J1553$-$5456 & WTN & $-$ & $-$ & $-$ \\
 J1555$-$3134 & WTN & $-$ & $-$ & $-$ \\
 J1557$-$4258 & PLRN & 10.7 & $-12.3^{+1.5}_{-1.2}$ & $8.7^{+5.0}_{-5.2}$ \\
 J1559$-$4438 & PLRN & 5.7 & $-10.8^{+0.7}_{-2.7}$ & $3.9^{+9.5}_{--0.7}$ \\
 J1559$-$5545 & PLRN & 221.5 & $-8.8^{+0.2}_{-0.2}$ & $4.9^{+1.6}_{-1.5}$ \\
 J1600$-$3053$^{\star}$ & WTN & $-$ & $-$ & $-$ \\
 J1600$-$5044 & PLRN & 270.2 & $-10.2^{+0.3}_{-0.4}$ & $6.1^{+3.4}_{-2.1}$ \\
 J1603$-$2531 & WTN & $-$ & $-$ & $-$ \\
 J1603$-$2712 & WTN & $-$ & $-$ & $-$ \\
 J1603$-$7202$^{\star}$ & WTN & $-$ & $-$ & $-$ \\
 J1604$-$4909 & PLRN & 133.1 & $-10.3^{+0.3}_{-0.5}$ & $5.4^{+2.9}_{-1.4}$ \\
 J1605$-$5257 & WTN & $-$ & $-$ & $-$ \\
 J1613$-$4714 & WTN & $-$ & $-$ & $-$ \\
         \hline
     \end{tabular}
     \renewcommand{\arraystretch}{}
 \end{table}

 \begin{table}
   \centering
   \contcaption{}
     \label{tab:timing_noise}
     \renewcommand{\arraystretch}{1.5}
     \begin{tabular}{llcccc}
         \hline
         \hline
         PSRJ & Model & $\ln(\mathcal{B})$ & $\log_{10}(A)$ & $\beta$ \\
              & & & ($\mathrm{yr}^{3/2}$) & \\
         \hline
 J1622$-$4950 & PLRN & 211.8 & $-4.9^{+0.6}_{-0.4}$ & $7.3^{+3.4}_{-3.6}$ \\
 J1623$-$0908 & WTN & $-$ & $-$ & $-$ \\
 J1623$-$4256 & PLRN & 21.2 & $-12.6^{+2.3}_{-1.6}$ & $19.9^{+0.0}_{-14.2}$ \\
 J1626$-$4537 & WTN & $-$ & $-$ & $-$ \\
 J1633$-$4453 & WTN & $-$ & $-$ & $-$ \\
 J1633$-$5015 & WTN & $-$ & $-$ & $-$ \\
 J1639$-$4604 & WTN & $-$ & $-$ & $-$ \\
 J1644$-$4559 & PLRN & 2519.7 & $-9.9^{+0.2}_{-0.2}$ & $6.2^{+1.0}_{-0.9}$ \\
 J1646$-$6831 & WTN & $-$ & $-$ & $-$ \\
 J1651$-$4246 & PLRN & 125.1 & $-12.8^{+2.5}_{-1.2}$ & $19.7^{+0.2}_{-12.9}$ \\
 J1651$-$5222 & WTN & $-$ & $-$ & $-$ \\
 J1651$-$5255 & PLRN & 23.6 & $-12.2^{+2.5}_{-1.3}$ & $12.6^{+7.3}_{-8.1}$ \\
 J1652$-$2404 & WTN & $-$ & $-$ & $-$ \\
 J1700$-$3312 & WTN & $-$ & $-$ & $-$ \\
 J1701$-$3726 & WTN & $-$ & $-$ & $-$ \\
 J1703$-$1846 & WTN & $-$ & $-$ & $-$ \\
 J1703$-$3241 & WTN & $-$ & $-$ & $-$ \\
 J1703$-$4851 & WTN & $-$ & $-$ & $-$ \\
 J1705$-$1906 & PLRN & 58.2 & $-10.5^{+0.4}_{-0.6}$ & $4.8^{+3.9}_{-2.1}$ \\
 J1705$-$3423 & PLRN & 27.3 & $-10.8^{+0.6}_{-3.6}$ & $6.2^{+13.7}_{-2.7}$ \\
 J1707$-$4053 & WTN & $-$ & $-$ & $-$ \\
 J1708$-$3426 & WTN & $-$ & $-$ & $-$ \\
 J1709$-$1640 & PLRN & 48.5 & $-9.8^{+0.2}_{-0.4}$ & $3.9^{+14.1}_{-1.4}$ \\
 J1709$-$4429 & PLRN & 504.8 & $-9.1^{+0.3}_{-0.4}$ & $8.0^{+2.6}_{-1.6}$ \\
 J1711$-$5350 & PLRN & 8.4 & $-12.6^{+2.9}_{-1.0}$ & $14.7^{+5.3}_{-11.0}$ \\
 J1715$-$4034 & WTN & $-$ & $-$ & $-$ \\
 J1717$-$3425 & PLRN & 10.9 & $-9.4^{+0.3}_{-2.9}$ & $6.2^{+13.7}_{-3.3}$ \\
 J1717$-$4054 & PLRN & 40.7 & $-11.0^{+1.6}_{-2.9}$ & $11.4^{+8.6}_{-6.6}$ \\
 J1720$-$1633 & WTN & $-$ & $-$ & $-$ \\
 J1720$-$2933 & WTN & $-$ & $-$ & $-$ \\
 J1722$-$3207 & PLRN & 6.6 & $-13.8^{+3.1}_{-0.8}$ & $18.0^{+2.0}_{-13.3}$ \\
 J1722$-$3712 & PLRN & 372.6 & $-9.0^{+0.2}_{-0.2}$ & $4.2^{+1.1}_{-0.9}$ \\
 J1727$-$2739 & WTN & $-$ & $-$ & $-$ \\
 J1730$-$2304$^{\star}$ & WTN & $-$ & $-$ & $-$ \\
 J1731$-$4744 & PLRN & 195.4 & $-9.5^{+0.2}_{-0.2}$ & $5.0^{+1.5}_{-1.4}$ \\
 J1733$-$2228 & WTN & $-$ & $-$ & $-$ \\
 J1736$-$2457 & WTN & $-$ & $-$ & $-$ \\
 J1739$-$2903 & PLRN & 19.7 & $-13.1^{+2.7}_{-1.1}$ & $19.5^{+0.4}_{-14.5}$ \\
 J1740$-$3015 & PLRN & 128.8 & $-8.9^{+0.3}_{-0.2}$ & $5.2^{+1.0}_{-1.4}$ \\
 J1741$-$3927 & PLRN & 159.5 & $-9.8^{+0.3}_{-0.3}$ & $6.3^{+2.5}_{-1.6}$ \\
 J1743$-$3150 & WTN & $-$ & $-$ & $-$ \\
         \hline
     \end{tabular}
     \renewcommand{\arraystretch}{}
 \end{table}

 \begin{table}
   \centering
   \contcaption{}
     \label{tab:timing_noise}
     \renewcommand{\arraystretch}{1.5}
     \begin{tabular}{llcccc}
         \hline
         \hline
         PSRJ & Model & $\ln(\mathcal{B})$ & $\log_{10}(A)$ & $\beta$ \\
              & & & ($\mathrm{yr}^{3/2}$) & \\
         \hline
 J1745$-$3040 & PLRN & 68.7 & $-14.3^{+3.3}_{-0.7}$ & $18.9^{+0.5}_{-5.2}$ \\
 J1751$-$4657 & WTN & $-$ & $-$ & $-$ \\
 J1752$-$2806 & PLRN & 292.9 & $-9.6^{+0.2}_{-0.1}$ & $2.9^{+0.8}_{-0.7}$ \\
 J1757$-$2421 & WTN & $-$ & $-$ & $-$ \\
 J1759$-$2205 & PLRN & 54.9 & $-10.1^{+0.3}_{-0.3}$ & $4.4^{+2.0}_{-1.6}$ \\
 J1759$-$3107 & WTN & $-$ & $-$ & $-$ \\
 J1801$-$0357 & WTN & $-$ & $-$ & $-$ \\
 J1801$-$2920 & WTN & $-$ & $-$ & $-$ \\
 J1803$-$2137 & PLRN & 41.3 & $-8.6^{+0.5}_{-0.5}$ & $17.9^{+2.0}_{-9.8}$ \\
 J1805$-$1504 & WTN & $-$ & $-$ & $-$ \\
 J1807$-$0847 & WTN & $-$ & $-$ & $-$ \\
 J1807$-$2715 & PLRN & 5.2 & $-12.2^{+2.1}_{-1.9}$ & $17.0^{+3.0}_{-12.8}$ \\
 J1808$-$0813 & WTN & $-$ & $-$ & $-$ \\
 J1809$-$2109 & WTN & $-$ & $-$ & $-$ \\
 J1810$-$5338 & WTN & $-$ & $-$ & $-$ \\
 J1816$-$2650 & WTN & $-$ & $-$ & $-$ \\
 J1818$-$1422 & WTN & $-$ & $-$ & $-$ \\
 J1820$-$0427 & PLRN & 99.8 & $-10.1^{+0.3}_{-0.5}$ & $5.7^{+3.3}_{-2.0}$ \\
 J1822$-$2256 & WTN & $-$ & $-$ & $-$ \\
 J1823$-$0154 & WTN & $-$ & $-$ & $-$ \\
 J1823$-$1115 & WTN & $-$ & $-$ & $-$ \\
 J1823$-$3106 & PLRN & 33.1 & $-10.3^{+0.3}_{-3.5}$ & $4.3^{+14.6}_{--4.9}$ \\
 J1824$-$0127 & WTN & $-$ & $-$ & $-$ \\
 J1824$-$1945 & PLRN & 327.7 & $-10.0^{+0.2}_{-0.2}$ & $6.1^{+1.2}_{-1.2}$ \\
 J1825$-$0935 & PLRN & 570.6 & $-9.0^{+0.2}_{-0.2}$ & $4.9^{+2.1}_{-1.0}$ \\
 J1827$-$0750 & PLRN & 30.5 & $-10.9^{+1.4}_{-1.4}$ & $18.8^{+1.1}_{-13.2}$ \\
 J1829$-$1751 & PLRN & 187.9 & $-9.6^{+0.2}_{-0.2}$ & $6.0^{+2.6}_{-1.7}$ \\
 J1830$-$1135 & WTN & $-$ & $-$ & $-$ \\
 J1832$-$0827 & PLRN & 67.8 & $-10.4^{+0.6}_{-1.4}$ & $6.1^{+7.9}_{-3.0}$ \\
 J1833$-$0338 & PLRN & 254.7 & $-9.6^{+0.2}_{-0.2}$ & $5.3^{+1.6}_{-1.1}$ \\
 J1833$-$0827 & PLRN & 28.4 & $-11.9^{+1.8}_{-1.7}$ & $15.4^{+4.5}_{-10.5}$ \\
 J1834$-$0426 & WTN & $-$ & $-$ & $-$ \\
 J1835$-$1020 & PLRN & 84.7 & $-8.9^{+0.3}_{-0.1}$ & $3.5^{+3.2}_{-1.3}$ \\
 J1836$-$0436 & WTN & $-$ & $-$ & $-$ \\
 J1836$-$1008 & PLRN & 85.4 & $-10.3^{+0.7}_{-0.9}$ & $7.8^{+3.5}_{-4.1}$ \\
 J1837$-$0653 & WTN & $-$ & $-$ & $-$ \\
 J1840$-$0809 & WTN & $-$ & $-$ & $-$ \\
 J1840$-$0815 & WTN & $-$ & $-$ & $-$ \\
 J1841$-$0425 & WTN & $-$ & $-$ & $-$ \\
 J1841$+$0912 & PLRN & 19.3 & $-9.8^{+0.4}_{-0.6}$ & $4.3^{+4.2}_{-2.5}$ \\
 J1842$-$0359 & WTN & $-$ & $-$ & $-$ \\
         \hline
     \end{tabular}
     \renewcommand{\arraystretch}{}
 \end{table}

 \begin{table}
   \centering
   \contcaption{}
     \label{tab:timing_noise}
     \renewcommand{\arraystretch}{1.5}
     \begin{tabular}{llcccc}
         \hline
         \hline
         PSRJ & Model & $\ln(\mathcal{B})$ & $\log_{10}(A)$ & $\beta$ \\
              & & & ($\mathrm{yr}^{3/2}$) & \\
         \hline
 J1843$-$0000 & PLRN & 6.3 & $-12.7^{+3.0}_{-1.0}$ & $19.0^{+0.9}_{-15.2}$ \\
 J1844$-$0433 & WTN & $-$ & $-$ & $-$ \\
 J1845$-$0743 & WTN & $-$ & $-$ & $-$ \\
 J1847$-$0402 & PLRN & 15.1 & $-10.5^{+0.5}_{-4.3}$ & $4.3^{+15.6}_{-1.8}$ \\
 J1848$-$0123 & PLRN & 46.7 & $-10.7^{+0.7}_{-3.3}$ & $5.7^{+12.0}_{--1.5}$ \\
 J1849$-$0636 & PLRN & 13.7 & $-12.7^{+2.6}_{-1.5}$ & $18.2^{+1.7}_{-14.3}$ \\
 J1852$-$0635 & WTN & $-$ & $-$ & $-$ \\
 J1852$-$2610 & WTN & $-$ & $-$ & $-$ \\
 J1857$+$0212 & WTN & $-$ & $-$ & $-$ \\
 J1900$-$2600 & WTN & $-$ & $-$ & $-$ \\
 J1901$+$0331 & PLRN & 277.3 & $-9.5^{+0.2}_{-0.2}$ & $4.4^{+1.3}_{-1.3}$ \\
 J1901$+$0716 & PLRN & 4.1 & $-11.4^{+2.3}_{-1.3}$ & $17.8^{+2.2}_{-13.9}$ \\
 J1901$-$0906 & WTN & $-$ & $-$ & $-$ \\
 J1902$+$0556 & WTN & $-$ & $-$ & $-$ \\
 J1902$+$0615 & WTN & $-$ & $-$ & $-$ \\
 J1903$+$0135 & PLRN & 74.4 & $-13.4^{+2.5}_{-0.8}$ & $20.0^{+0.0}_{-12.4}$ \\
 J1903$-$0632 & PLRN & 10.2 & $-10.1^{+0.3}_{-4.0}$ & $2.7^{+16.4}_{--8.1}$ \\
 J1905$-$0056 & WTN & $-$ & $-$ & $-$ \\
 J1909$+$0007 & PLRN & 64.3 & $-10.3^{+0.3}_{-0.5}$ & $6.6^{+10.9}_{--1.6}$ \\
 J1909$+$0254 & WTN & $-$ & $-$ & $-$ \\
 J1909$+$1102 & PLRN & 183.9 & $-10.6^{+0.5}_{-0.5}$ & $7.9^{+3.5}_{-2.5}$ \\
 J1909$-$3744$^{\star\dagger}$ & WTN & $-$ & $-$ & $-$ \\
 J1910$-$0309 & WTN & $-$ & $-$ & $-$ \\
 J1910$+$0358 & WTN & $-$ & $-$ & $-$ \\
 J1913$-$0440 & PLRN & 175.6 & $-10.9^{+0.6}_{-0.7}$ & $7.5^{+5.0}_{-2.4}$ \\
 J1913$+$1400 & WTN & $-$ & $-$ & $-$ \\
 J1915$+$1009 & WTN & $-$ & $-$ & $-$ \\
 J1916$+$0951 & PLRN & 19.7 & $-13.2^{+2.7}_{-1.2}$ & $16.5^{+3.5}_{-12.2}$ \\
 J1916$+$1312 & PLRN & 111.7 & $-9.3^{+0.3}_{-0.3}$ & $6.0^{+3.3}_{-1.8}$ \\
 J1917$+$1353 & PLRN & 74.0 & $-10.1^{+0.3}_{-0.2}$ & $3.7^{+1.9}_{-1.4}$ \\
 J1919$+$0021 & WTN & $-$ & $-$ & $-$ \\
 J1926$+$0431 & WTN & $-$ & $-$ & $-$ \\
 J1932$+$1059 & PLRN & 206.5 & $-10.4^{+0.3}_{-0.2}$ & $5.4^{+2.5}_{-1.6}$ \\
 J1932$-$3655 & WTN & $-$ & $-$ & $-$ \\
 J1935$+$1616 & PLRN & 31.3 & $-10.8^{+0.3}_{-2.4}$ & $4.5^{+15.5}_{-1.4}$ \\
 J1941$-$2602 & WTN & $-$ & $-$ & $-$ \\
 J1943$-$1237 & WTN & $-$ & $-$ & $-$ \\
 J1945$-$0040 & WTN & $-$ & $-$ & $-$ \\
 J1946$-$2913 & WTN & $-$ & $-$ & $-$ \\
 J2006$-$0807 & WTN & $-$ & $-$ & $-$ \\
 J2033$+$0042 & WTN & $-$ & $-$ & $-$ \\
         \hline
     \end{tabular}
     \renewcommand{\arraystretch}{}
 \end{table}

 \begin{table}
   \centering
   \contcaption{}
     \label{tab:timing_noise}
     \renewcommand{\arraystretch}{1.5}
     \begin{tabular}{llcccc}
         \hline
         \hline
         PSRJ & Model & $\ln(\mathcal{B})$ & $\log_{10}(A)$ & $\beta$ \\
              & & & ($\mathrm{yr}^{3/2}$) & \\
         \hline
 J2038$-$3816 & WTN & $-$ & $-$ & $-$ \\
 J2046$-$0421 & WTN & $-$ & $-$ & $-$ \\
 J2046$+$1540 & WTN & $-$ & $-$ & $-$ \\
 J2048$-$1616 & WTN & $-$ & $-$ & $-$ \\
 J2051$-$0827$^{\star}$ & WTN & $-$ & $-$ & $-$ \\
 J2053$-$7200 & WTN & $-$ & $-$ & $-$ \\
 J2116$+$1414 & WTN & $-$ & $-$ & $-$ \\
 J2129$-$5721$^{\star}$ & WTN & $-$ & $-$ & $-$ \\
 J2144$-$3933 & WTN & $-$ & $-$ & $-$ \\
 J2145$-$0750$^{\star}$ & PLRN & 33.1 & $-11.3^{+0.3}_{-0.3}$ & $4.8^{+3.3}_{-2.7}$ \\
 J2155$-$3118 & WTN & $-$ & $-$ & $-$ \\
 J2222$-$0137$^{\star}$ & WTN & $-$ & $-$ & $-$ \\
 J2241$-$5236$^{\star}$ & PLRN & 8.7 & $-12.1^{+0.3}_{-0.3}$ & $0.4^{+2.0}_{-0.4}$ \\
 J2248$-$0101 & WTN & $-$ & $-$ & $-$ \\
 J2324$-$6054 & WTN & $-$ & $-$ & $-$ \\
 J2330$-$2005 & WTN & $-$ & $-$ & $-$ \\
 J2346$-$0609 & PLRN & 49.1 & $-12.7^{+2.4}_{-1.3}$ & $19.1^{+0.9}_{-13.7}$ \\
         \hline
     \end{tabular}
     \renewcommand{\arraystretch}{}
 \end{table}


\bsp    
\label{lastpage}
\end{document}